\documentclass[12pt,twoside]{article}

\usepackage{a4wide}
\usepackage{amsmath}
\usepackage{amssymb}
\usepackage[dvips]{graphicx}
\newcommand{\beq}{\begin{equation}}
\newcommand{\eeq}{\end{equation}}
\newcommand{\bdm}{\begin{displaymath}}
\newcommand{\edm}{\end{displaymath}}
\newcommand{\beqa}{\begin{eqnarray}}
\newcommand{\eeqa}{\end{eqnarray}}
\newcommand{\beqan}{\begin{eqnarray*}}
\newcommand{\eeqan}{\end{eqnarray*}}

\newcommand{\trace}[1]{Tr\left[#1\right]}                      
\newcommand{\real}[1]{\mbox{${\cal R}e\,#1$}~\!\!}                 
\newcommand{\imag}[1]{\mbox{${\cal I}m\,#1$}~\!\!}                
\newcommand{\expt}[1]{\left\langle #1 \right\rangle}	
\newcommand{\D}{\displaystyle}

\begin{document}

\begin{center}
{\Large \bf Hadronic Spectral Functions in Nuclear Matter}
\footnote{Work supported by DFG and GSI Darmstadt.}
\bigskip
\bigskip

{M. Post, S. Leupold and U. Mosel}

\bigskip
\bigskip
{ \it
Institut f\"ur Theoretische Physik, Universit\"at Giessen,\\
D-35392 Giessen, Germany
}
\end{center}

\bigskip
\bigskip


\begin{abstract}
We study the in-medium properties of mesons $(\pi,\eta,\rho)$ and baryon resonances in cold nuclear matter within a coupled-channel analysis. The meson self energies are generated by particle-hole excitations. Thus multi-peak spectra are obtained for the mesonic spectral functions. In turn this leads to medium-modifications of the baryon resonances. Special care is taken to respect the analyticity of the spectral functions and to take into account effects from short-range correlations both for positive and negative parity states. Our model produces sensible results for pion and $\Delta$ dynamics in nuclear matter. We find a strong interplay of the $\rho$ meson and the $D_{13}(1520)$, which moves spectral strength of the $\rho$ spectrum to smaller invariant masses and leads to a broadening of the baryon resonance. The optical potential for the $\eta$ meson resulting from our model is rather attractive whereas the in-medium properties modifications of the $S_{11}(1535)$ are found to be quite small.  \newline



\end{abstract}


\newpage

\section{Introduction}

A wealth of experimental indications for in-medium modifications of hadrons has been accumulated over the last years. Besides serving as a suitable testing ground for the understanding of hadronic interactions in the nuclear medium, the search for medium modifications has been stimulated by the works of \cite{brownrho,sumrulehatsuda}, which -- based on very general arguments concerning the restoration of chiral symmetry -- predict dropping hadron masses at finite density. Dilepton spectra measured in heavy-ion collisions by the NA45 \cite{aga1,aga2,lenkeit,wessels} and the HELIOS \cite{masera} collaboration indicate that the spectral function of the $\rho$ meson may undergo a significant reshaping in a hot and dense environment with spectral strength moving down to smaller invariant masses. In nuclear reactions medium modifications of the $P_{33}(1232)$, the $D_{13}(1520)$ and the $S_{11}(1535)$ have been studied. An analysis of pion- and photo-induced reactions has established the need to introduce a spreading potential for the $P_{33}(1232)$, yielding a moderate broadening of about $80$ MeV for this state \cite{hirata}. The disappearance of the second resonance region in photoabsorption reactions on the nucleus as observed in \cite{bianchi1, bianchi2, frommhold} has been interpreted in terms of a broadening of the $D_{13}(1520)$ in nuclear matter \cite{kondrat}. Finally, data of $\eta$ photoproduction on nuclei \cite{metageta,yorita} have opened up the possibility to study the in-medium properties of the $S_{11}(1535)$. A prime source of information of pions and $\eta$ mesons in nuclear matter is the study of pionic \cite{gal} and $\eta$-mesic atoms \cite{hayano}.

At the same time numerous theoretical models have been developed in order to arrive at an understanding of the observed phenomena. Concerning the in-medium properties of the $\rho$ meson, the dilepton spectra reported in \cite{aga1,aga2,lenkeit,wessels,masera} have triggered the development of a variety of hadronic models. For a comprehensive review of these models see \cite{rwrev}. Although these works differ quite substantially in details, as a general picture a shift of spectral strength down to smaller invariant masses is found in most of them. In one type of models \cite{fripir,postrho1,postrho2,lutzvector,lutzvector1} this shift is generated by the excitation of resonance-hole pairs in the nuclear medium. The formation of these states leads to additional branches of the spectral function. Coupling the $\rho$ to the $D_{13}(1520)N^{-1}$ state moves a lot of spectral strength down to small invariant masses \cite{postrho1,postrho2}. Another class of models \cite{herrmann,rw,rwurban,klinglweise,osetrho} takes into account the effects of the renormalization of the pion cloud generated by the strong interaction of pions and nucleons and finds a broadening of the $\rho$ peak. Besides offering an appealing interpretation of the dilepton spectra, a shift of spectral strength as offered by most hadronic models is also required by QCD sum rules \cite{sumrulehatsuda,klinglweise,sumrulestefan}. 

The in-medium properties of the $P_{33}(1232)$ resonance have been studied extensively in the literature \cite{salcedo,helgesson1,helgesson2,wehrberger,kim,xia,lutzreso}. While operating on different levels of sophistication, in most of these models the in-medium self energy is due to a change of the dispersion relation of pions in a nuclear environment. For a quantitative description of the resonance properties, a consistent inclusion of short-range correlations (SRC) is necessary \cite{salcedo,helgesson1,helgesson2}. For the $D_{13}(1520)$ and the $S_{11}(1535)$ much less work has been done. In an attempt to explain nuclear photoabsorption data reported in \cite{bianchi1,bianchi2,frommhold}, a large broadening of about $300$ MeV for the $D_{13}(1520)$ has been obtained in a resonance fit in \cite{kondrat}. The later works of \cite{effeabs,effeabs2,rwphotoabs} have given further support to the conjecture that an in-medium broadening of this state yields a possible explanation of the photoabsorption data. As alternative mechanisms effects from Fermi motion and a change of interference patterns in the nuclear medium have been pointed out in the analysis of \cite{hirata2}, thus questioning the direct connection between the data and a broadening of the $D_{13}(1520)$. As a result in that analysis a smaller broadening of about $100$ MeV for this state is found. In the microscopic models of \cite{postrho1,lutzreso} a substantial broadening of the $D_{13}(1520)$ has been generated dynamically, based either on the coupling to the $N\rho$ channel \cite{postrho1} or on the coupling to the $N\pi$ channel \cite{lutzreso}. Concerning the properties of the $S_{11}(1535)$ in nuclear matter, the existing models \cite{oseteta,oseteta2,weiseeta} suggest rather moderate in-medium effects. This finding is well supported from data on photoproduction of $\eta$ off nuclei \cite{lehreta}.

From a theoretical point of view it would be desirable to describe as many in-medium effects as possible within one model in order to arrive at a combined understanding of these phenomena. For example, reshuffling the spectral strength of the $\rho$ meson (as suggested from dilepton spectra) might have an immediate impact on the width of the $D_{13}(1520)$ \cite{postrho1} and can help to explain the nuclear photoabsorption data \cite{effeabs,effeabs2,rwphotoabs}. Similarly, a quantitative analysis of the optical potential of the $\eta$ meson is constrained from the fact that recent data on $\eta$ photoproduction \cite{metageta,yorita} suggest that the in-medium modifications experienced by the $S_{11}(1535)$ are relatively small. To this end we have set up a model which generates the in-medium modifications of mesons and baryon resonances within a self-consistent coupled channel analysis. The mesons are dressed by the excitation of resonance-hole loops and a remarkably complicated spectrum with various peak structures is found for the mesonic spectral functions. In a second step the in-medium self energy of the baryon resonances arising from the dressing of the mesons is calculated. The corresponding set of coupled-channel equations is then solved iteratively. In the course of the iterations one leaves the regime of the low-density theorem \cite{dover}, which relates
the in-medium self energy to vacuum scattering amplitudes. It is least reliable for systems close to threshold, where already small changes of the available phase space can lead to large modifications of the resonance and therefore the meson as well.  A well-known example is the $\Lambda(1405)$ coupling to the ${\bar K}\,N$ channel, see for example \cite{kochkaon,lutzkaon}. Another case is the $\rho N D_{13}(1520)$ system: in a previous publication \cite{postrho1}, a first step in this direction was done and strong effects from the interplay of $\rho$ and $D_{13}(1520)$ were reported, modifying both the $\rho$ spectral function and that of the baryon resonance. We have extended the model presented in \cite{postrho1} in several ways: in order to guarantee the normalization of the vacuum and the in-medium spectral functions, we employ dispersion relations to generate the real part of the self energies. Since most baryon resonances couple strongly to the pion, a complete analysis of their in-medium properties requires also a dressing of the pion. In order to obtain reliable estimates for the $S_{11}(1535)$, which couples dominantly to the $\eta N$ channel, the $\eta$ meson is included as well. Finally, stimulated by the fact that the in-medium width of the $P_{33}(1232)$ needs to be protected by repulsive short-range terms, we have developed a framework that allows for the incorporation of such effects for resonances with negative parity, such as the $D_{13}(1520)$ and the $S_{11}(1535)$.

The paper is organized as follows: in Section \ref{vacself} we discuss the vacuum self energies of the $\rho$ meson and the included baryon resonances. Special emphasis is put on the effect of dispersion relations on the baryonic spectral functions. Section \ref{exp} discusses the current theoretical and experimental status concerning the coupling of the $\rho$ meson to baryon resonances, in particular the $D_{13}(1520)$. In Section \ref{itscheme} we set up the general framework for the discussion of the in-medium self energies of mesons and baryons. The theoretical concepts for the inclusion of  short-range correlations (SRC) are given in Section \ref{nrint}, with details presented in Appendix \ref{srcdetails}. The results obtained for the mesons $\pi,\,\eta$ and $\rho$, as well as for the resonances $P_{33}(1232)$, $D_{13}(1520)$ and $S_{11}(1535)$ are discussed in Section \ref{results}. In Section \ref{conclusions} we summarize our findings. In four Appendices we discuss some necessary technical issues.

\section{Meson and Baryon Resonances in Vacuum}	
\label{vacself}

In this Section we discuss the vacuum spectral functions of the $\rho$ meson and baryon resonances, which are denoted by ${\cal A}(q)$ and $\rho(k)$, respectively.
The spectral function is defined as the imaginary part of the retarded propagator, see Appendix \ref{analytic} and \cite{bd}. In terms of the self energies $\Pi_{vac}^+(q)$ and $\Sigma_{vac}^+(k)$ they are given by:
\beqa
\label{specvac}
	{\cal A}(q)&=&-\frac{1}{\pi}\frac{\imag{\Pi_{vac}^+(q)}}
	{(q^2-m_M^2-\real{\Pi_{vac}^+(q)})^2+\imag{\Pi_{vac}^{+\,2}(q)}}\\
	\rho(k)&=&-\frac{1}{\pi}\frac{\imag{\Sigma_{vac}^+(k)}}
	{(k^2-m_R^2-\real{\Sigma_{vac}^+(k)})^2+\imag{\Sigma_{vac}^{+\,2}(k)}}
			\nonumber \quad.
\eeqa
Throughout this paper, we will denote the four-momentum of meson $M$ by $q=(q_0,{\bf q})$ and that of resonance $R$ by $k=(k_0,{\bf k})$. Note that our ansatz for $\rho(k)$ does not take into account the full Dirac structure of the self energy. A detailed discussion of this topic can be found in \cite{postrho2}.

Both ${\cal A}(q)$ and $\rho(k)$ are normalized quantities. In order to guarantee this within our model, we calculate the retarded self energies $\Pi_{vac}^+$ and $\Sigma_{vac}^+$ in the following way:
\begin{itemize}
	\item calculate the imaginary part of the self energy $\imag{\Pi_{vac}^+}$ and $\imag{\Sigma_{vac}}$ for $q_0 > 0$ by means of Cutkosky's cutting rules
	\item for mesons use the antisymmetry $\imag{\Pi_{vac}^+(-q_0)}=-\imag{\Pi_{vac}^+(q_0)}$
	      (see Appendix A.1) and apply a dispersion relation to obtain the real part of the 					self energy .
	\item as outlined in Appendix \ref{analytic}, self energy and propagator of baryons are symmetric in the vacuum, but not in the nuclear medium. Therefore we neglect the contribution from negative energies to the dispersion integral already in the vacuum.
\end{itemize}
A further discussion of this topic can be found in \cite{leupoldtest}.
The issue of how to obtain normalized spectral functions is of relevance to us since within our coupled channel analysis the spectral function of any state is allowed to influence the spectral function of any other state. This implies that even rather small violations of the normalization can lead to uncontrollable errors in the calculation.

Let us make a purely technical note: throughout this work we will encounter various traces, arising from the spin summation at the meson-nucleon-resonance vertices. From these traces we only keep the leading non-relativistic contribution. In \cite{postrho2} it was
shown that a non-relativistic approach leads to a very good approximation of the fully relativistic results, as long as the kinematical quantities are evaluated in the rest frame of the resonance. A non-relativistic reduction simplifies the expressions for the in-medium self energies, see Chapter \ref{itscheme}. In particular, a consistent relativistic description of the short-range interactions is a formidable task \cite{leinson,lutzshortrange} and - albeit in principle desirable - beyond the scope of this work. 


\subsection{$\rho$ Meson} 
\label{gamma1}

\begin{figure}[t]
\centering
\includegraphics[scale=0.75]{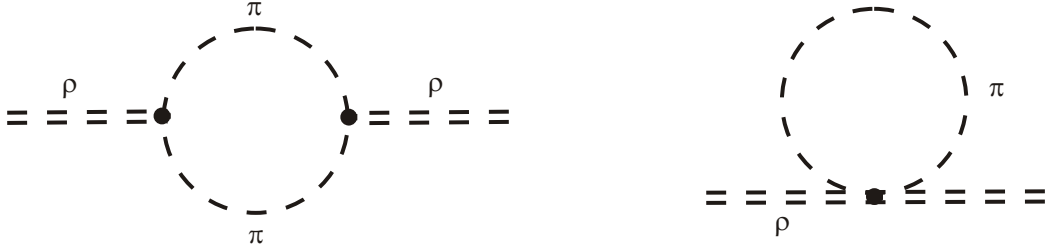}
	\caption{\label{rhoselfvac}Self energy $\Pi_{vac}$ resulting from the coupling of the $\rho$ to pions. The diagram on the left corresponds to the usual decay of a $\rho$ to two pions. The diagram on the right gives rise to an energy independent real mass shift.}
\end{figure}

The propagator of the $\rho$ meson in vacuum has both a four-transverse and a four-longitudinal part, of which in the presence of current conserving couplings only the former gets dressed by a self energy $\Pi_{vac}^+$ \cite{rwrev,herrmann,gale}:
\beqa
	{\cal D}_\rho^{\mu\nu,+}(q) &=& \frac{1}{q^2-(m_\rho^0)^2-\Pi_{vac}^+(q)}\,P_T^{\mu\nu} + \frac{q^\mu\,q^\nu}{q^2}\,\frac{1}{(m_\rho^0)^2} \quad.	
\eeqa
Here $P_T^{\mu\nu}$ is the four-transverse projector:
\beqa
\label{4trans}
P^T_{\mu\nu}(q) &=& g_{\mu\nu}-\frac{q_\mu\,q_\nu}{q^2} \quad.
\eeqa
The self energy $\Pi_{vac}^+(q)$ arises from the coupling of the $\rho$ to pions, which is described by the following Lagrangian \cite{herrmann}:
\beqa
\label{lagrhopi}
	{\cal L}_{\rho\pi} &=& (D_\mu\,\pi)^\star\,(D^\mu\,\pi) - m_\pi^2 \pi^\star \pi - \frac 1 4 \rho_{\mu\nu}\,\rho^{\mu\nu} + \frac 1 2 (m_\rho^0)^2 \rho_\mu\,\rho^\mu \nonumber\\
	\rho_{\mu\nu} &=& \partial_\mu\,\rho_\nu - \partial_\nu\,\rho_\mu \quad,\quad
	D_\mu \,\,=\,\, \partial_\mu + i g_\rho \rho_\mu \nonumber \quad.
	\eeqa
From ${\cal L}_{\rho\pi}$ one derives two Feynman diagrams for $\Pi_{vac}^+(q)$, see Fig. \ref{rhoselfvac}. Concerning the evaluation of these diagrams we use the results from \cite{herrmann}, where the divergent integrals have been treated by a Pauli-Villars regularization in order to preserve the gauge invariance of the self energy:
\beqa
\label{rhoself2}
	\real{\Pi_{vac}^+}(q) &=& -\frac{g_\rho^2}{24 \pi^2}q^2 \left[ {\cal G}(q,m_\pi)- {\cal G}(q,\Lambda) + 4\,(\Lambda^2-m_\pi^2)/q^2 + ln \frac{\Lambda}{m_\pi} \right]
		\\
	\imag{\Pi_{vac}^+}(q) &=& - sgn(q_0)\,\frac{g_\rho^2}{48 \pi} q^2 \left[ 
	\theta(q^2-4\,m_\pi^2)\,\left( 1-\frac{4\,m_\pi^2}{q^2}\right)^{3/2} - \right. \\ && \left. \quad  \quad - 
	\theta(q^2-4\,\Lambda^2)\,\left( 1-\frac{4\,\Lambda^2}{q^2}\right)^{3/2}  \right] 
	\nonumber \quad.
\eeqa
Here $\Lambda$ is a regularization parameter. The function ${\cal G}$ is defined as
\beqa
	{\cal G}(q,m) &=& \left\{ 
	\begin{array}{ll}
	\D{y^{3/2}\,\mbox{arctan}(1/\sqrt{y})} & \mbox{for}\, y\,>\,0 \\ &  \\
	\D{-\frac 1 2 (-y)^{3/2}\,\mbox{ln}\left|\frac{\sqrt{-y}+1}{\sqrt{-y}-1} \right|} 
	& \mbox{for}\, y\,<\,0 
	\end{array} \right. \\
	y &=& \frac{4 m^2}{q^2}-1 \quad. \nonumber 
\eeqa
Note that the real part of the self energy is related to $\imag{\Pi}^+_{vac}$ by a subtracted dispersion relation:
\beqa
\label{realrhovac}
	\real{\Pi_{vac}^+(q)} &=& q^2\,{\cal P}\,\int_{4 m_\pi^2}^\infty\frac{dq^{\prime\,2}}{\pi}\frac{\imag{\Pi_{vac}^+(q^\prime)}}{q^{\prime\,2}(q^2-q^{\prime\,2})} \quad.
\eeqa
The subtraction is made at the point $q=0$ in order to satisfy the condition $\real{\Pi_{vac}^+(q=0)}=0$ as required from gauge invariance \cite{herrmann,klinglweise}.
For the imaginary part the Pauli-Villars prescription acts like a form factor which improves the convergence of the dispersion integral. 
There are three free parameters in this expression, $m_\rho^0$, $g_\rho$ and $\Lambda$, which are determined by fitting the phase shift of $\pi\,\pi$ scattering in the vector-isovector channel and the pion electromagnetic form factor, with the result:
\beq
	m_\rho^0 = 0.875 \mbox{GeV} \,\,,\,\, g_\rho = 6.05 \,\,,\,\, \Lambda = 1 \mbox{GeV} \quad.
\eeq	
As shown in \cite{herrmann}, with these parameters a good fit of both observables is obtained. Summarizing, this model for the $\rho$ meson in vacuum is a good starting point for an in-medium calculation since it provides a spectral function, which is both normalized and in good agreement with observables.


\subsection{Baryon Resonances}
\label{gammabarres}

\begin{figure}[t]
\centering
\includegraphics[scale=0.5]{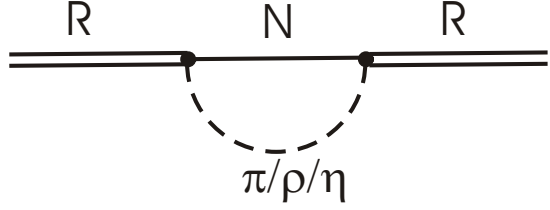}
	\caption{\label{resself}Self energy of a baryon resonance from the decay into a $\pi$, $\eta$ or $\rho$ meson.}
\end{figure}

The self energy of a baryon resonance arises from the coupling to meson-nucleon channels, see Fig. \ref{resself}. After calculating the corresponding decay width, we obtain
$\real{\Sigma_{vac}^+}$ by a dispersion analysis. 

Following standard Feynman rules, the decay width of a nucleon resonance with invariant mass $\sqrt{k^2}$ into a pseudoscalar meson $\varphi = \pi,\,\eta$ of  mass $m_\varphi$ is in the resonance rest frame given by:
\beqa
\label{gamres1}
	\Gamma_{N\varphi}(k) &=&  \frac{1}{2\,j+1}
	\,I_\Sigma\,\left(\frac{f}{m_                      \varphi}\right)^2 \,F^2(k,m_\varphi)\,\frac{q_{cm}}{8\pi k^2}\,\Omega^\varphi \nonumber  \quad.
\eeqa
The coupling constants $f$ are obtained by fits to the corresponding hadronic partial decay widths and $j$ denotes the spin of the decaying resonance. For a complete list of the included resonances see Table \ref{restable}. 
By $q_{cm}$ we denote the momentum of the decay products in the rest frame of the resonance. The isospin factor $I_\Sigma$ is derived form the isospin part of the Lagrangian which is given in Eq. \ref{lagiso} in Appendix \ref{applag}. One finds that $I_\Sigma = 1$
for $\Delta$ resonances with isospin $\frac 3 2$ and $I_\Sigma=3$ for $N^*$ resonances with isospin $\frac 1 2$ if the decay into an isovector $\pi$ meson is considered. For the decay into the isoscalar $\eta$ this factor is $1$ and there is no coupling to $\Delta$ resonances. The quantity $\Omega^\varphi$ is the non-relativistic trace
arising from spin summation at the meson-nucleon-resonance vertices. 
Explicit expressions for $\Omega^{\varphi}$, listed according to the quantum numbers of the resonances considered in this work, can be found in Table \ref{ntraces}, Appendix \ref{lagnrel}. 

When considering the decay into one stable and one unstable particle, an integration over the spectral function of the unstable particle is necessary. For a resonance with fixed mass $\sqrt{k^2}$ and spin $j$ decaying into the $N\rho$ channel, for example, one finds:
\beqa
\label{gamres2}
 \Gamma_{N\rho}(k) &=& \int\limits_{4 m_\pi^2}^{\left(\sqrt{k^2}-m_N\right)^2}\!\!\! dq^2\,
 \Gamma_{N\rho}(k,q)\,{\cal A}_\rho(q) \\
 \Gamma_{N\rho}(k,q) &=& \frac{1}{2\,j+1}
 \,I_\Sigma\,\left(\frac{f}{m_\rho}\right)^2 \,F^2(k,q)\,\frac{q_{cm}}
 {8\pi k^2}\,(2\,\Omega^T+\Omega^L)  \nonumber \quad.
\eeqa
where $\Gamma_{N\rho}(k,q)$ stands for the width of a resonance for decay into a $\rho$ meson with invariant mass $\sqrt{q^2}$. The isospin factor $I_\Sigma$ is the same as for the pions. Explicit expressions for the spin-traces $\Omega^{T/L}$ are found in Table \ref{ntraces}, Appendix \ref{lagnrel}. 

\renewcommand{\arraystretch}{2}
\begin{table}
\bdm
\begin{array}{|c||cc||ccccc|cc|cc||c|}  \hline
            &m              &\Gamma_{tot}     &\Gamma_{N\pi} &\Gamma_{N_\rho}
            &\Gamma_{\Delta\pi} &\Gamma_{N\eta} &\Gamma_{N\omega} 
            &J              &I              	 &l_\varphi       &l_V   &\Lambda_\rho^s\\   
 \hline \hline
P_{11}(1440)     &1.462          &0.391             &0.270          &0              &0.088                          &0              &0                 &1/2           &1/2            &p          &p  &0.8
\\ \hline \hline
P_{33}(1232)&1.232          &0.12             &0.12          &0              &0                            &0              &0                &3/2           &3/2            &p          &p  &0.8
\\ \hline
P_{13}(1720)    &1.717          &0.121            &0.011         &0.11           &0             
           &0              &0                &3/2           &1/2            &p          &p  &1.0
\\ \hline
P_{13}(1879)    &1.879          &0.498            &0.13          &0.217          &0
           &0              &0.151            &3/2           &1/2            &p          &p  &1.1
\\ \hline \hline
F_{15}(1680)     &1.684          & 0.139           &0.096         &0.011          &0.014      
           &0              &0                &5/2           &1/2            &f         &p   &0.9
\\ \hline
F_{35}(1905)& 1.881         &0.329            &0.041         &0.282          &0.006            
           &0              &0                &5/2           &3/2            &f         &p   &1.4
\\ \hline
F_{15}(2000)    &1.903          &0.494            &0.039         &0.369          &0.086          
           &0              &0                &5/2           &1/2            &f         &p   &1.4
\\ \hline \hline
S_{11}(1535)    &1.534          &0.151            &0.077         &0.005          &0          
           &0.066          &0                &1/2           &1/2            &s         &s   &0.8
\\ \hline
S_{31}(1620)&1.672          &0.154            &0.014         &0.044          &0.095          
            &0              &0                &1/2           &3/2            &s         &s  &0.9
\\ \hline 
S_{11}(1650)    &1.659          &0.173            &0.154         &0.005          &0.008          
           &0.006          &0                &1/2           &1/2            &s         &s   &0.9
\\ \hline 
S_{11}(2090)    &1.928          &0.415            &0.043         &0.203          &0.167          
           &0.002          &0                &1/2           &1/2            &s         &s   &1.5
\\ \hline \hline
D_{13}(1520)    &1.524          &0.124            &0.073         &0.026          &0.025           
           &0              &0                &3/2           &1/2            &d         &s  &0.9
\\ \hline
D_{33}(1700)&1.762          &0.598            &0.081         &0.046          &0.471
            &0              &0                &3/2           &3/2            &d         &s  &1.3
\\ \hline
D_{33}(1940)&2.057          &0.460            &0.081         &0.162          &0.217        
            &0              &0                &3/2           &3/2            &d         &s  &1.8
\\ \hline 
D_{13}(2080)    &1.804          &0.447            &0.104         &0.114          &0.229       
           &0              &0                &3/2           &1/2            &d         &s  &1.6
\\ \hline \hline                   
\end{array}
\edm
\caption{\label{restable} List of all resonances which are taken into account in our calculation. Apart from mass and width into the individual decay channels, we also give spin and isospin as well as the lowest orbital angular momentum needed in pseudoscalar ($\varphi$) or vector ($V$) meson scattering on a nucleon to form the resonance. All quantities are given in GeV. In those cases where the given partial widths do not add up to the full decay width, the remaining width is assigned to the $\Delta\pi$ channel. In the last row we denote the cutoff of the form factor $F_s(k^2)$ of Eq. \ref{ff1} at the $\rho\,N\,R$ vertex.}
\end{table}
\renewcommand{\arraystretch}{1}

In Table \ref{restable} we give a list of all resonances and their decay channels. For most states the sum $\Gamma_{\pi\,N}+\Gamma_{\eta\,N}+
\Gamma_{\rho\,N}$ does not exhaust the total width. As an approximation we put the remaining width into the $\Delta\pi$ channel and take the energy dependence to be $s$-wave for negative parity states and $p$-wave for positive parity states. In contrast to the other decay channels we do not modify $\Gamma_{\Delta\pi}$ when going to the nuclear medium.
The corresponding Lagrangians are given in Appendix \ref{lagnrel} and lead to traces $\Omega^\Delta$, which we do not explicitly denote here. In analogy to the $N\,\rho$ width, we find for $\Gamma_{\Delta\pi}$:
\beqa
\label{gamres3}
 \Gamma_{\Delta\pi}(k) &=& \int\limits_{(m_N+m_\pi)^2}^{(\sqrt{k^2}-m_\pi)^2}\!\!\! dm^2\,
 \Gamma_{\Delta\pi}(k,m)\,{\rho}_\Delta(m)\\
 \Gamma_{\Delta\pi}(k,m) &=& \frac{1}{2\,j+1}
 \,I_\Sigma\,\left(\frac{f}{m_\Delta}\right)^2 \,F^2(k,m_\pi)\,\frac{q_{cm}}
 {8\pi k^2}\,\Omega^{\Delta}  \nonumber \quad.
\eeqa
The isospin factor $I_\Sigma$ is $1$ both for the decay of isospin-$\frac3 2$ and isospin-$\frac1 2$ states and $\Gamma_{\Delta\pi}(k,m)$ stands for the decay into a pion and a $\Delta$ with invariant mass $m$. Since the $\Delta$ resonance is a broad particle, we need to integrate over its spectral function $\rho_\Delta(m)$.

\begin{figure}[t]
\centering
\includegraphics[scale=1.0]{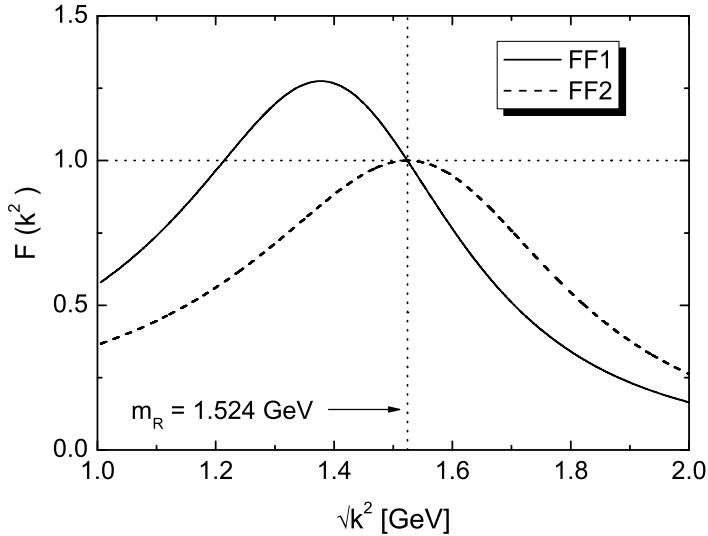}
\caption{\label{formfac} Form factor $FF1$ of Eq. \ref{ff1} (solid line) and form factor $FF2$ of Eq. \ref{ff2} (dashed line) for the $D_{13}(1520)$ resonance. The cutoff parameter in $FF2$ is taken to be $\Lambda=1$ GeV, the parameters $s_0$ and $\Lambda$ of form factor $FF1$ are listed in Table \ref{restable}.}
\end{figure}

We write the form factor $F(k,q)$ at the resonance-meson-nucleon vertex in the following form:
\beqa
\label{ffts}
	F(k,q) &\equiv& F_s(k) \, F_t(q) \quad.
\eeqa
The form factor $F_t(q)$ is a usual monopole form factor:
\beqa
		\label{fft} F_t(q) &=& \frac{\Lambda_M^2-m_M^2}{\Lambda_M^2-q^2} \quad.
\eeqa
The values taken for $\Lambda_M$ are listed in Table \ref{param} in Appendix \ref{parameters}. We multiply the resonance-nucleon-meson vertex with a monopole type form factor since this vertex is also used in baryon-baryon interactions, where the large space-like $4$-momenta acquired by the exchange particle need to be cut off. For the decay of a resonance into a stable final state we have $q^2=m_M^2$ and therefore $F_t(q)=1$. When going to the nuclear medium, we will be forced to evaluate $F_t$ also for time like $4$-momenta $q^2 \approx \Lambda_M^2$. In order to avoid poles of $F_t(q)$, we put the form factor equal to unity for $q^2 \ge m_M^2$. This does not affect the action of $F_t$ in the space-like region. For $F_s(k)$ we take different parameterizations at the $RN\rho$ and the $RN\varphi$ vertices.
When considering an $RN\rho$ vertex we choose \cite{gregorpi}
\beqa
	\label{ff1} F_s(k) &=& \frac{\Lambda^4+\frac 1 4\,(s_0-m_R^2)^2}
	{\Lambda^4+(k^2-\frac 1 2 (s_0+m_R^2))^2} 	\quad,
\eeqa
while at the $RN\varphi$ vertex we take \cite{gregorpi,scholten}:
\beqa
	\label{ff2} F_s(k) &=& \frac{\Lambda^4}{\Lambda^4+(k^2-m_R^2)^2} \quad.
\eeqa
In the following we will refer to the form factor of Eq. \ref{ff1} as $FF1$ and to the form factor of Eq. \ref{ff2} as $FF2$.

We have plotted both $FF1$ (solid line) and $FF2$ (dashed line) in Fig. \ref{formfac}.
As a function of $k^2$ the form factor of Eq. \ref{ff1} is asymmetric with respect to the resonance mass $m_R$. It is equal to unity both at $k^2=m_R^2$ and at the decay threshold $k^2=s_0<m_R^2$, larger than unity within the interval $\{s_0,m_R^2\}$ and smaller outside. 
The exact shape depends on the cutoff $\Lambda$ and the threshold parameter $s_0$.
We give values for the cutoff $\Lambda$ in Table \ref{restable} and take $s_0=(m_N+2 m_\pi)^2$. In the following Section we will give arguments in support of the somewhat unconventional form factor Eq. \ref{ff1}. Asymmetric form factors like Blatt-Weisskopf type form factors are quite commonly used in the literature, e. g. in \cite{manley1}.

For positive energies, the imaginary part of the self energy $\imag{\Sigma^+_{vac}}$ is obtained from the decay width via 
\beqa
	\imag{\Sigma^+_{vac}(k^2)} &=& -\sqrt{k^2}\,\Gamma(k^2) \quad.
\eeqa
The real part of the self energy is calculated using a dispersion integral:
\beqa
\label{resodispvac}
	\real{\Sigma^+_{vac}(k)} &=& 
	{\cal P}\,\int\limits_{\omega_{min}}^\infty\,\frac{d\omega}{\pi} 
	\frac{\imag{\Sigma^+(\omega,{\bf k})}}{\omega-k_0} - c_{vac}({\bf k}) \\ \mbox{with} && \nonumber \\
	c_{vac}({\bf k}) &=& {\cal P}\,\int\limits_{\omega_{min}}^\infty\,\frac{d\omega}{\pi} 
	\frac{\imag{\Sigma^+(\omega,{\bf k})}}{\omega-\sqrt{m_R^2+{\bf k}^2}}  \nonumber \quad.
\eeqa
Here ${\cal P}$ denotes the principal value. The energy $\omega_{min}=\sqrt{(m_N+m_\pi)^2+{\bf k}^2}$ follows from the threshold for the decay into $N\pi$. By $m_R$ the mass of the resonance is denoted. 
As can be inferred from Fig. \ref{d13p33width}, the suppression from the form factor $F(k,q)$ is sufficient to produce a decreasing width, such that the dispersion integral converges. The subtraction is convenient to ensure that the physical mass of the resonance is recovered.
In principle the dispersion integral extends over negative energies as well.
We omit this contribution since in the nuclear medium no symmetry exists which relates $\imag{\Sigma^+}(k_0)$ to $\imag{\Sigma^+}(-k_0)$. This issue is addressed in Appendix \ref{analytic}. We have checked that in the vacuum the contributions from negative energies to the dispersion integral can safely be neglected. Also, in cold nuclear matter we do not expect that antibaryons are important.


\subsection{Results for $\real{\Sigma}$ and $\imag{\Sigma}$}
\label{ressigmavac}

In this Section we present results for $\Sigma_{vac}$ and compare two resonances --  the $P_{33}(1232)$ and the $D_{13}(1520)$ resonance, both of which have according to the PDG \cite{pdg} an on-shell width of about $120$ MeV. 

\begin{figure}[h!]
\centering
\includegraphics[scale=1.4]{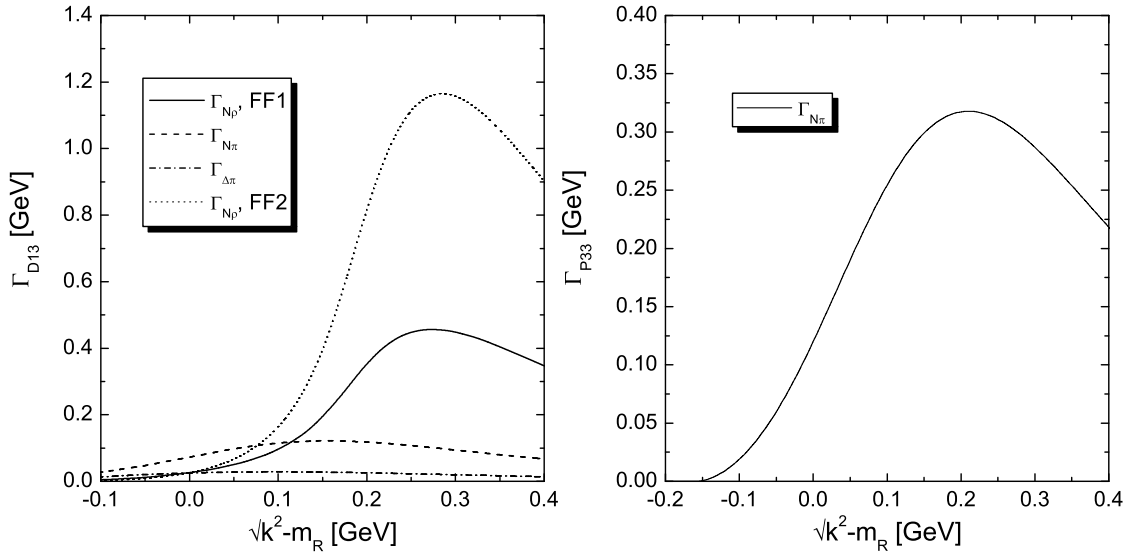}
\caption{\label{d13p33width}Left: Partial decay widths of the $D_{13}(1520)$. The $N\rho$ width $\Gamma_{N\rho}$ as following from form factor $FF1$ of Eq. \ref{ff1} is indicated by the solid line, the result from form factor $FF2$ of Eq. \ref{ff2} by the dotted line. Right: Width of the $P_{33}(1232)$ resonance in vacuum. The $\rho$ component is negligible over the energy interval shown here and has not been plotted.}
\end{figure}

In Fig. \ref{d13p33width} we show the decay width of both resonances as a function of their invariant mass. Note the different scales on the $y$ axes.
The $N\rho$ width (solid line) of the $D_{13}$ state displays a strong energy dependence, which is of kinematical origin. The resonance is below the nominal threshold for the decay at $m_N+m_\rho = 0.938 + 0.77 \approx 1.7$ GeV. Therefore the decay into this channel can only proceed via the low mass tail of the $\rho$ meson, which in turn generates a steep increase of the width as the available phase space opens up. 
\begin{figure}
\centering
\includegraphics[scale=1.25]{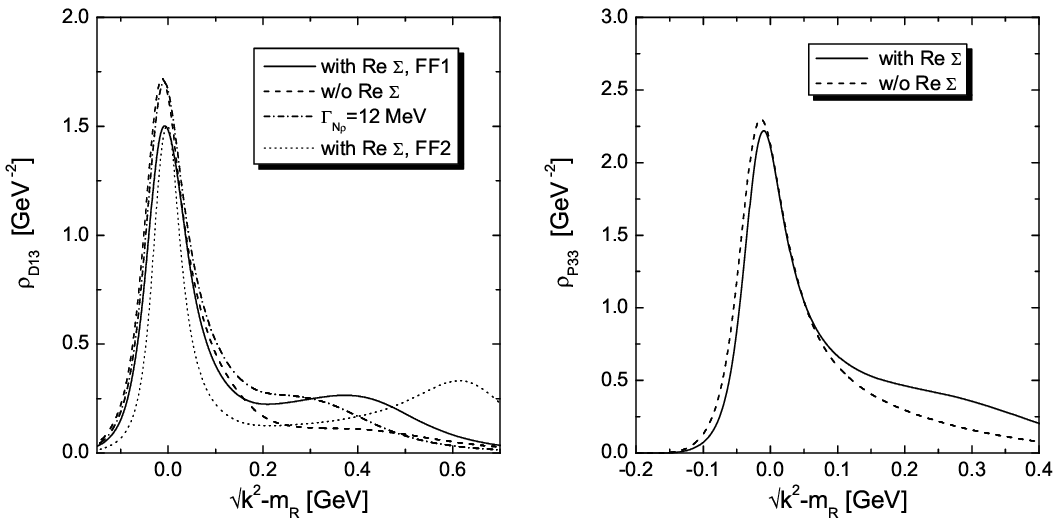}
\caption{\label{d13specnorm} Spectral function $\rho$ of the $D_{13}(1520)$ (left) and the $P_{33}(1232)$ resonances. In both cases the solid lines indicate the results as following from using $FF1$ for $\Gamma_{N\rho}$. The dashed line shows the spectral function without $\real{\Sigma}$. In the left figure, the dash-dotted curve is the spectral function of the $D_{13}(1520)$ if a value of $12$ MeV for $\Gamma_{N\rho}$ is used. In the dotted curve the form factor $FF2$ of Eq. \ref{ff2} is taken at the $RN\rho$ vertex.}
\end{figure}

\begin{figure}[t]
\centering
\includegraphics[scale=0.9]{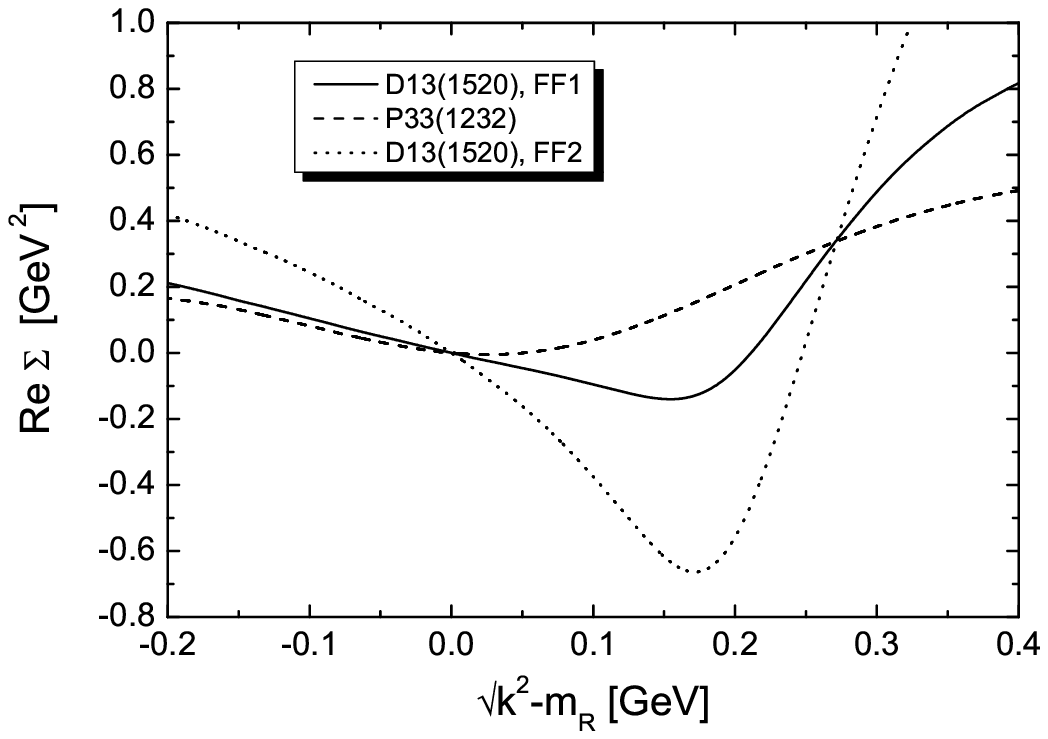}
\caption{\label{d13p33resig} Real part of the self energy of the $D_{13}(1520)$ resonance (solid and dotted lines) and the $P_{33}(1232)$ resonance (dashed line). The solid line is obtained using form factor $FF1$ of Eq. \ref{ff1} and the dotted line follows from form factor $FF2$ of Eq. \ref{ff2}.}
\end{figure}

Let us now turn to the results for the self energy and the spectral function, which are depicted in Figs. \ref{d13specnorm} and \ref{d13p33resig}. From Fig. \ref{d13specnorm} we find that around the resonance peak the spectral functions of the $D_{13}(1520)$ and the $P_{33}(1232)$ (solid lines) obtained by a full calculation including the real part of the self energy do not differ much from those obtained by neglecting $\real{\Sigma}$ (dashed lines). In both cases we observe a slight squeezing of the resonance peak in the spectral function due to the real part, which is more pronounced for the $D_{13}(1520)$. Going away from the resonance peak, we observe an additional shoulder for the $D_{13}(1520)$.
This shoulder has most probably no direct influence on observables since it is located far away from the resonance peak and is likely to be overshadowed by the contribution of other resonances. Hence we are not concerned about this structure. The real part of the self energy of both states -- indicated by the solid and the dashed lines in Fig. \ref{d13p33resig} -- is comparable around resonance, with a slightly larger energy variation for the $D_{13}(1520)$. Away from the resonance peak we observe a strong energy variation in the self energy of the $D_{13}(1520)$ which is responsible for the additional shoulder found in the spectral function. The dashed-dotted line in Fig. \ref{d13specnorm} shows the spectral function of the $D_{13}(1520)$ as resulting from the use of a smaller partial decay width $\Gamma_{N\rho}=12$ MeV instead of $\Gamma_{N\rho}=26$ MeV. Using the smaller width the shoulder nearly disappears. We will come back to the issue of the proper choice for $\Gamma_{N\rho}$ in the next Section.

Next we address the question as to why form factor $FF2$ should be discarded at
the $RN\rho$ vertex. Therefore consider the results for $\imag{\Sigma}$, $\real{\Sigma}$ and the spectral function $\rho$, which are depicted by the dotted lines in Figs. \ref{d13p33width}, \ref{d13specnorm} and \ref{d13p33resig}. All three curves display unsatisfying features: the $N\rho$ decay width obtained with $FF2$ rises very quickly to values above $1$ GeV, around the resonance peak the real part of the self energy has a strong energy dependence $\partial \real{\Sigma}/ \partial k^2$ and we observe a significant squeezing of the resonance peak in the spectral function $\rho$. The sum of these effects provides enough evidence to abandon form factor $FF2$ and take form factor $FF1$ instead.

These three effects are connected to each other in the following way: the rapid increase of $\Gamma_{N\rho}$ translates into a strong energy dependence of $\real{\Sigma}$. If $\imag{\Sigma}$ is nearly constant around the pole of the resonance, one expects $\real{\Sigma}$ to be small since the contributions from below and above the pole approximately cancel. Turning this argument around implies that a rapid variation with energy leads to a sizeable $\real{\Sigma}$.
This leads to a squeezing of the peak which can be understood by expanding $\real{\Sigma}$ to first order in $k^2$, thus producing the quasi-particle approximation. One gets for the spectral function:
\beqa
\label{zfactor}
	\rho(k^2) &\approx& \frac{-1}{\pi}\frac{z^2\imag{\Sigma}}
	{(k^2-m_R^2)^2+z^2\imag{\Sigma}^2} \\
	z &=& \left.\left(1-\frac{\partial \real{\Sigma}}{\partial k^2}\right)^{-1}\right|_{k^2=m_R^2} \nonumber .
\eeqa
The factor $z<1$ effectively measures the influence of $\real{\Sigma}$ and indicates that -- depending on the energy variation of $\real{\Sigma}$ -- strength is shifted away from the resonance peak to larger invariant masses. This explains the pronounced peak at invariant resonances masses $\sqrt{k^2}$ around $m_R + 0.6$ GeV. Now one can also understand why form factor $FF1$ can cure this problem: its functional form limits the energy variation of the $N \rho$ width of the $D_{13}(1520)$, which we have identified as the main source of trouble. 
 
It follows from Eq. \ref{zfactor} that one could recover the nominal width of the resonance peak by increasing $\imag{\Sigma}$. The price to pay is that then the peak height is reduced, because in any case the area under the peak is reduced since $z<1$. A priori it is not clear how this problem should be handled, since in principle peak height and peak width can be additionally influenced by interference with background terms. In this work we have decided to preserve the peak height. Adjusting instead to the peak width leads to further 
complications for the coupled channel problem at hand: considering for example the $D_{13}(1520)$, the $z$ factor is essentially generated by the $\rho N$ width. We have checked numerically that by increasing the $N\rho$ width -- leading to a smaller $z$ factor -- one cannot restore the original peak width. Instead, one had to change the $\pi N$ or the $\pi\Delta$ partial width (or both of them). In a numerical simulation we have found that both had to be multiplied by a factor of $1.7$ in order to restore the original peak width. Clearly, this would influence the results for the $\pi$ self energy, since a modification of $\Gamma_{\pi N}$ leads to a new coupling constant at the $\pi\,N\,D_{13}(1520)$ vertex. A reliable solution of this problem mandates a complete analysis of $\pi\,N$ scattering taking into account effects from $\real{\Sigma}$.

The results for the spectral shape of the $D_{13}(1520)$ presented in this work show a strong similarity to the textbook case of a stable state, whose mass is below the multi-particle threshold \cite{peskin}. In the absence of interactions, all the spectral strength sits in the particle peak. After the interactions have been turned on, however, strength appears at masses above the threshold. For the spectral function to remain normalized, this implies that strength has to be removed from the quasi-particle peak.

The large decay width of $\Gamma_{N\rho}=300$ MeV for the $P_{13}(1720)$ given in \cite{manley2} also leads to problems when calculating $\real{\Sigma}$, requiring a very small cutoff value $\Lambda$. Since such a large decay width seems questionable for a resonance close to the nominal $N\rho$ threshold and coupling in a $p$-wave to this channel, we instead use the PDG estimate $\Gamma_{N\rho}=110$ MeV for this resonance. See also the discussion in Chapter \ref{exp}.

\section{$\rho\,N$ Scattering and Experiment}
\label{exp}

Experimental information on the coupling of mesons to a baryon resonance enters into our model via the coupling constants $f_M$, which are adjusted to the experimentally observed partial decay width of that resonance into the respective meson-nucleon channel. 
Whereas for the $N\pi$ and $N\eta$ final states those branching ratios are rather well known, some uncertainty prevails for the decay width into the $N\rho$ channel.
This is due to the principal difficulties associated with the experimental identification of a $\rho$ meson at energies below or only slightly above the nominal threshold, as occurring in the decay of, e. g., the $D_{13}(1520)$ or the $P_{13}(1720)$.
In the following we discuss the experimental information about the decay of the $D_{13}(1520)$ in some detail and also touch on some of the uncertainties concerning a few other states.


\subsection{The $\boldsymbol{D_{13}}$ Amplitude}
\label{expd13}

A major part of the results of this work hinges on the coupling of the $D_{13}(1520)$ resonance to the $N \rho$ channel. According to the PDG \cite{pdg}, one has $\Gamma_{N \rho}=24$ MeV. 
This is close to the value suggested by Manley \emph{et al} \cite{manley2}, where $\Gamma_{N \rho}=26$ MeV is found. In this work we adopt the value of \cite{manley2} and give some motivation for our choice in the following paragraphs. 

The experimental information on $\Gamma_{N\rho}$ is primarily derived from a partial wave analysis of $\pi N \to \pi\pi N$ scattering presented in \cite{manley1}, where a very clear resonance structure is found in the $\pi N \rightarrow \rho N$ channel of the $D_{13}$ partial wave amplitude for energies around $1.5$ GeV. Similar results are reported in Herndon \emph{et al}\cite{herndon} and Dolbeau \emph{et al} \cite{dolbeau}. Only in the work of Brody \cite{brody} no coupling of the $D_{13}(1520)$ to the $N\rho$ channel is found, since below $\sqrt{s}=1.6$ GeV the $N \rho$ contribution is set to zero by hand. In subsequent analyses, the partial decay width $\Gamma_{N\rho}$ has been extracted from a resonance fit of this partial wave amplitude.
Both the work of Manley \emph{et al} \cite{manley2} and Vrana \emph{et al}\cite{vrana} achieve a reasonable fit by assigning a relatively large value (in view of the available phase space) to $\Gamma_{N\rho}$. In \cite{manley2} a partial width of $26$ MeV is found, whereas the analysis of \cite{vrana} reports a value of $12$ MeV.

Further support for a rather large value for $\Gamma_{N\rho}$ comes from
a complementary experiment, where photoproduction of pion pairs on the nucleon has been studied \cite{langgartner}. The two-$\pi$ invariant mass spectra - measured at photon energies just below the nominal threshold for $\rho$ production -  follow the expected phase space distribution in the isoscalar channel $(\pi^0 \pi^0)$. In the isovector channel a systematic asymmetry favouring larger invariant masses is reported. An appealing interpretation of this finding assigns this asymmetry to the $\rho$ meson, which does not couple to the isoscalar channel. In a subsequent theoretical analysis \cite{osetgam2pi}, this conjecture has been put on a more solid basis. There, a coupling of the $D_{13}(1520)$ to the $N\rho$ channel in line with the PDG values is necessary for a successful description of the data.

In \cite{lutzvector1, lutzvector} meson-nucleon scattering is described in terms of $4$-point interactions. After iterating the interaction, the resonant structures seen in experiment emerge dynamically. Fitting these structures with a Breit-Wigner type ansatz, width and mass can be compared the results of other analyses. For $\Gamma_{N\rho}$ a value of about $6$ MeV is found in \cite{lutzvector1}, smaller than the results from \cite{manley2,vrana}. This is claimed to be due to the fact that a direct fit of $\pi\,N \rightarrow \rho\,N$ data from \cite{brody} is performed, thus leaving the region around the $D_{13}(1520)$ essentially unconstrained. In \cite{lutzvector} an even smaller coupling is obtained after the inclusion of photo-induced data to the coupled channel analysis. There the direct constraint of photo data to the hadronic $\rho\,N$ vertex results from the assumption of strict vector meson dominance. The $\rho$ coupling strength is then determined rather indirectly by the isovector part of the photon-coupling and not from hadronic data.

Although we cannot exclude the possibility of such a weak coupling to the $N\rho$ channel, we believe that there is enough experimental evidence supporting a rather large width $\Gamma_{N\rho}$ of the $D_{13}(1520)$. We base our calculations on $\Gamma_{N\rho}=26$ MeV as suggested in \cite{manley2}. In order to get some feeling for the sensitivity of our results on $\Gamma_{N\rho}$ we present also calculations using $\Gamma_{N\rho}=12$ MeV reported in \cite{vrana}.


\subsection{Other Partial Waves}
\label{exprest}

Uncertainties concerning resonance parameters not only exist for the $D_{13}(1520)$ state, but for many of the high lying resonances included in this work. Fortunately, in most cases it turns out that the results are not too sensitive to changes in the parameters. This is not true, however, for the $P_{13}(1720)$ and the $D_{33}(1700)$ states, which have a large energy overlap with the $N\rho$ system.

For both resonances a large branching ratio into the $2\pi N$ final state is well established in the literature, see for example \cite{manley2, vrana, gregorpi, gregorphoto}. However, so far no agreement has been reached both for the total width of the resonance and for the relative strength of $N\rho$ and $\Delta\pi$ contributions.
Note first, that the decay of the $P_{13}(1720)$ to $N\,\rho$ is strongly suppressed from phase space, if one takes into account that the coupling is $p$-wave. In this light, the huge partial decay width assigned in \cite{manley2} of about $300$ MeV seems questionable.  Therefore we have opted to take the PDG value $\Gamma_{N\rho}=110$ MeV \cite{pdg} for this channel, which is in agreement with the findings in \cite{vrana}. 
For the $D_{33}(1700)$ in \cite{manley2} a value of $46$ MeV is found, whereas the PDG suggests a value of $120$ MeV. Here we follow the results of \cite{manley2}. This way we arrive at a conservative estimate concerning the influence of both states for in-medium effects.

\section{Mesons and Baryons in the Nuclear Medium}
\label{itscheme}

Up to now we have discussed aspects of the meson-nucleon and baryon-nucleon interaction in the vacuum. We are now in a position to consider these interactions in the nuclear medium. The goal is to achieve a coupled-channel analysis of the in-medium properties of pions, $\eta$ and $\rho$ mesons as well as baryon resonances. 
We therefore need to calculate the spectral function of all particles under consideration:
\beqa
\label{propmed}
	{\cal A}^{med}_M(q)&=&-\frac{1}{\pi}\imag{\D{\frac{1}
	{q^2-m_M^2-\Pi_{vac}^+(q)-\Pi^+_M(q)}}}  \\
	\rho^{med}(k)&=&-\frac{1}{\pi}\imag{\D{\frac{1}{k^2-m_R^2-\Sigma_{med}^+(k)}}}
			\nonumber \quad,
\eeqa
where $M$ stands for the meson under consideration. Here we denote the full in-medium self energy of meson $M$ by the sum $\Pi_{vac}^+(q)+\Pi^+_M(q)$, where $\Pi^+_M$ contains the contribution of all processes that take place only in the medium. 
For resonances this splitting is not reasonable within our model, since the vacuum and in-medium self energies are generated by the same type of diagrams.

For the mesons we consider effects from resonant meson-nucleon scattering, where the nucleon is provided by the surrounding nuclear medium. This leads to the excitation of particle-hole pairs, see Fig. \ref{particlehole}. In the baryon sector, the in-medium decay width of resonances is affected by two mechanisms: Pauli blocking reduces the width, whereas resonance-nucleon scattering leads to a broadening. We generate collisional broadening -- being due to the exchange of mesons -- by replacing the vacuum meson propagator $D_M^{vac}$ by its in-medium counterpart $D_M^{med}$ in the resonance self energy Fig. \ref{resself}. Applying Cutkosky's cutting rules to the modified diagram yields the collisional broadening. For an illustration of these rules compare the self energy diagram of Fig. \ref{resoselfmed} and the cuts of Fig. \ref{gammamed}.

We thus have set the stage for a typical self consistency problem. Starting with a model for the in-medium self energy of mesons, we are led to modify also the resonance self energy, which in turn serves as input for an improved calculation of the meson self energy. By this iterative procedure diagrams of higher order in the density are generated. A resonance self energy of order $\rho^1$ produces a meson self energy of order $\rho^2$, since the meson effectively interacts with two nucleons (albeit not at the same point).

In this Section we present the underlying theoretical framework, which is akin to our program in the vacuum. Starting point is the calculation of the imaginary part of the in-medium self energy. The dispersive real part follows then from a dispersion relation over the energy, i.e. at fixed $3$-momentum. We repeat our statement from the introduction to Chapter \ref{vacself} that we keep only the leading non-relativistic contribution from  the traces arising at the meson-nucleon-resonance vertices.


\subsection{The in-medium Self Energy of Mesons}
\label{mesoself}

Compared to the vacuum, the meson self energies exhibit two distinct properties in nuclear matter. As a consequence of the fact that nuclear matter constitutes a suitable reference frame, which allows for the definition of a transverse and longitudinal polarization, we obtain two independent self energies for the $\rho$ meson, $\Pi_{\rho}^T$ and $\Pi_{\rho}^L$. These are obtained by contracting the three-transverse and three-longitudinal projectors $T^{\mu\nu}$ and $L^{\mu\nu}$, respectively, with the self energy tensor $\Pi_{\mu\nu}$
\beqa
	\Pi_\rho^T &=& \frac 1 2\,T^{\mu\nu}\,\Pi_{\mu\nu} \\
	\Pi_\rho^L &=& L^{\mu\nu}\,\Pi_{\mu\nu} \nonumber \quad,
\eeqa
where the projectors read \cite{gale}:
\beqa
L^{\mu\nu}(q) &=& -\frac{q^2}{(n q)^2-n^2q^2} \, 
							\left(n_\mu-q_\mu\,\frac{n q}{q^2} \right)																							\left(n_\nu-q_\nu\,\frac{n q}{q^2}  \right) \\
	T^{\mu\nu}(q)	&=& P^T_{\mu\nu}(q) - L_{\mu\nu}(q)	\nonumber \quad .
\eeqa
Here $n$ characterizes the nuclear medium and we choose $n=(m_N,0)$. The four-transverse projector $P^T_{\mu\nu}$ has been introduced in Chapter \ref{vacself}, Eq. \ref{4trans}. As a second consequence of the presence of nuclear matter, the in-medium self energy depends on both variables $q_0$ and $|{\bf q}|$ independently.

\begin{figure}
\centering
\includegraphics{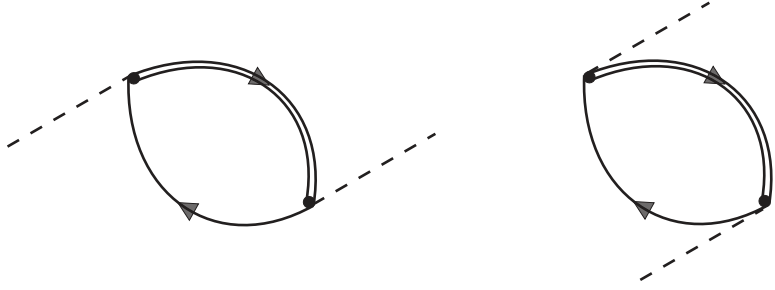}
\caption{\label{particlehole}Feynman diagram representing the resonance hole excitation. Left: $s$-channel contribution. Right: $u$-channel contribution. The double-line stands for any of the resonances or a nucleon. The meson lines represent any of $\pi$, $\eta$ or $\rho$ meson.}
\end{figure}

By using standard Feynman rules and the Lagrangians given in Appendix \ref{applag}, we arrive at the following expressions for the imaginary part of the self energy of a pseudoscalar meson $\varphi$ due to the excitation of a resonance-hole loop:
\beqa
\label{scalarself}
	\imag{\Pi_{\varphi}(q_0,\bf q)} &=& I_\Pi\,\left(\frac{f}{m_\varphi}\right)^2
	\int \frac{d^3 p}{(2\,\pi)^3}\frac{\theta(p_F-|{\bf p}|)}{2\,E_N({\bf p})}
	\imag{\D{\frac{\Omega^\varphi}
	{k_0^2-E_R^2({\bf k})-\Sigma_{vac}^+(k)}} } \quad. 
\eeqa
Similarly, for $\rho$ mesons we find:
\beqa
\label{vectorself}
	\imag{\Pi_{\rho}^{T/L}(q_0,\bf q)} &=& I_\Pi\,\left(\frac{f}{m_V}\right)^2
	\int \frac{d^3 p}{(2\,\pi)^3}\frac{\theta(p_F-|{\bf p}|)}{2\,E_N({\bf p})}
	\imag{\D{\frac{\Omega^{T/L}}
	{k_0^2-E_R^2({\bf k})-\Sigma_{vac}^+(k)}}} \quad. 
\eeqa
In the above expression we have $k=p+q$, where $p=(p_0,{\bf p})$ is the $4$-momentum of a nucleon. The on-shell energies of nucleon and resonances are given by $E_N$ and $E_R$.
By $p_F$ the Fermi momentum is denoted. The factor $\frac{1}{2 E_N}$ comes from the hole part of the relativistic nucleon propagator. Explicit expressions for the traces $\Omega$ can be found in Appendix \ref{applag}. The isospin factor $I_\Pi$ is calculated from the isospin part of the Lagrangian and is $I_\Pi=2$ for isospin-$\frac 1 2$ and $I_\Pi = 4/3$ for isospin-$\frac 3 2$ resonances if pions or $\rho$ mesons are considered. For the isoscalar $\eta$ this factor is $2$.
In addition, we multiply the form factor $F(k,q)$ (not displayed explicitly in Eqs. \ref{scalarself} and \ref{vectorself}), see Chapter \ref{vacself}, Eq. \ref{ffts}. At sufficiently small densities the above expressions can be obtained from the low density theorem \cite{dover}, which relates the self energy of meson $M$ to the nuclear density $\rho$ and the meson nucleon forward scattering amplitude ${\cal T}_{NM}$ as:
\beqa
\label{lowdens}
	\Pi_M(q_0,\bf q) &=& \frac{\rho}{8 m_N}\,{\cal T}_{MN}(q_0,\bf q) \quad.
\eeqa

The complete self energy is given as the sum of all individual resonance contributions. A complete list of the resonances which are taken into account is given in Table \ref{restable}. We include all resonances which have a sizeable coupling to either of the mesons that are considered in this work. As for the parameters mass and decay width we follow with one exception the results of Manley \emph{et al} \cite{manley2}. This exception is the $P_{13}(1720)$ resonance as discussed in Section \ref{exprest}. 

Apart from the resonance excitations, we also take into account the conversion of mesons into nucleon-hole loops. This is done in a non-relativistic manner. Since the nucleon is stable, the integration over the Fermi distribution can be performed analytically and one finds for meson $M$:
\beqa
\label{pinn}
	\Pi_M^N(q_0,{\bf q}) &=& 
	4\,{\bf q}^2\, \left(\frac{f_{NNM}}{m_M}\right)^2\,U_N(q)\,F_t^2(q) \quad.
\eeqa
The form factor $F_t(q)$ is defined in Eq. \ref{fft}, Chapter \ref{vacself}. The cutoff parameters $\Lambda$ as well as the coupling constants $f_{MNN}$ are listed in Table \ref{param} in Appendix \ref{parameters}. The Lindhard function $U_N(q)$, consisting of $s$ and $u$ channel contributions, is given explicitly in \cite{ericsonweise}. Following a suggestion made in \cite{oseteta2},
for the case of $\eta$ and $\pi$ mesons, we multiply $U_N$ with a recoil factor:
\beqa
	U_N(q_0,{\bf q}) &\to& U_N(q_0,{\bf q})\,\left(1-\frac{q_0}{2 m_N}\right)^2 \quad,
\eeqa
which is a ${\cal O}\left(\frac{p}{m_N}\right)$ correction from the relativistic pseudo-vector coupling \cite{ericsonweise}.

We obtain the real part of the in-medium self energy of meson $M$ by means of an unsubtracted dispersion relation:
\beqa
\label{realpi}
	\real{\Pi^+_{M}(q_0,{\bf q})} &=& {\cal P}\int\limits_{0}^\infty \frac{d\omega^2}{\pi}\frac{\imag{\Pi^+_{M}(\omega,{\bf q})}}{\omega^2-q_0^2} \quad,
\eeqa
which can be written down as an integral over positive energies due to the antisymmetry 
of the imaginary part of the meson self energy, see Appendix \ref{analytic}. Note that due to the antisymmetry of $\imag{\Pi^+}$, in this way we also generate the $u$ channel contributions depicted in the right hand side of Fig. \ref{particlehole}: changing the meson energy from $q_0 $ to $-q_0$ in an $s$-channel diagram yields exactly the corresponding $u$-channel diagram since an incoming meson with negative energy $-q_0$ is nothing but an outgoing one with positive energy $q_0$. We find that in comparison to a direct calculation of this quantity from the Feynman diagram of Fig. \ref{particlehole}, the dispersion relation leads to a more pronounced resonant structure of the self energy. Note that a direct computation violates analyticity since the form factors $F_s$ and $F_t$ have poles in the complex energy plane.

\begin{figure}[t]
\centering
\includegraphics[scale=.9]{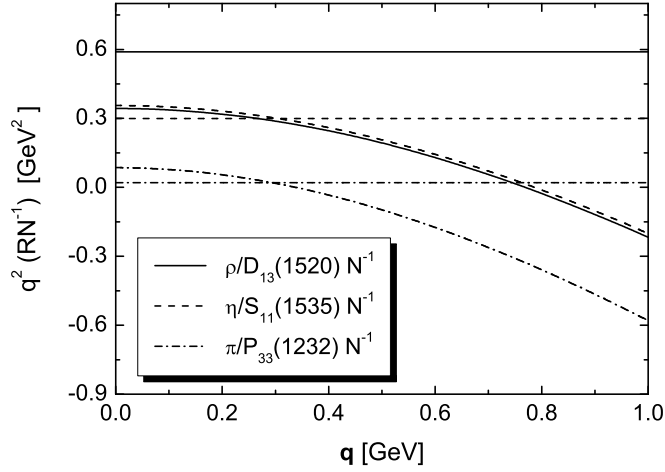} 
\caption{\label{branches} Invariant mass squared $q^2$ (see Eq. \ref{posbranch}) as function of the 3-momentum ${\bf q}$ of the following resonance-hole excitations: $D_{13}(1520) N^{-1}$ (solid), $S_{11}(1535) N^{-1}$ (dashed) and $P_{33}(1232) N^{-1}$ (dashed-dotted). 
Also indicated are the masses of $\rho$, $\eta$ and pion (straight).}
\end{figure}

The meson spectral function generated by the self energies of Eqs. \ref{scalarself}, \ref{vectorself}, \ref{pinn} and \ref{realpi} is characterized by the formation of additional peaks arising from the excitation of particle-hole loops. At a given $3$-momentum the energy of the peak from a resonance-hole state with resonance mass $m_R$ is approximately determined by the equation:
\beqa
\label{posbranch}
	q_0^2-{\bf q}^2+m_N^2+2 m_N q_0 &=& m_R^2 \quad.
\eeqa
In Fig. \ref{branches} we have indicated the solution of this equation for some particle-hole states. Also indicated are the meson peaks. Due to the interaction between the various branches of the spectral function, one observes level repulsion and the actual position of the peaks is slightly rearranged in comparison to the solution of Eq. \ref{posbranch}. The use of dispersion relations enhances this effect.

As indicated by the low density theorem, Eq. \ref{lowdens}, the self energy of on-shell mesons in nuclear matter is constrained from the meson nucleon phase shifts, while the off-shell dynamics remains largely model dependent. However, direct information on the phase shifts is only available for the pion. As we have explained in Section \ref{ressigmavac}, the inclusion of $\real{\Sigma}$ leads to a reshaping of the resonance spectral function, which in turn also influences -- and actually worsens -- the description of $\pi\,N$ phase shifts away from the pole. A better description would necessitate the inclusion of non-resonant background terms. We would like to stress, however, that the main factors determining the importance of the resonance contribution -- coupling strength and resonance mass -- are not affected by these problems.


\subsection{The in-medium Self Energy of Baryon Resonances}
\label{resoself}

\begin{figure}[t]
\centering
\includegraphics[scale=0.55]{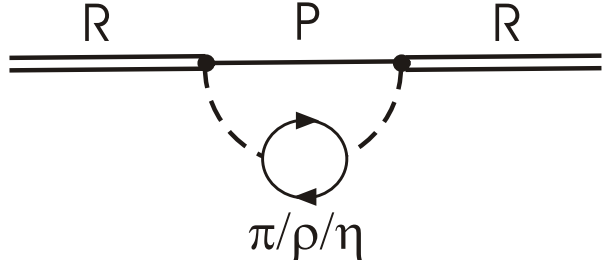}
\caption{\label{resoselfmed} Feynman diagram representing the in-medium decay of a baryon resonance into a nucleon and a dressed meson. The particle-hole loops stands for both for nucleon and resonance excitations and it is understood that it represents a complete resummation of particle-hole insertions. The symbol $P$ indicates that Pauli blocking is taken into account.}
\end{figure}

We calculate the in-medium broadening of baryon resonances by replacing the vacuum meson propagator by the in-medium one in the meson-nucleon self energy loops. The corresponding Feynman diagram is shown in Fig. \ref{resoselfmed}, where the self energy insertion is to be understood as a complete resummation of particle-hole loops. The $\Delta\pi$ channel is not modified. Applying Cutkosky's cutting rules, one finds for the width of a nucleon resonance with spin-$\frac{j}{2}$ and invariant mass $\sqrt{k^2}$ decaying into a pseudoscalar meson $\varphi$ or a $\rho$ meson:
\beqa
\label{reswidthmed}
	\imag{\Sigma_{\varphi,med}^+(k_0,{\bf k})} &=& -\frac{I_{\Sigma}}{2j+1}\left(\frac{f}{m_\varphi}\right)^2     \!\!\!	\int\limits_{p_F}^{\sqrt{k_0^2-m_N^2}} \!\!\!\frac{dp\,p^2}{8\,\pi\,E_N} 
	\times \nonumber \\ &&\hspace{0cm} \times
	\int\limits_{-1}^{+1}dx\,F^2(k,(k-p))\,
	\Omega^\varphi\,{\cal A}_\varphi^{med}(k-p) \\ && \nonumber \\
		\imag{\Sigma_{\rho,med}^+(k_0,{\bf k})} &=& -\frac{I_{\Sigma}}{2j+1}\left(\frac{f}{m_\rho}\right)^2     \!\!\!	\int\limits_{p_F}^{\sqrt{k_0^2-m_N^2}} \!\!\!\frac{dp\,p^2}{8\,\pi\,E_N}  \times \nonumber \\ &&\hspace{0cm} \times \int\limits_{-1}^{+1}dx\, F^2(k,(k-p))\,
	(2\,\Omega^T\,{\cal A}_{\rho,T}^{med}(k-p)+
	\Omega^L\,{\cal A}_{\rho,L}^{med}(k-p)) \nonumber \quad.
\eeqa
The spectral function of meson $M$ is denoted by ${\cal A}_M^{med}$.
The isospin factor $I_\Sigma$ is $1$ for $\Delta$ resonances with isospin $\frac 3 2$ and $3$ for $N^*$ resonances with isospin $\frac 1 2$. For the decay into $N\eta$  one finds $I_\Sigma=1$. 
The integration variable $p$ refers to the nucleon momentum with respect to the rest frame of nuclear matter, $E_N$ is the on-shell energy of a nucleon and we have for the Fermi energy $E_F=\sqrt{m_N^2+p_F^2}$.
This explains the lower integration bound, reflecting Pauli blocking. The upper integration limit follows from the condition that the meson energy $q_0 > 0$.
By $x$ we denote the cosine of the polar angle between ${\bf k}$ and ${\bf p}$.
Note that these expressions are similar to what one obtains in the vacuum for the decay into one stable and one unstable particle, see Eqs. \ref{gamres2} and \ref{gamres3} in Chapter \ref{gammabarres}. For practical reasons we perform the phase space integration in Eq. \ref{reswidthmed} in the rest frame of nuclear matter and not in the rest frame of the resonance.

Note that in our approach $\imag{\Sigma_{M,med}^+(k_0,{\bf k})}$ vanishes for energies $k_0 < E_F$. This is strictly speaking only true for decay processes, whereas for energies below the Fermi energy one might also think of the formation of resonances from the decay of a nucleon sitting in the Fermi sphere into a (far off-shell) resonance and a meson. Such a process formally follows from the same Feynman diagram and enters into the imaginary part of the self energy with the opposite sign. Contributions like this are responsible for the fact that the imaginary part of the self energy is not symmetric in $k_0$ (cf. Appendix \ref{analytic}). Since they only affect the self energy of the baryon resonances in the far off-shell region, we can safely neglect such effects in this work.

In analogy to the meson sector, the presence of nuclear matter leads to spectral functions for baryon resonances, which depend on the polarization of the resonance. Since in nuclear matter both the realization of a polarized resonance state and its detection are impossible, we average over the spin of the resonance and obtain only one spectral function. Note that the self energy depends both on $k_0$ and $|{\bf k}|$.

As in the vacuum case, we calculate the real part of the self energy in the medium by means of a dispersion relation, which guarantees that the spectral function of the resonance remains normalized:
\beqa
\label{resrealmed}
	\real{\Sigma^+_{med}(k_0,{\bf k})} &=&  
	\,{\cal P}\,\int\limits_{E_F}^\infty\,\frac{d\omega}{\pi} 
	\frac{\imag{\Sigma^+_{med}(\omega,{\bf k})}}{\omega-k_0} - c_{vac}({\bf k}) \quad.
\eeqa
The lower integration bound is the Fermi energy $E_F$ below which $\imag{\Sigma^+_{med}(\omega,{\bf k})}$ vanishes in our calculation, see discussion after 
Eq. \ref{reswidthmed}. As pointed out there, contributions from below the Fermi surface exist. However, since they are far away from the resonance pole we expect only minor repercussions on the dispersion relation from neglecting these terms.
The subtraction constant $c_{vac}({\bf k})$ has been introduced in Eq. \ref{resodispvac}. We have not considered processes like $ R N \to R^\prime R^{\prime\prime}$, which on-shell are either closed or suppressed by phase space, because the relevant coupling constants are in most cases completely unknown.

Note that the resonance self energy of Eq. \ref{resrealmed} is defined with respect to the nucleon. In addition one might also think of mean field contributions (generated by $\sigma$ and $\omega$ exchange) to the self energy, giving rise to binding of the resonances and the nucleon. The effect of these mean-field terms can be sizeable \cite{larry}, but except for the nucleon their size is only poorly known. Therefore we have omitted such effects in the present calculation.

\begin{figure}[t]
\centering
\includegraphics[scale=0.6]{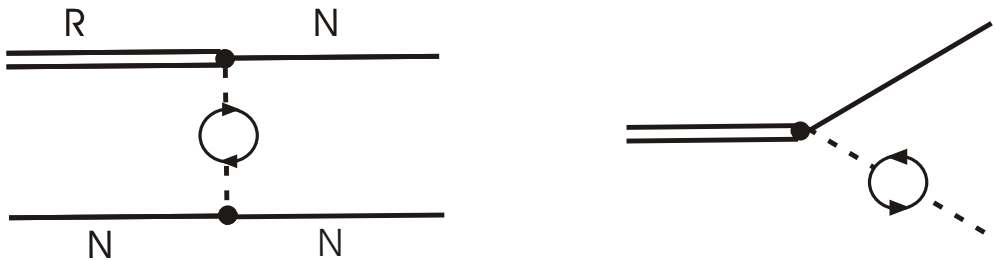}
\caption{\label{gammamed} Interpretation of the in-medium self energy as a sum of a collision term and the decay into a nucleon and a dressed meson.}
\end{figure}

The imaginary part of the resonance self energy contains two contributions, which are depicted in Fig. \ref{gammamed}. The left graph describes resonance-nucleon scattering processes with the exchange of (dressed) pions, $\rho$ mesons and $\eta$ mesons and arises from the particle-hole branches in the meson spectral function.
This way elastic scattering $RN \to NR$ or inelastic processes $RN \to NR^\prime$ and $RN \to NN$ are produced. The relative weight of these final states may be important in the interpretation of experimental results \cite{lehr}. 
The right graph in Fig. \ref{gammamed} describes the decay of the resonance into a nucleon (with Pauli-blocking) and a dressed meson. This contribution is generated from the meson peak of the spectral function. Note that the strength in this peak is smaller than one and that its dispersion relation changes compared to the vacuum. This implies that the in-medium width can be smaller than the vacuum one, even before Pauli-blocking is taken into account. In the result section we will find examples for that.
In the actual calculations many of the individual branches overlap, which makes a clear separation of the contributions impossible.

The self consistency problem outlined before is tackled with an iterative procedure. Having calculated the in-medium self energy of the resonances according to Eqs. \ref{reswidthmed} and \ref{resrealmed}, we can improve the meson self energy by replacing $\Sigma^+_{vac} \rightarrow \Sigma^+_{med}$ in Eqs. \ref{scalarself} and \ref{vectorself}:
\beqa
\label{mesonselfit}
\imag{\Pi_{\pi/\eta}^{2}(q_0,\bf q)} &=& I_\Pi\,\left(\frac{f}{m_\varphi}\right)^2
	\int \frac{d^3 p}{(2\,\pi)^3}\frac{\theta(p_F-|{\bf p}|)}{2\,E_N({\bf p})}
	\imag{\D{\frac{\Omega^{\varphi}}
	{k_0^2-E_R^2({\bf k})-\Sigma_{med}^+(k)}} } \\ 
\imag{\Pi_{\rho}^{T/L,2}(q_0,\bf q)} &=& I_\Pi\,\left(\frac{f}{m_\rho}\right)^2
	\int \frac{d^3 p}{(2\,\pi)^3}\frac{\theta(p_F-|{\bf p}|)}{2\,E_N({\bf p})}
	\imag{\D{\frac{\Omega^{T/L}}
	{k_0^2-E_R^2({\bf k})-\Sigma_{med}^+(k)}}} \nonumber \quad. 
\eeqa
Thus an improved meson spectral function ${\cal A}_{med}^{2}$ is generated, leading to a new guess for the resonance self energy $\Sigma_{med}^2$ and so forth. As it will turn out in the result section, we find convergence after at most four iterations.

\begin{figure}[t]
\centering
\includegraphics[scale=0.55]{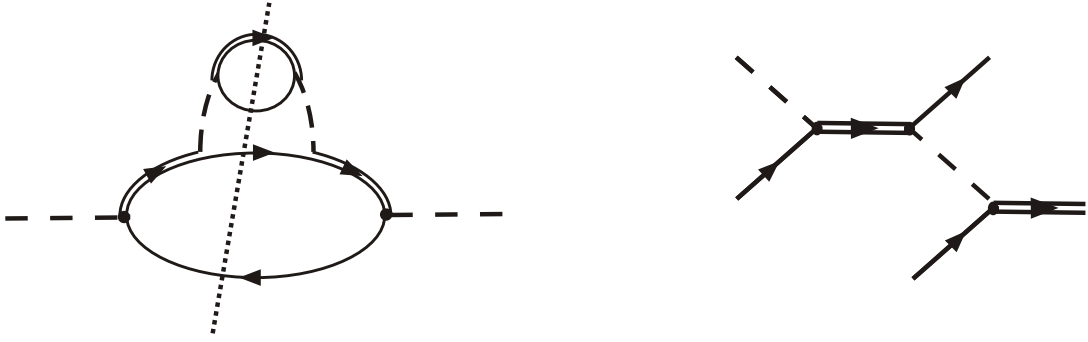}
\caption{\label{iterationcut} Interpretation of the in-medium self energy as a sum of a collision term and the decay into a nucleon and a dressed meson.}
\end{figure}

In Fig. \ref{iterationcut} we show the Feynman diagram corresponding to the second iteration Eq. \ref{mesonselfit}. The diagram on the left is the self energy as generated by plugging in the in-medium width of the resonance. The diagram on the right is obtained by cutting the self energy diagram and displays the physical scattering amplitude leading to the self energy. One sees that the iteration generates reactions of the incoming meson with more than one nucleon.

\section{The $RN$ interaction}
\label{nrint}

This Chapter is concerned about short-range correlations (SRC), which constitute an important ingredient of the resonance-nucleon interaction. The need for adding a short-range term to the conventional meson exchange potential is readily seen when considering the central part of the pion exchange contribution to the nucleon-nucleon potential. It contains an unphysical constant in momentum space or, equivalently, a $\delta$ contribution in position space:
\beqa
\label{vcorr1}
	V(q_0=0,{\bf q}) &=& -\frac{1}{3}\,\left( \frac{f_{NN}}{m_\pi}\right)^2\,\frac{{\bf q}^2}{{\bf q}^2+m_\pi^2}\,{\boldsymbol\sigma}_1\cdot{\boldsymbol\sigma}_2 \\  
	&=& -\frac{1}{3}\,\left( \frac{f_{NN\pi}}{m_\pi}\right)^2\,\left(1-\frac{m_\pi^2}{{\bf q}^2+m_\pi^2}\right)\,{\boldsymbol\sigma}_1\cdot{\boldsymbol\sigma}_2 	\nonumber \quad.
\eeqa
This can be removed by adding a short-range piece to the potential:
\beqa
\label{vcorr2}
	V_C(q_0,{\bf q}) &=& g\,\left(\frac{f_{NN\pi}}{m_\pi}\right)^2\,{\boldsymbol\sigma}_1\cdot{\boldsymbol\sigma}_2 
\eeqa
with $g=\frac 1 3$. Note that by adding $V_C$ a large correction of the potential for small momenta ${\bf q}$ is introduced. Central for the ${\bf q}^2$ in the numerator of Eq. \ref{vcorr1} is the $p$-wave coupling. Such a potential is not only encountered in the nucleon-nucleon case, but also for $RN$ scattering, provided the resonance has positive parity and spin-$\frac 1 2$ or spin-$\frac 3 2$. Indeed, it is well known that the inclusion of short-range terms is crucial for an understanding of $N N \rightarrow N N \pi$ scattering, where the $P_{33}(1232)$ plays a dominant role, see e.g. \cite{schaefer,larry} and references therein. Even more important for our work, a realistic description of the in-medium properties of the $P_{33}(1232)$ can only be achieved by including such corrections, which greatly reduce the in-medium broadening \cite{salcedo,helgesson1, helgesson2, lutzreso}. We will discuss this issue in Chapter \ref{results}.

This quenching of the in-medium broadening serves as a motivation for us to study SRC not only for positive parity states - where apart from the strength $g$ of the SRC no principal problems arise - but also for negative parity states like the $D_{13}(1520)$ and the $S_{11}(1535)$ resonance, which have spin-$\frac32$ and spin-$\frac12$, respectively.

We describe the SRC by means of contact interactions. These are derived both from the exchange of a pion and a $\rho$ meson, leading to ${\cal L}_C^\pi$ and ${\cal L}_C^\rho$ respectively:
\beqa
\label{lagc}
	{\cal L}_C^\pi &=& c_\pi\,J^\mu\,J_\mu	\quad,\\
  {\cal L}_C^\rho &=& c_\rho\,B^{\mu\,\nu}\,B_{\mu\,\nu} \nonumber \quad.
\eeqa
Then a non-relativistic reduction of these interactions is performed.
The current $J_\mu$ is obtained from the $\pi\,N\,R$ Lagrangians given in Appendix \ref{applag}, Eq. \ref{rlagpin}. The tensor $B_{\mu\nu}$ is defined by rewriting the $RN\rho$ Lagrangians given in Appendix \ref{applag}, Eq. \ref{rlagrhon} as:
\beqa
	{\cal L} &=& B^{\mu\,\nu}\,\partial_\mu\,\rho_\nu \quad.
\eeqa
To give a specific example, for the case of a $J^\pi=\frac32^+$ state, the current $J_\mu$ and the tensor $B_{\mu\nu}$ read:
\beqa
	J^\mu &=&\frac{f}{m_\pi}{\bar \psi_R}^\mu\,\psi_N  \quad, \\
	B^{\mu\nu} &=& i\,\frac{f}{m_\rho}\,
	\left({\bar \psi_R}^\mu\,\gamma^\nu\,-\,{\bar \psi_R}^\nu\,\gamma^\mu \right)
	\gamma_5\,\psi_N \nonumber \quad.
\eeqa
One might think of these contact interactions as arising from the exchange of a heavy meson with the quantum numbers $J^\pi=1^+$ $({\cal L}^\pi_C)$ or $J^\pi=2^+$ $({\cal L}^\rho_C)$. Sending the mass of the exchange particle to infinity, the contact interactions of Eq. \ref{lagc} result \cite{schaefer}.

Unfortunately, in contrast to positive parity resonances, not much is known about the strength of the contact interactions for $P=-1$ states. In a first attempt to obtain estimates for $c_\pi$ and $c_\rho$, we follow an approach which was originally introduced in \cite{osetweisecorr}. There correlations are generated by means of a correlation function $C({\bf r})$, such that $V(q_0,{\bf r}) \rightarrow V(q_0,{\bf r})\, C({\bf r})$. Taking $C(r)=1-j_0(q_c r)$, where $j_0(q_c r)$ is the lowest order Bessel function, the correlations produce the following additional term $V_C$ in momentum space:
\beqa
\label{corrint}
	V(q_0,{\bf q}) &\rightarrow& V(q_0,{\bf q}) - \frac{2\,\pi^2}{q_c^2}\,\int \frac{d^3q^\prime}{(2\pi)^3}\delta(|q-q^\prime|-q_c)\,V(q_0,{\bf q^\prime}) \\
	&=& V(q_0,{\bf q}) - V_C(q_0,{\bf q})\quad.\nonumber
\eeqa
The resulting integral is readily solved with an appropriate angular average \cite{osetweisecorr}. Within this model the strength of the short-range interactions is determined by the meson coupling constant (which is known) and the parameter $q_c$,
which is choosen to be $q_c = m_\omega$.
Thus, once $q_c$ is fixed the theory is essentially parameter free. The assumption is that $q_c$ itself does not change when considering different types of potentials. It has been demonstrated in \cite{osetweisecorr} that applying this method to the well studied cases nucleon and $P_{33}(1232)$ leads to reasonable estimates.

Let us give an argument as to why SRC are not necessarily negligible for $s$-wave potentials, which typically look like
\beqa
\label{vswave}
	V(q_0,{\bf q}) &=& f^2\,\frac{1}{q_0^2-{\bf q}^2-M^2} \quad.	
\eeqa
Although $V$ contains no $\delta$ type contribution in position space, it can be modified to the correlation integral Eq. \ref{corrint}. As will become clear from the discussion in the Appendices \ref{pospar} and \ref{negpar}, one obtains (see also \cite{osetswave}): 
\beqa
	V(q_0,{\bf q}) &=& f^2\,\frac{1}{q_0^2-{\bf q}^2-M^2} - f^2\,\frac{1}{q_0^2-{\bf q}^2-q_c^2-M^2} \quad.
\eeqa
Comparing this with the result from Eq. \ref{vcorr2} for the $p$-wave correlations shows that the corrections induced from $V_C$ with $q_c=m_\omega$ are relatively small at low momenta. However, with a sufficiently large value for the coupling constant $f$, $V_C$ may still be sizeable. Then the effects of SRC become important when the interactions are iterated in the nuclear medium. Therefore a discussion of the effects from SRC is also important for resonances with negative parity. In this work we stick to a non-relativistic treatment for simplicity.

For the $\eta$ the treatment of SRC is done in exactly the same way as for the pion. Therefore we do not mention this case specifically in the following.

In principle a relativistic description of the SRC is of great interest, in particular when considering the resonance-nucleon interaction for heavy resonances, where the exchanged $4$-momenta can become sizeable. For positive parity states first attempts towards a fully relativistic description, based on Lagrangians of the type ${\cal L}_C^\pi$, have recently been proposed in \cite{leinson,lutzshortrange}. However, structures as those generated by ${\cal L}_C^\rho$ have not yet been considered in both works. A consistent treatment is furthermore complicated by the fact that the accepted phenomenological range for values of $g^p$ needs to be rejected if the relativistic corrections are sizeable and new values have to be obtained by fits to observables.


\subsection{Contact Interactions}

As detailed in Appendix \ref{srcdetails}, starting from Eq. \ref{lagc} and performing a non-relativistic reduction, we find the following contact interactions for positive parity states:
\beq
\label{lagconplus}
\begin{array}{rclcc}
{\cal L}_C &=& \D{g^p\,\left(\frac{f}{m_\pi}\right)^2\,
	\left({\psi}_R^\dagger\,\sigma^i\,\psi_N \right)
	\left({\psi}_N^\dagger\,\sigma_i\,\psi_R \right)}
	&\mbox{for}&J^\pi=\frac{1}{2}^+ \\ \\
{\cal L}_C  &=& \D{g^p\,\left(\frac{f}{m_\pi}\right)^2\,
	\left({\psi}_R^\dagger\,S^{i\,\dagger}\,\psi_N \right)
	\left({\psi}_N^\dagger\,S_i\,\psi_R \right)}
	&\mbox{for}&J^\pi=\frac{3}{2}^+ 
\end{array}
\eeq
As shown in Appendix \ref{pospar}, for $P=+1$ states ${\cal L}_C^\pi$ and ${\cal L}_C^\rho$ have the same form in the non-relativistic limit and can be combined to give a single contact interaction. Concerning the strength parameter $g^p$, we adopt the following philosophy: rather than strictly following the results obtained from the correlation integral via the Eqs. \ref{vcorr3} and \ref{gprmatch1} in Appendix \ref{pospar}, for the nucleon and the $P_{33}(1232)$ resonance we determine $g^p$ from the requirement that the in-medium properties of the $P_{33}(1232)$ are described reasonably well. The resulting values $g^p$ are given in Table \ref{param}, Appendix \ref{parameters}. For the other $p$-wave resonances we then take the same strength parameters.

Spin-$\frac52$ resonances are treated in a more phenomenological manner: we have not attempted a construction of contact interactions from pion or from $\rho$ meson exchange. For the pion propagation in nuclear matter we neglect the effects from SRC in the $J^\pi=\frac52^+$ sector altogether. For the $\rho$ meson we carry over the results obtained in the $J^\pi=\frac12^+$ and $J^\pi=\frac32^+$ sectors. This ad-hoc prescription is motivated by the fact that subjecting the $\rho$ meson part of the exchange potential to the correlation integral leads to exactly the same results in all three sectors. This is due to the fact that the non-relativistic potentials are $p$-wave. For negative parity resonances we get the following Lagrangians:
\beq
\label{lagconminus}
\begin{array}{rclcc}
{\cal L}_C^\pi &=& \D{g_\pi^s\,\left(\frac{f}{m_\pi}\right)^2\,
	\left({\psi}_R^\dagger\,\psi_N \right)
	\left({\psi}_N^\dagger\,\psi_R \right) }
	&\mbox{for}&J^\pi=\frac{1}{2}^- \\ \\
	{\cal L}_C^\pi &=& \D{g_\pi^d\,\left(\frac{f}{m_\pi}\right)^2\,
	\bigg({\psi}_R^\dagger\,S^{i\,\dagger}\,\frac{\sigma_k\,\partial_k}
	{2 m_N}\,\psi_N \bigg) \bigg({\psi}_N^\dagger\,
	\frac{\stackrel{_\gets}{\partial_k} \,\sigma_k}{2 m_N}\,S_i\,\psi_R \bigg) }
	&\mbox{for}&J^\pi=\frac{3}{2}^- \\ \\
	{\cal L}_C^\rho &=& \D{g_\rho^s\,\left(\frac{f}{m_\rho}\right)^2\,
	\left({\psi}_R^\dagger\,\sigma^{i}\,\psi_N \right)
	\left({\psi}_N^\dagger\,\sigma_i\,\psi_R \right) }
	&\mbox{for}&J^\pi=\frac{1}{2}^- \\ \\
	{\cal L}_C^\rho &=& \D{g_\rho^s\,\left(\frac{f}{m_\rho}\right)^2\,
	\left({\psi}_R^\dagger\,S^{i\,\dagger}\,\psi_N \right)
	\left({\psi}_N^\dagger\,S_i\,\psi_R \right)}
	&\mbox{for}&J^\pi=\frac{3}{2}^- 
\end{array}
\eeq
As discussed in Appendix \ref{negpar}, in contrast to the $P=+1$ resonances, here pion and $\rho$ induced contact interactions are of a different structure and need to be treated separately. By $s/d$ we denote the angular momentum of the underlying meson-nucleon interaction and $\pi/\rho$ explain whether the contact interactions are derived from pion or from $\rho$ meson exchange. Estimates for the strength parameters are obtained from the correlation approach. Since for $P=+1$ states this method leads to reasonable results for the strength of the short-range interactions, we are confident that it is also applicable for negative parity states. Details are given in Appendix \ref{negpar} and explicit values for the strength parameters can be found in Table \ref{param}, Appendix \ref{parameters}.

In Appendix \ref{mixing} we discuss the possibility that -- mediated by the short-range interactions -- resonances with different quantum numbers can mix. The corresponding Lagrangians are obtained from those given in Eqs. \ref{lagconplus} and \ref{lagconminus}.The mixing of $J^\pi=\frac12^+$ or $J^\pi=\frac32^+$ states and $J^\pi=\frac32^-$ states is described by:
\beq
\label{lagconmix}
\begin{array}{rcl}
	{\cal L}_C &=&\D{g_\pi^{dp}\,\left(\frac{f}{m_\pi}\right)^2\,
	\bigg({\psi}_{R1}^\dagger\,S^{i\,\dagger}\,\frac{\sigma_k\,\partial_k}
	{2 m_N}\,\psi_N \bigg) \bigg({\psi}_N^\dagger\,S_i\,\psi_{R2} \bigg) } 
	\,\,+\,\, h.c.\\ \\
	{\cal L}_C &=&\D{g_\pi^{dp}\,\left(\frac{f}{m_\pi}\right)^2\,
	\bigg({\psi}_{R1}^\dagger\,S^{i\,\dagger}\,\frac{\sigma_k\,\partial_k}
	{2 m_N}\,\psi_N \bigg) \bigg({\psi}_N^\dagger\,\sigma_i\,\psi_{R2} \bigg) } \,\,+\,\, h.c.
\end{array}
\eeq
Here $dp$ indicates that the Lagrangian describes the mixing of $p$ and $d$ waves and the index $\pi$ implies that mixing takes place only in the pion sector. The coupling $f^2$ stands for the product of the coupling constants of both involved resonances.
Similarly, the mixing of $J^\pi=\frac12^+$ and $J^\pi=\frac32^+$ states as well as the mixing in the $\rho$ sector of $J^\pi=\frac12^-$ and $J^\pi=\frac32^-$ states follows from the Lagrangian:
\beq
\label{lagconmix2}
\begin{array}{rcl}
	{\cal L}_C &=& \D{g\,\left(\frac{f}{m_\pi}\right)^2\,
	\left({\psi}_{R1}^\dagger\,\sigma^i\,\psi_N \right)
	\left({\psi}_N^\dagger\,S_i\,\psi_{R2} \right)} \,\,+\,\, h.c.
\end{array}
\eeq
Here $g$ stands either for $g^p$ (positive parity states) or for $g_\rho^s$ (negative parity states).


\subsection{Effect on the self energies $\Pi_{med}$ and $\Sigma_{med}$}
\label{resoselfci}

In this Section we discuss how the previous results for the in-medium self energies of mesons and baryon resonances, Eqs. \ref{reswidthmed} and \ref{mesonselfit}, are modified in the 
presence of short range interactions.

For notational convenience, let us first introduce a new quantity for nucleon resonances:
\beqa
\label{chi}
	\chi_M(q_0,{\bf q}) &=& I_\Pi\,\left(\frac{f}{m_M}\right)^2
	\int \frac{d^3 p}{(2\,\pi)^3}\frac{n(\bf p)}{2\,E_N({\bf p})}
	\D{\frac{F_s^2(k)\,\Omega^{red}}
	{k_0^2-E_R^2({\bf k})-\Sigma_{med}(k)}} \quad,
\eeqa
where $k=p+q$ is the resonance 4-vector. The form factor $F_s(k)$ has been introduced around Eq. \ref{ff1}. The trace $\Omega^{red}$ arises from tracing the contact interactions. Results for $\Omega^{red}$ are given in Appendix \ref{applag} in Table \ref{ntracesred}.
For the nucleon we define in analogy:
\beqa
\label{chinn}
	\chi_M^{N}(q_0,{\bf q}) &=&  4\, \left(\frac{f_{NNM}}{m_M}\right)^2\,
	U_N(q) \quad.
\eeqa

Since the short-range interactions couple resonances of various quantum numbers to each other, the resulting coupled channel problem ought to be written down in a matrix formulation. 
Before proceeding we define the quantities 
\beqa
\label{grhop}
g_\pi^p &\equiv& g^p \\
g_\rho^p &\equiv& \frac{f_{RN\pi}^2}{f_{RN\rho}^2}\,\frac{m_\rho^2}{m_\pi^2}\,g^p \nonumber \quad.
\eeqa
and similarly if the interaction of two different resonances is considered. Then we can introduce the following matrices:
\beqa
	{\boldsymbol \chi}_\pi &=& \left( 
			\begin{array}{cccc} 
			   {\boldsymbol \chi}_\pi^p & 0 & 0 & 0
			\\ 0 & {\boldsymbol \chi_\pi^d} & 0 & 0 
			\\ 0 & 0 & {\boldsymbol \chi_\pi^s} & 0 
			\\ 0 & 0 & 0 & {\boldsymbol \chi_\pi^f} 
			\end{array} \right) \quad,\\ &&\nonumber \\ &&\nonumber \\
  {\bf g}_\pi &=& \left( 
  \begin{array}{cccc} 
     {\bf g}_\pi^p & {\bf g}_\pi^{dp} & 0 & 0
  \\ {\bf g}_\pi^{dp} & {\bf g}_\pi^d & 0 & 0
  \\ 0 & 0 & 	{\bf g}_\pi^s  & 0
  \\ 0 & 0 &  0 & 0 
  \end{array} \right) \quad, \\ &&\nonumber \\ &&\nonumber \\
  {\boldsymbol \Pi}_\pi^0 &=& \left( 
			\begin{array}{cccc} 
			   {\boldsymbol \Pi}_\pi^p & 0 & 0 & 0 
			\\ 0 & {\boldsymbol \Pi_\pi^d} & 0 & 0 
			\\ 0 & 0 & {\boldsymbol \Pi_\pi^s} & 0 
			\\ 0 & 0 & 0 & {\boldsymbol \Pi_\pi^f} 
			\end{array} \right) \quad.
\eeqa
Each entry is a matrix itself. For example, ${\boldsymbol \chi_\pi^p}$ is a diagonal matrix consisting of all the states to which the pion couples in a $p$-wave and similarly for 
${\boldsymbol \chi}_\pi^d$ and ${\boldsymbol \chi}_\pi^s$. The diagonal elements of the matrix ${\boldsymbol \Pi}_\pi^0$ contain the self energies, Eqs. \ref{mesonselfit} and \ref{realpi}, grouped in the same order as in the matrix ${\boldsymbol \chi}_\pi$.
Since we allow for different $g_\pi^p$ parameters, we find for the matrix ${\bf g}_\pi^p$ assuming that $n$ resonances couple in a $p$-wave:
\beqa
	{\bf g}_\pi^p = \left( \begin{array}{cccc}
	g_\pi^{p,NN} & g_\pi^{p,R_1N} & \cdots & g_\pi^{p,R_n N} \\ g_\pi^{p,R_1N} & g_\pi^{p,R_1 R_1} & & \\
	\vdots & & \ddots& \\ g_\pi^{p,R_n N} & &  & g_\pi^{p,R_n R_n} \end{array} \right) \quad.
\eeqa
In the following we take all entries but $g_\pi^{NN}$ to be the same, see Table \ref{param}.
The matrices ${\bf g}_\pi^s$  and ${\bf g}_\pi^d$ describe the coupling of $s$- and $d$-wave states to themselves. The structure of the matrix ${\bf g}_\pi$ indicates the mixing of resonances with different quantum numbers as discussed above. 
Thus, we have a coupling of $p$- and $d$-wave states represented by ${\bf g}_\pi^{dp}$, whereas the $s$-wave states couple only to themselves. We take the same entries for $g_\pi^{d}$ and $g_\pi^{dp}$. For spin-$\frac52$ states, which couple in an $f$-wave to $N\pi$, we have no short range correlations, which is indicated by the entry $0$ for $f$-wave resonances in ${\bf g}_\pi$. In Table \ref{param} we give explicit values for the short-range parameters.

For the $\rho$ meson one can introduce similar matrices ${\boldsymbol \chi}_\rho$ and ${\bf g}_\rho$ which are constructed in exactly the same way:
\beqa
	{\boldsymbol \chi}_\rho &=& \left( 
			\begin{array}{cc} {\boldsymbol \chi}_\rho^p & 0  \\ 0 & 
			{\boldsymbol \chi_\rho^s}  
			\end{array} \right) \quad, \\ &&\nonumber \\ &&\nonumber \\
  {\bf g}_\rho &=& \left( 
  \begin{array}{cc} {\bf g}_\rho^p & 0 \\ 0 & {\bf g}_\rho^s  \end{array} \right)\quad,
  \\ &&\nonumber  \\ &&\nonumber \\
  {\boldsymbol \Pi}_\rho^{0,T/L} &=& \left( 
  \begin{array}{cc} {\boldsymbol \Pi}_\rho^{p,T/L} & 0 \\ 0 & {\boldsymbol \Pi}_\rho^{s,T/L}  \end{array} \right) \quad.
\eeqa
The matrix ${\bf g}_\rho^p$ looks formally identical to ${\bf g}_\pi^p$. For simplicity, we relax the relation between $g_\pi^p$ and $g_\rho^p$ Eq. \ref{grhop} for the two $J^\pi=\frac32^+$ states $P_{13}(1720)$ and $P_{13}(1878)$ and take the same values $g_\rho^p$ as for the $P_{33}(1232)$. This is reasonable, since the vacuum properties of these states are only poorly known. In the case of the  $J^\pi=\frac12^+$ state $P_{11}(1440)$ this problem does not arise since in our model it does not couple to the $\rho$ meson. For $J^\pi=\frac52^+$ states we take $g_\rho^p=g_\rho^{p,\Delta\,N}$.
The matrix ${\boldsymbol \Pi_\rho}^{0,T/L}$ contains the self energies from Eqs. \ref{mesonselfit} and \ref{realpi}.

For the $\eta$ meson the corresponding matrices ${\boldsymbol \chi}_\eta$, ${\bf g}_\eta$
and ${\boldsymbol \Pi_\eta^0}$ have the same form as for the $\rho$. Here the only state coupling in a $p$-wave is the nucleon, all other resonances couple in an $s$-wave, see also Table \ref{restable}.

Let us finally introduce the vector ${\bf v}$ as 
\beqa
	{\bf v}^T &=& \underbrace{\left(\begin{array}{cccc} 1&1&\ldots &1 \end{array}\right)}_{N\times} \quad,
\eeqa
where $N$ is number of resonances included in the calculation.

\begin{figure}[t]
\centering
\includegraphics[width=9cm]{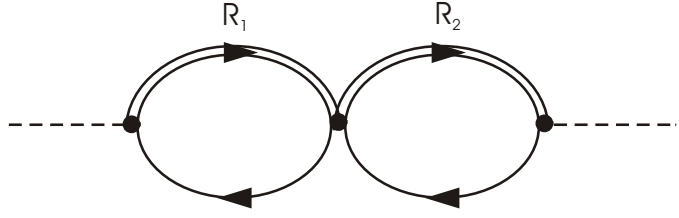}
\caption{\label{resholegpr} Lowest order correction of the self energy from the contact interaction. The double-line stands for any of the resonances or the nucleon and does not need to represent the same state in both loops. }
\end{figure}

Having defined these quantities, we can now proceed and present the results for the self energies. In Fig. \ref{resholegpr} we display the lowest order correction to the meson self energy. Summing up to all orders leads to the following result for the self energy of meson $M$ \cite{helgesson1}:
\beqa
\label{pici}
	\Pi_M(q_0,{\bf q}) &=& F^2(k,q)\,{\bf v}^T\,\frac{1}{1-{\bf g}_M\,{\boldsymbol \chi}_M}\, {\boldsymbol \Pi}_{M}^0\,{\bf v} \quad.
\eeqa 

\begin{figure}[t]
\centering
\includegraphics[scale=0.5]{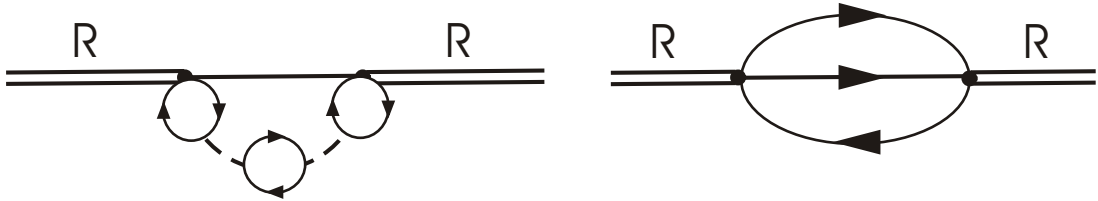}
\caption{\label{collbroadgpr}The resonance self energy in the presence of short-range correlations (SRC). The loop insertion stands for any particle-hole state discussed in the text. Note that the displayed diagrams are (only) typical examples for loop insertions in the propagators and vertices. In the calculations these insertions are of course iterated according to Dyson's equation, cf. Eqs. \ref{respici}, \ref{resetaci}, \ref{resrhoci}.  }
\end{figure}

Taking into account the contact interactions, the resonance self energy consists of two diagrams, see Fig. \ref{collbroadgpr}. In terms of scattering processes these diagrams correspond to the coherent sum of a scattering with meson exchange and scattering  via a contact interaction. Note that each loop stands for either the fully iterated vertex correction or the full in-medium self energy of the meson according to Eq. \ref{pici}. Thus for vanishing short-range correlations the diagram of Fig. \ref{resoselfmed} is obtained. Iterating the particle-hole insertions generates vertex corrections at the resonance-nucleon-meson vertices. We split the resonance self energy into three parts:
\beqa
	\imag{\Sigma(k_0,{\bf k})} &=&  \imag{\Sigma_\pi(k_0,{\bf k})} +\imag{\Sigma_\rho(k_0,{\bf k})} +\imag{\Sigma_\eta(k_0,{\bf k})}
\eeqa
according to whether the resonance has decayed into a medium-modified pion, $\eta $ or $\rho$ meson. Throughout the following formulae $\Omega_\varphi$ and $\Omega_{T/L}$ as well as $\Omega^{red}$ have to be chosen according to the quantum numbers of the resonance.

For the imaginary part of the resonance self energy  one obtains a matrix \cite{helgesson1}. The diagonal elements of this matrix yield the imaginary part of the self energy for the individual resonances. Thus we obtain for the pionic decay mode:
\beqa
\label{respici}
	\imag{{\boldsymbol \Sigma}_\pi(k_0,{\bf k})} &=&
	-\frac{I_{\Sigma}}{2j+1}\left(\frac{f}{m_\pi}\right)^2 \,\, \int \frac{dp\,p^2}{8\pi^2}
	\int\limits_{-1}^{+1}dx\,\frac{1}{E_N} \times \\ && \hspace{-3cm}\times 
	{\cal I}m\,\left[\D{\Omega\,D_\pi\,F^2(k,(k-p))\frac{1}{1-{\bf g}_\pi\,{\boldsymbol \chi}_\pi}
	\,{\bf v}\,{\bf v}^T\, \frac{1}{1-{\boldsymbol \chi}_\pi\,{\bf g}_\pi} 
	+ \Omega^{red}\,F_s^2(k)\,{\bf g}_\pi\,\frac{1}{1-{\boldsymbol \chi}_\pi\,{\bf g}_\pi} }\right]						 \nonumber \quad.
\eeqa
For the decay into an $\eta$ meson one obtains the same result, if the appropriate matrices are chosen:
\beqa
\label{resetaci}
	\imag{{\boldsymbol \Sigma}_\eta(k_0,{\bf k})} &=&
	-\frac{I_{\Sigma}}{2j+1}\left(\frac{f}{m_\eta}\right)^2 \,\, \int \frac{dp\,p^2}{8\pi^2}
	\int\limits_{-1}^{+1}dx\,\frac{1}{E_N} \times \\ && \hspace{-3cm}\times 
	{\cal I}m\,\left[\D{\Omega\,D_\eta\,F^2(k,(k-p))\frac{1}{1-{\bf g}_\eta\,{\boldsymbol \chi}_\eta}
	\,{\bf v}\,{\bf v}^T\, \frac{1}{1-{\boldsymbol \chi}_\eta\,{\bf g}_\eta} 
	+ \Omega^{red}\,F_s^2(k)\,{\bf g}_\eta\,\frac{1}{1-{\boldsymbol \chi}_\eta\,{\bf g}_\eta} }\right]						 \nonumber \quad.
\eeqa
Finally, for the decay into a $\rho$ meson we find:
\beqa
\label{resrhoci}
	\imag{{\boldsymbol \Sigma}_\rho(k_0,{\bf k})} &=&
	-\frac{I_{\Sigma}}{2j+1}\left(\frac{f}{m_\rho}\right)^2 \,\, \int \frac{dp\,p^2}{8\pi^2}
	\int\limits_{-1}^{+1}dx\,\frac{1}{E_N} \times \\ && \hspace{-2cm} \times \,
	{\cal I}m\,\left[\D{(2\,\Omega^T\,D_\rho^T+\Omega^L\,D_\rho^L)\,F^2(k,(k-p))
	\,\frac{1}{1-{\bf g}_\rho\,{\boldsymbol \chi}_\rho}
	\,{\bf v}\,{\bf v}^T\, \frac{1}{1-{\boldsymbol \chi}_\rho\,{\bf g}_\rho} + }
	\right. \nonumber \\ && \hspace{-1cm} \left. \D{
	+ \,3\,(2)\Omega^{red}\,F_s^2(k)\,{\bf g}_\rho\,\frac{1}{1-{\boldsymbol \chi}_\rho\,{\bf g}_\rho} }\right]						 \nonumber \quad.
\eeqa
Since a longitudinal $\rho$ meson does not couple to $P=+1$ resonances, we have a factor of $2$ instead of $3$ in the second term for these states.

\section{Results}
\label{results}

When discussing the results for the meson and baryon resonances, we refer to a calculation according to Eqs. \ref{scalarself} and \ref{vectorself} (mesons) or Eqs. \ref{reswidthmed} and \ref{resrealmed} as "first iteration". Corrections to these results arising from the self-consistent iteration, see Eq. \ref{mesonselfit}, are referred to as "second iteration", "third iteration" and so on and the results achieved after convergence is reached are denoted by "self consistent (SC)".


\subsection{Mesons}

Let us begin the discussion of our results for the mesons $\pi$, $\eta$ and $\rho$ with some general remarks. The excitation of particle-hole loops leads to remarkably rich structures in the meson spectral functions. In principle, each particle-hole loop generates an additional branch. The invariant mass squared $q^2$ of this branch moves down to smaller values as the $3$-momentum increases, eventually reaching space-like kinematics $q_0^2 < {\bf q}^2$ \cite{postrho1}. We have plotted the invariant mass squared of various resonance-hole excitations as a function of their $3$-momentum ${\bf q}$ in Fig. \ref{branches}. It follows that if enough spectral strength is sitting in these branches, one can expect a considerable population of states with small or even negative squared invariant masses. In the following three sections we will study this question in detail. Note that this issue is of particular interest for the $\rho$ meson, where dilepton spectra indicate a shift of spectral strength to smaller invariant masses \cite{rwrev}.

\subsubsection{$\rho$ Meson}
\label{rhores}

\begin{figure}[h!]
\centering
\includegraphics[width=15cm]{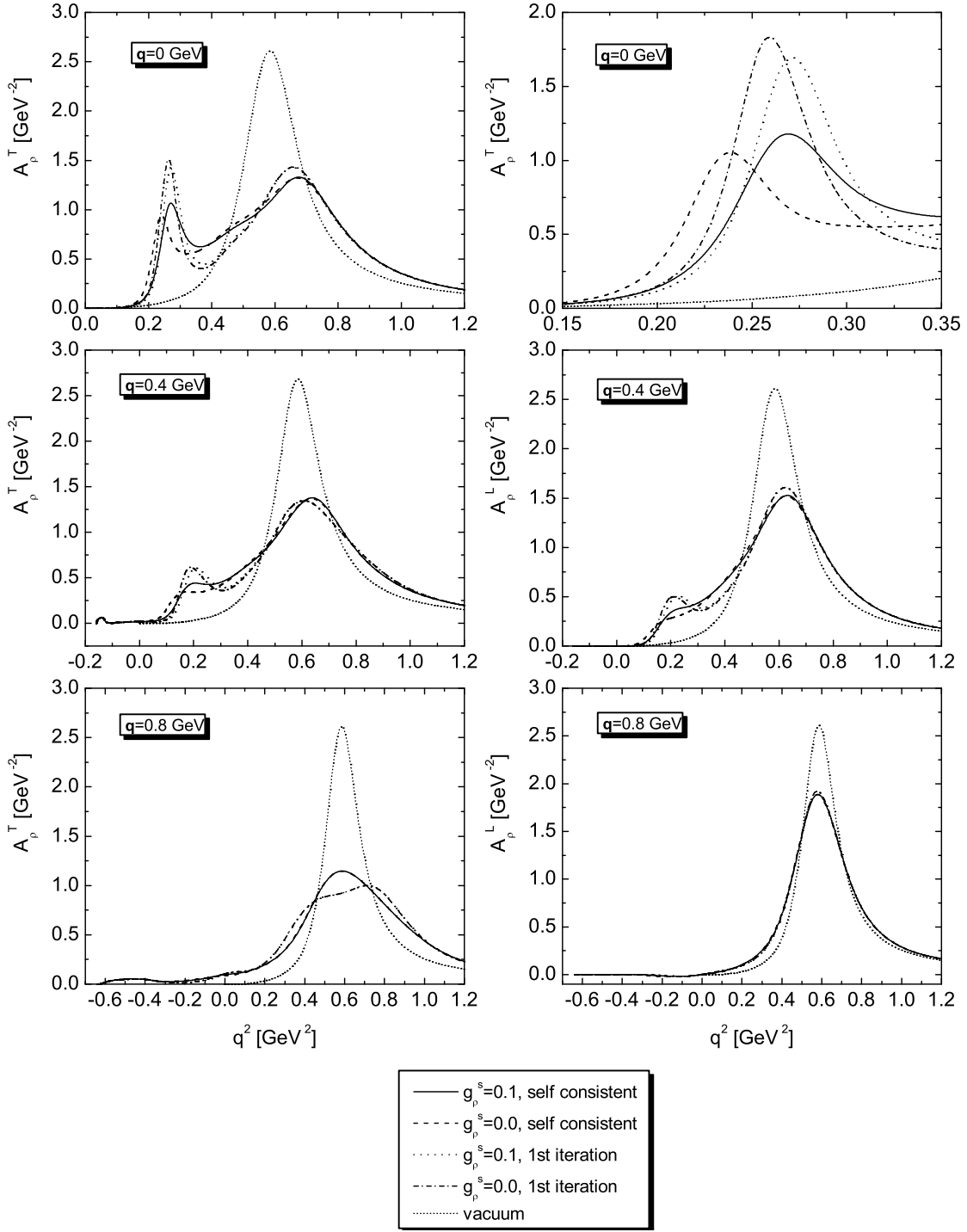} 
\caption{\label{rhospec2d}Spectral function of the $\rho$ meson at normal nuclear matter density. Shown are the transverse and longitudinal spectral functions $A_\rho^T$ and $A_\rho^L$, which are degenerate at ${\bf q}=0$ GeV. Shown are the effects of iterating the spectral function and of varying the short-range parameter $g_\rho^s$. The picture in upper right corner represents a zoom around the $D_{13}(1520)N^{-1}$ peak for ${\bf q}=0$ GeV.}
\end{figure}

In Fig. \ref{rhospec2d} we show the spectral functions $A_\rho^T$ and $A_\rho^L$ for momenta $0$, $0.4$ and $0.8$ GeV at normal nuclear matter density $\rho_0=0.15\,\textrm{fm}^{-3}$. Let us first discuss the general features of the results before turning to the details. Note that as far as the first iteration is concerned the results presented here are in quantitative agreement with our previous calculation \cite{postrho2}. A self-consistent scheme was not pursued there. We begin with the transverse channel $A_\rho^T$. Due to the large width of many of the involved resonances and their close overlap with the broad $\rho$ meson, the individual peaks cannot be identified for all resonances. However, the structure coming from the $D_{13}(1520)$ is clearly seen. This state couples in a relative $s$-wave and is responsible for the peak at invariant masses below the $\rho$ seen at small momenta. If the momentum increases, the relative importance of this state is reduced due to the fact that it moves away from the $\rho$ pole. Then it is the sum of a few higher lying $p$-wave states like the $F_{35}(1905)$ or the $P_{13}(1720)$ which mostly affects $A_\rho$ \cite{fripir,postrho2}. Because they couple in a $p$-wave, these resonances are not seen at small momenta. Thus in the transverse channel the following general picture emerges: at small momenta the spectrum is dominated by the excitation of the $D_{13}(1520)$ state, leading to a pronounced double-peak structure. Increasing the momentum, the additional peak diminishes, but a sizeable broadening of the original $\rho$ peak persists, which varies from $\Gamma_{med}=\imag{\Pi_{\rho}^{T}(q^2=m_\rho^2)}/m_\rho \approx 130$ MeV at ${\bf q}\approx 0.4$ GeV to $\Gamma_{med}=250$ MeV at ${\bf q}=0.8$ GeV. One should not conclude that the broadening keeps increasing with momentum. Within our model, there is a certain momentum above which all the resonances have passed the point $q^2=m_\rho^2$. Beyond that momentum the particle-hole excitations move away from the resonance and their influence on the $\rho$ spectral function becomes less important.

The different momentum dependence of $A_\rho^T$ and $A_\rho^L$, which is expected from 
the existence of a preferred Lorentz frame in the presence of nuclear matter
(see Chapter \ref{mesoself}), is nicely displayed in the results Fig. \ref{rhospec2d}. Whereas at small momenta up to $0.4$ GeV both quantities develop a similar behaviour, for large momenta around $0.8$ GeV $A_\rho^L$ is much less modified and starts resembling the vacuum spectral function. This is due to the fact that the $p$-wave resonances - responsible for the broadening of the meson peak in $A_\rho^T$ - do not couple to a longitudinal $\rho$ meson and that - apart from the $D_{13}(1520)$ - the $s$-wave states do not couple strongly enough to cause a large effect. Furthermore, their coupling is proportional to $q^2$ and gets therefore smaller when the momentum increases until it eventually vanishes in the vicinity of the photon point $q^2= 0$, which is approached by states with a mass around $1.5$ GeV at a momentum of $0.8$ GeV.

In Fig. \ref{rhospec2d}  we also compare the effects from higher order corrections in the density and from inclusion of the short-range correlations (SRC) for $s$-wave states. 
For $p$-waves, the SRC are always taken into account.
Concerning the iterations, we obtain good convergence after maximal four iterations. At low momenta the main effect of the iterations is to smear out the region around the peak generated by the $D_{13}(1520)$ (see especially top right plot in Fig. \ref{rhospec2d}). 
This is due to a broadening of that state, which is discussed below in Section \ref{resd13}. The strength sitting in the $\rho$ peak remains stable. At larger momenta one observes a minor shift of the $\rho$ peak down to smaller invariant masses. This shift is due to a combined effect of the in-medium modifications of the higher lying resonances and is therefore not easily disentangled. 
In our previous publication \cite{postrho1}, the iterations led to a structureless $\rho$ spectral function due to a large in-medium broadening of the baryon resonances. We find that this discrepancy results from the use of different form factors. In the previous publication the form factor did not depend on the invariant energy $k^2$ of the baryon resonance, but on the $3$-momentum ${\bf q}$ of the $\rho$ meson relative to nuclear matter, leading to much larger self energies away from the resonance-hole peak. This way the in-medium broadening of baryon resonances was allowed to generate large effects far away from the resonance peak, in particular in the vicinity of the $\rho$ peak. Using an $k^2$ dependent form factor, the effects of resonance broadening are confined to the region around the resonance peak. Thus the width of the $\rho$ peak remains smaller than before and at large momenta the $\rho$ regains a quasi-particle structure. This effect does not depend on the exact shape of the form factor as long as it leads to a suppression at large energies. The form factor at the meson-nucleon-resonance vertex does not only affect the meson self energy $\Pi_{\rho}^{T/L}$, but also the resonance self energy. There such a suppression is necessary to achieve that $\imag{\Sigma}$ does not grow faster than $\sqrt{s}$ for large values of $s$, which is mandatory if one insists on the spectral function to be normalized, see Appendix A in \cite{leupoldtest}. We conclude that with the new form factor the effects from self-consistency are estimated more reliably.

The SRC for $s$-wave states have mainly the effect of moving the $D_{13}$ peak slightly up, leaving the gross structure of the results untouched (see again top right plot in Fig. \ref{rhospec2d}).
The repulsive nature of the SRC is well known. Putting the resonance width to zero, the contribution of the $D_{13}(1520)$ to the transverse self energy can be cast into the form (cf. also \cite{henning}):
\beqa
	\Pi_{\rho}^T(q_0,{\bf q})&=& q_0^2\,\frac{\chi_s}{1-g_\rho^s\,\chi_s} \\
	                       &=& q_0^2\,\frac{C}{q_0^2-{\bar E}^2-g_\rho^s\,C} \nonumber \quad,
\eeqa
where we have introduced ${\bar E}= E_R-m_N$, simulating the kinematical situation of a $\rho$ meson scattering on a nucleon at rest. The constant $C>0$ is proportional to the density and the coupling constant. One sees that the inclusion of SRC acts like a repulsive mass shift of the resonance. This effect is enhanced by the fact that the attractive in-medium shift of the peak of the spectral function of the $D_{13}(1520)$ is less pronounced once the short-range interactions are switched on, see Section \ref{resd13}. Since $s$-wave states are less important at large momenta, we find virtually no influence of $g_\rho^s$ on the results. Therefore only three curves can be distinguished in the bottom of Fig. \ref{rhospec2d}.
Summarizing, the spectral function is rather stable with respect to SRC in the $s$-wave sector, which mainly influence the details around the $D_{13}$ peak at small momenta.

\begin{figure}[t]
\centering
\includegraphics[scale=1.1]{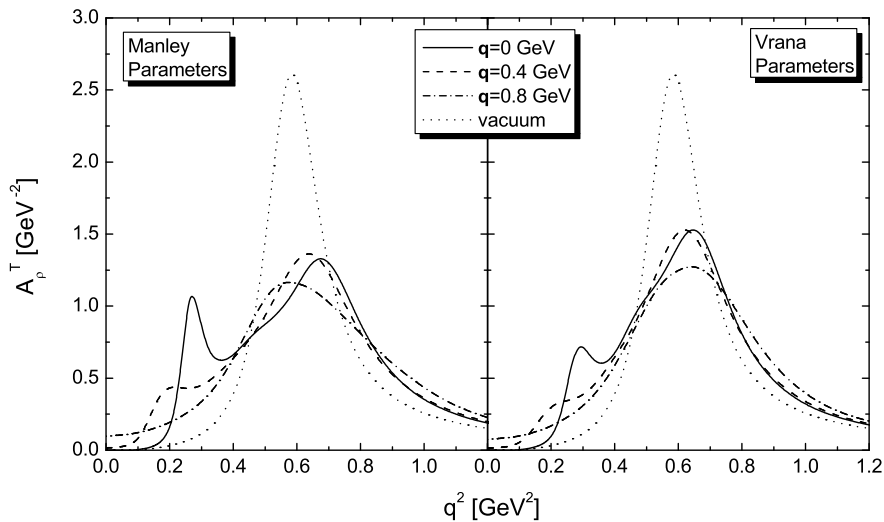} 
\caption{\label{manvrana}Comparison of a calculation of $A_\rho^T$ employing the parameter set of Manley \cite{manley1}(left) and that of Vrana \cite{vrana} (right). The calculations are done at a density $\rho=\rho_0$. Note that we only show the timelike part of the spectrum since there are no differences in the parameter sets for the $P_{33}(1232)$ and the nucleon. For the SRC in the $s$-wave sector we take $g_\rho^s=0.1$.}
\end{figure}

In Fig. \ref{manvrana} we study the influence of the resonance parameters as extracted from \cite{manley1} and \cite{vrana}. As explained in Chapter \ref{exp}, those analyses differ in that they assign different strength to the $N\,\rho$ channel, the analysis \cite{manley1} favouring larger values for the partial decay widths. As can be seen, the differences between both results are most pronounced around the $D_{13}(1520)$ peak, where the smaller coupling leads to a reduced influence of that state. However, at lot of strength is still removed from the $\rho$ peak. At larger momenta the differences are quite small and the broadening of the $\rho$ peak remains untouched.

\paragraph{Comparison with other models:} 
Let us now compare our results with those obtained from other models for the $\rho$ meson. In the works of \cite{rw,osetrho} the main source of in-medium modifications is due to the renormalization of the $\rho\pi\pi$ decay in the nuclear medium, generated by the coupling of pions to nucleon-hole and $\Delta$-hole states. On the level of scattering amplitudes this corresponds to a consideration of background terms of the $\rho N$ scattering amplitude.
The overall picture emerging from such works is a substantial broadening of the $\rho$ peak, accompanied by a slight repulsion. In addition, the effects from coupling the $\rho$ to $D_{13}(1520) N^{-1}$ holes have been estimated in \cite{osetrho} and the peak structure reported here and in our previous publications has been qualitatively confirmed. 
In \cite{rwurban} the momentum dependence of the spectral function inherent to such models has been studied and found to be small. This is in clear contrast to the finding in our work. As shown in Fig. \ref{rhospec2d} the resonance-hole loops create a sizeable dependence on ${\bf q}$.
In the work of \cite{klinglweise} the $\rho N$ forward scattering amplitude has been calculated based on a combination of vector meson dominance (VMD) and heavy baryon chiral perturbation theory. There a strong broadening of the $\rho$ in combination with attractive mass shift is reported. As compared to our scheme the models \cite{rw,osetrho,klinglweise,rwurban} are clearly more elaborate concerning the in-medium $\rho\pi\pi$ decay. Qualitatively, however, such effects only lead to an additional broadening and shift of the $\rho$ peak. On the other hand, the gross features of the spectral function -- especially the rich peak structure -- is given by the resonance-hole excitations studied here with great sophistication. 

Closer in spirit to our approach are the works of \cite{fripir,lutzvector,lutzvector1}. In \cite{fripir} the effects of coupling the $\rho$ to two $p$-wave resonances, the $P_{13}(1720)$ and the $F_{35}(1905)$, are considered, which only contribute at finite momenta. This model predicts the existence of additional peaks in the spectral function and a broadening of the original $\rho$ peak. In \cite{lutzvector,lutzvector1} the $\rho N$ scattering amplitude is generated as a solution of a coupled-channel Bethe-Salpeter equation. This way resonant structures are formed dynamically. This analysis is restricted to small momenta since no $p$-wave states are incorporated in the model. For a $\rho$ at rest the in-medium modifications are found to be much smaller than in our work, owing to a much smaller coupling of $\rho N$ in the $D_{13}$ channel (see also discussion in Section \ref{expd13}). For a more detailed overview of the different models we refer the reader to \cite{rw}.

Let us finally comment on contact or tadpole diagrams, which generate a shift of the peak position of the $\rho$ meson, but do not lead to an additional broadening. Typical examples are the $\rho \rho N N$ contact interaction \cite{herrmann,osetrho} or the tadpole diagram from $\sigma$ exchange \cite{frisoy}. The former can be reliably estimated to be repulsive and in the order of $10$ MeV at $\rho_0$ \cite{herrmann,osetrho}, while the $\sigma$-tadpole diagram is subject to large model dependencies. In \cite{frisoy} it gives rise to an attraction of about $130$ MeV at $\rho_0$. Due to these uncertainties we have decided not to include these diagrams in our calculation, although they might be of relevance for the numerical results.

Despite their strongly differing model ingredients, most state of the art models -- including the present one -- agree in that they predict a shift of spectral strength down to smaller invariant masses. This is also required by the QCD sum rules \cite{sumrulehatsuda,klinglweise,sumrulestefan} and offers a possible explanation for the evident shift in the dilepton spectra measured in heavy ion collisions.

\begin{figure}[t]
\centering
\includegraphics[scale=1.0]{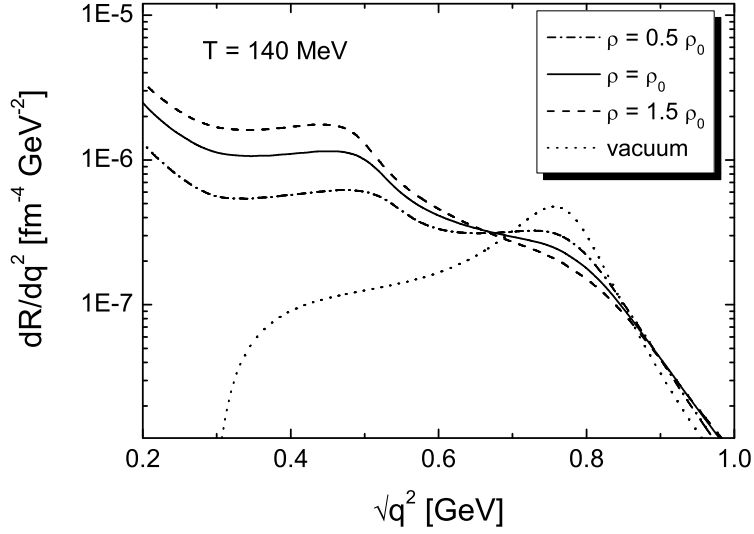} 
\caption{\label{dilep} Momentum integrated dilepton rates at a temperature $T=140$ MeV for three different densities, $0.5\,\rho_0$ (dashed-dotted), $\rho_0$ (solid) and $1.5\,\rho_0$ (dashed). In all curves SRC with $g_\rho^s=0.1$ and $g_\pi^d=0.4$ are included. }
\end{figure}

\paragraph{Dilepton Spectra:}
Much of the interest in the in-medium properties of the $\rho$ meson has been triggered by dilepton spectra of the CERES/NA45 \cite{aga1,aga2,lenkeit,wessels} and the HELIOS \cite{masera} collaboration, indicating an enhancement of spectral strength below the free $\rho$ mass.
We turn therefore to a computation of momentum integrated dilepton rates as resulting from our model for the $\rho$ meson. This rate is defined as \cite{rwrev,gale}:
\beqa
	\frac{dR_{e^+e^-}}{dq^2}(q) &=& \int \frac{d^3q}{2 q_0}\,
	\frac{d^4R_{e^+e^-}(q_0,{\bf q})}{dq_0\,d^3q} \quad.
\eeqa
Using the assumption of strict Vector Meson Dominance (VMD), the four-fold momentum differential rate is directly related to the transverse and longitudinal spectral functions of the $\rho$ meson:
\beqa
\label{dRd4q}
	\frac{d^4R_{e^+e^-}}{dq_0 \,d^3q} &=& \frac{\alpha^2}{\pi^2\,q^2}\,n_B(q_0,T)\,\frac{m_\rho^{0,4}}{g_\rho^2}\,\left[\frac23\,{\cal A}_{\rho}^{T}(q_0,{\bf q}) + \frac13\,{\cal A}_{\rho}^L(q_0,{\bf q}) \right] \quad.
\eeqa
Here $\alpha$ is the electromagnetic fine-structure constant and $g_\rho=6.05$ denotes the coupling strength of the $\rho$ to the photon. The thermal Bose occupation factor reads:
\beqa
\label{bose}
	n_B(q_0,T) &=& \frac{1}{e^{q_0/T}-1} \quad.
\eeqa
In Fig. \ref{dilep} we present results for the momentum integrated dilepton rates $\frac{dR_{e^+e^-}}{dq^2}(q)$ for the densities $\rho=0.5\,\rho_0$, $\rho=\rho_0$ and $\rho=1.5\,\rho_0$. For the temperature we take $T=140$ MeV, which should be typical for SPS energies \cite{rwrev,fripir}. For comparison we have also plotted the dilepton spectrum as resulting from using the vacuum $\rho$ propagator in Eq. \ref{dRd4q}. No experimental acceptance cuts are taken into account.

The overall picture is that our model produces a substantial reduction of strength around the free $\rho$ peak and leads to a strong enhancement of the dilepton yield at small invariant masses around $300-600$ MeV, where the resonance-hole contributions dominate the dilepton spectrum. The strong population of small invariant masses is due to the
excitation of the resonance-hole pairs, for example the $D_{13}(1520)$ and the $P_{13}(1720)$ states. It is further enhanced by the 
factor $1/q^2$ as well as the thermal Bose distribution factor in Eq. \ref{dRd4q}. 
The $D_{13}(1520)N^{-1}$ excitation leaves a clear trace in the peak structure seen at invariant masses of about $500$ MeV.
Probably we overestimate the effects resulting from dressing the $\rho$ meson by using the strict VMD picture where the photon coupling of hadrons is directly related to the hadronic coupling. In \cite{postomega} it was demonstrated explicitly, that within such an approach the electromagnetic coupling of the baryon resonances is overestimated by about a factor of $2$. Still the qualitative picture remains valid also with more elaborate versions of VMD.

The spectral function ${\cal A}_\rho^{T/L}$ is calculated at zero temperature. This is a reasonable approximation for resonance-hole states. Finite temperature effects only slightly rearrange the nucleon distribution function and also the small overestimation of Pauli-blocking -- which is not important for the $D_{13}(1520)$ -- should leave the results shown in Fig. \ref{dilep} qualitatively intact. 
We do not consider scattering processes of the $\rho$ meson on pions present at finite temperatures lead to a further broadening of the $\rho$ peak  of about $80$ MeV at $T=140$ MeV \cite{rwrev}. Albeit non-negligible, inclusion of this effect would not lead to qualitative changes of the results in Fig. \ref{dilep}.

\subsubsection{$\pi$ Meson}
\label{pires}

The properties of the pion in nuclear matter have been exhaustively studied within the $\Delta$-hole model, see for example \cite{osetreview}, where the pion is allowed to couple to the $P_{33}(1232)$ resonance and the nucleon. Our model goes beyond that by explicitly including resonances with higher mass, like the $P_{11}(1440)$ or the $s$-wave states $S_{11}(1535)$ and $S_{11}(1650)$. In addition, we generate corrections which are due to the self-consistent iteration of resonance and meson spectral functions. In that way different mesons influence each other to some extent.

\begin{figure}[h!]
\centering
\includegraphics[width=16cm]{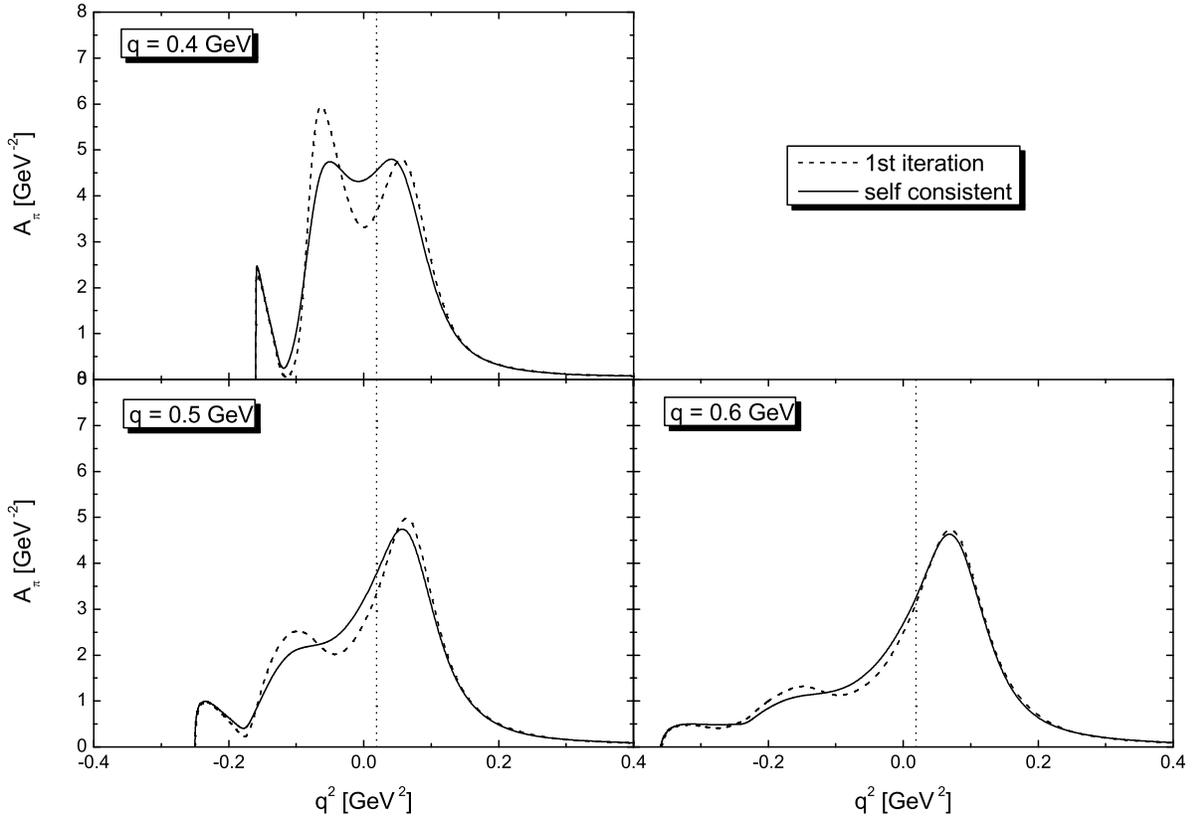} 
\caption{\label{pimed1}Spectral function of the pion at $\rho=\rho_0$. Shown is the spectral function at three momenta,  $0.4$ GeV, $0.5$ GeV and $0.6$ GeV after the first (dashed line) and fourth (solid line) iteration. Also indicated is the position of the free pion peak (dotted vertical line).}
\end{figure}

In Fig. \ref{pimed1} we show results for the spectral function ${\cal A}_\pi$ at three momenta, $0.4$ GeV $0.5$ GeV and $0.6$ GeV, after the first (dashed lined) and the fourth (solid line) iteration at density $\rho_0$. One can clearly see the multi-peak structure of the spectral function due to the excitation of nucleon-hole (left most peak) and $P_{33}(1232)$-hole states (peak in the middle). In the momentum range under consideration the kinematics are such that the pion branch is above the $P_{33}(1232)$ and the nucleon branches, see also Fig. \ref{branches}.
Due to level repulsion the position of the pion peak is therefore shifted to larger invariant masses. Note that a substantial amount of spectral strength is sitting at space-like four-momenta. For momenta ${\bf q}\ge 0.6$ GeV the additional peaks from nucleon and $P_{33}(1232)$ become much less pronounced since they are too far away from the pion pole and the pion starts resembling a good quasi-particle. The results presented here are in qualitative agreement with those of other analyses, see for example the results presented in \cite{oseteta2}, where on the basis of a model that is close to ours after the first iteration the spectral function is plotted for a momentum of ${\bf q}=0.5$ GeV.

\begin{figure}[h!]
\centering
\includegraphics[width=16cm]{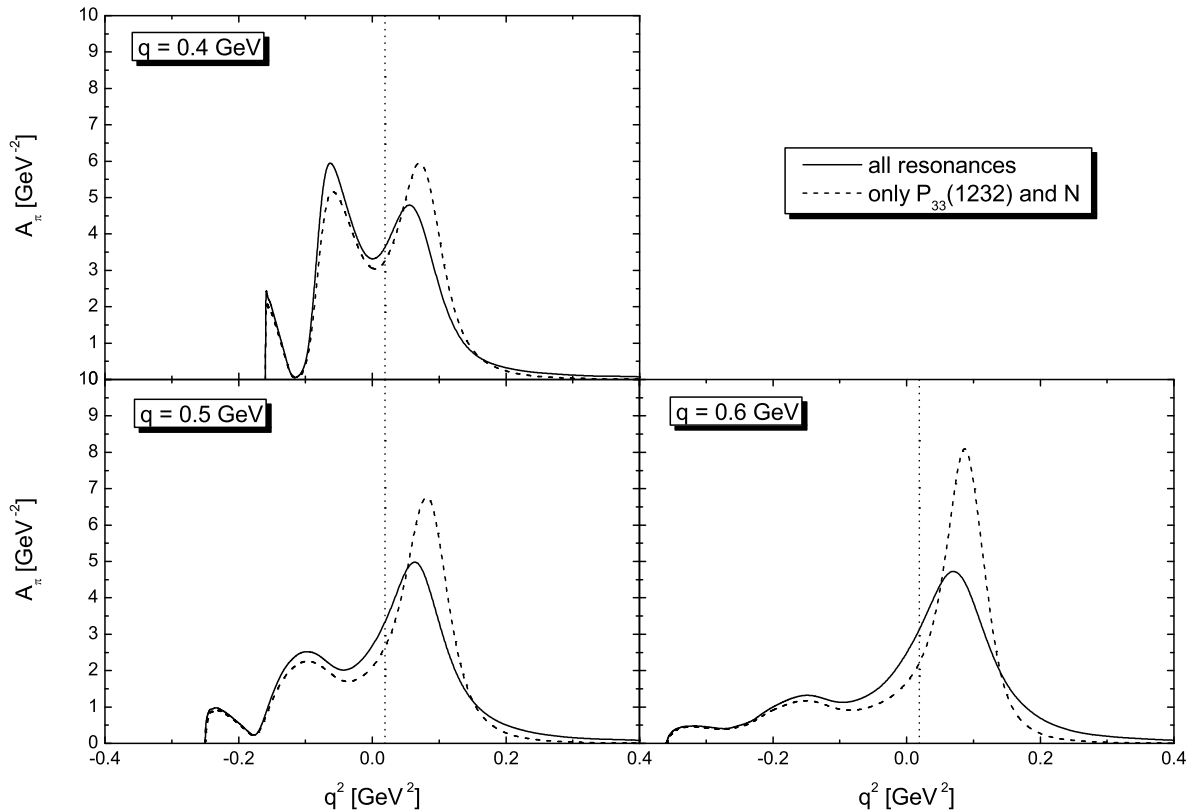} 
\caption{\label{pimed2}Spectral function of the pion at $\rho=\rho_0$. Shown is the spectral function at three momenta,  $0.4$ GeV, $0.5$ GeV and $0.6$ GeV after the first iteration. The solid line indicates the results if all resonances are included, for the dashed line only the $P_{33}(1232)$ and the nucleon are coupled to the pion.}
\end{figure}

The effects of the iteration lead to a smearing of pion and the $P_{33}(1232)$ peaks, while leaving the nucleon contribution unchanged. More importantly, however, there is no significant shift of spectral strength down to smaller invariant masses by the iterations. This point will become important when we discuss the properties of the $P_{33}(1232)$ in nuclear matter.

In Fig. \ref{pimed2} we display the influence of the remaining resonances besides the $P_{33}(1232)$ on the pion spectral function. 
We show the results after the first iteration in order to avoid mixing up effects from the inclusion of these additional states and from the iterations.
We find a modest influence of two $s$-wave states, the $S_{11}(1535)$ and the $S_{11}(1650)$,
and the $P_{11}(1440)$, coupling in a $p$-wave. Although these states have sizeable decay widths into the $N\pi$ channel, the large available phase space prevents a strong coupling of these resonances and the coupling is not sufficient to produce distinct peaks in the spectral function. Nonetheless, these states serve as a background contribution and smear out the pion peak. The somewhat smaller repulsion in the pion peak is due to the attractive interaction generated by heavy resonance-hole states due to level repulsion.
Their impact is most pronounced at $3$-momenta ${\bf q} \ge 0.6$ GeV where the energy of the corresponding resonance-hole states is close to that of the pion, see Fig. \ref{branches}. Effects from the $D_{13}(1520)$ state are suppressed by the $d$-wave coupling. Thus only at large momenta we see the influence of that state.

At very small 3-momenta ${\bf q}$, the $\pi N$ scattering amplitude receives contributions from non-resonant $s$-wave terms \cite{meissner}, which are not included in our model. We expect this approximation to be justified, since the leading $s$-wave contribution, the Weinberg-Tomozawa term, vanishes in isospin symmetric nuclear matter \cite{ericsonweise}.
Nonetheless, our results are not completely reliable in this kinematical regime and we cannot comment on the problem of $s$-wave repulsion demanded by data on pionic atoms \cite{gal}. Due to level repulsion, our model gives a small attraction of the pion since all the $s$-wave resonance-hole pairs have energies larger than $m_\pi$.
This shortcoming at small momenta has no sizeable effects on the results of the iterative scheme presented in this work, however, since the regime of small $3$-momenta is hardly tested in the decay of baryon resonances due to Pauli-blocking. Apart from these details, the spectral function of the pion is at low momenta dominated by the pion peak since both the nucleon-hole and the $P_{33}(1232)$-hole excitation are $p$-wave and open up only at finite momenta.


\subsubsection{$\eta$ Meson}
\label{etares}

In the discussion of the in-medium properties of the $\eta$ meson we focus on the question of $\eta$-mesic nuclei, where the $\eta N$ interaction is tested at small relative momenta. It is well known from coupled-channel analyses of $\pi N$ scattering, that close to threshold the $\eta N$ interaction is dominated by the $S_{11}(1535)$ resonance, see for example \cite{gregorpi,osetetan}. As was already pointed out in e. g. \cite{oseteta2, weiseeta, nagahiro}, the presence of the $S_{11}(1535)$ in the $\eta N$ interaction provides an attractive optical potential $U_\eta$ since the resonance pole is about $50$ MeV above the $\eta N$ threshold. In terms of the self energy $\Pi_\eta$, the optical potential reads close to threshold:
\beqa
	U_\eta(m_\eta,{\bf 0},\rho) &=& \frac{\Pi_\eta(q_0,{\bf 0},\rho)}{2\,m_\eta} \quad.
\eeqa

Let us first focus on the vacuum scattering amplitude, which via the low-density theorem Eq. \ref{scalarself} yields a first estimate for $\Pi_\eta$. In our model the $S_{11}(1535)$ resonance has a total width of about $151$ MeV with $\Gamma_{N\eta} = 66$ MeV, corresponding to a branching ratio of $44\%$. In the approach described in \cite{oseteta2} this state is generated dynamically and the resulting resonance parameters are quite different \cite{osetetan}: for the total width a value of $94$ MeV is found whereas the partial width $\Gamma_{N\eta}$ is the same as in our model. To be more quantitative, let us compare results for the scattering length $a_{\eta N}$: we find $a_{\eta N} = (0.43\,+\,i\,0.32)$ fm. The model of \cite{osetetan} produces $a_{\eta N} = (0.26\,+\,i\,0.24)$ fm, whereas in \cite{gregorpi} a value of $a_{\eta N} = (0.991\,+\,i\,0.347)$ fm is reported and in \cite{lutzvector} a scattering length $a_{\eta N} = (0.43\,+\,i\,0.21)$ fm results. We conclude that our model and that of \cite{oseteta2} yield results for the elementary $\eta N$ amplitude, which -- albeit different -- are well within the commonly accepted range.

\begin{figure}[t]
\centering
\includegraphics[width=16cm]{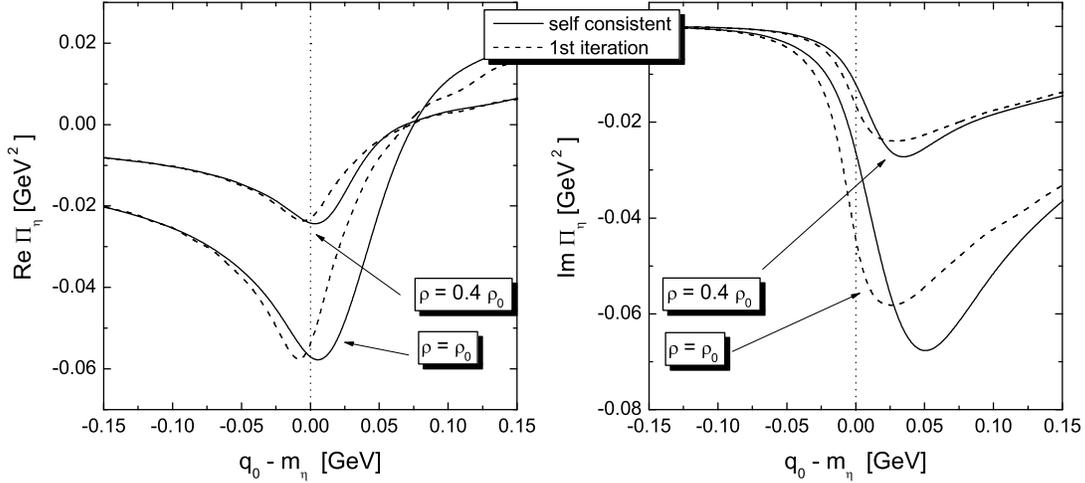} 
\caption{\label{etamed1} Real (left) and imaginary (right) part of the in-medium self energy of the $\eta$ meson. Shown are calculations at densities $\rho=\rho_0$ and $\rho=0.4\,\rho_0$. The dashed lines indicate the results following from the first iteration Eq. \ref{scalarself} and the solid incorporate all the effects from dressing the $S_{11}(1535)$. }
\end{figure}

The different resonance parameters have a direct effect on the self energy $\Pi_{\eta}$ of the $\eta$ meson calculated via Eq. \ref{scalarself}, which is indicated with dashed lines in Fig. \ref{etamed1} for the densities $0.4\,\rho_0$ and $\rho_0$, where $\rho_0$ is the normal nuclear matter density $\rho_0$. Comparing our results
with those obtained in \cite{oseteta2}, one finds that for the imaginary part the peak value is smaller in our case whereas off-shell we obtain larger values. Both findings are an immediate consequence of the larger total width in our model. 

The most important in-medium correction is generated from Pauli-blocking the $N\eta$ 
width of the $S_{11}(1535)$. At normal nuclear matter density and for a resonance at rest the Pauli-blocked width is zero, however even if excited by a $\eta$ meson at rest the resonance has a finite 3-momentum ${\bf k} \approx 0.2$ GeV due to the Fermi motion of the nucleons. This weakens the effects of Pauli-blocking and the approximation to put $\Gamma_{N\eta}=0$ as done in \cite{nagahiro} is not accurate. On top of Pauli-blocking there are additional mechanisms that influence mass and width of the $S_{11}(1535)$. This is discussed in Section \ref{ress11} and we find an additional broadening of about $30$ MeV for the $S_{11}(1535)$ relative to the Pauli-blocked width at these small momenta, accompanied by a small repulsive mass shift of the $S_{11}(1535)$. Concentrating on the point $q_0=m_\eta$ (indicated by the dotted vertical line in Fig. \ref{etamed1}) as appropriate for the optical potential, we find that the mass shift leads to a depletion of $\imag{\Pi_\eta}$ while leaving $\real{\Pi_\eta}$ nearly unaffected. 
The peak of $\imag{\Pi_\eta}$ is shifted upwards whereas the height of the peak remains essentially the same. In the analysis of \cite{oseteta2} this is different since the relative weight of Pauli-blocking is enhanced due to the larger branching ratio for this channel and an enhancement of the peak of the self energy in the nuclear medium results. 

\begin{figure}[t]
\centering
\includegraphics[width=16cm]{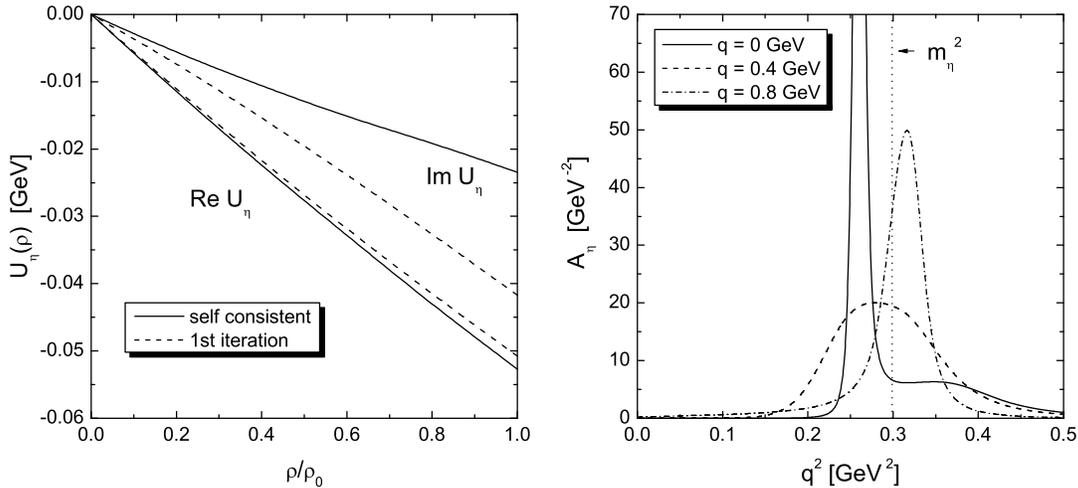} 
\caption{\label{etamed2} Left: Optical potential $U_\eta$ of an $\eta$ meson at rest as a function of the nuclear density $\rho$. The solid lines indicate the results including in-medium corrections, for the dashed lines the vacuum scattering amplitude has been used. 
Right: Spectral function ${\cal A}_\eta$ of the $\eta$ meson in nuclear matter at $\rho=\rho_0$. Shown are the results for three momenta: ${\bf q}=0$ GeV, ${\bf q}=0.4$ GeV and ${\bf q}=0.8$ GeV.}
\end{figure}

In Fig. \ref{etamed2} we plot the optical potential as a function of the density. As expected, the low-density approximation Eq. \ref{scalarself} shows a linear behaviour in the density both for the real and imaginary part of $U_\eta$. When the in-medium corrections are included, we find deviations from this linearity for the imaginary part, which is strongly reduced as was already indicated in the discussion in the preceding paragraph. At normal nuclear matter density we find $U_\eta=(-50-i\,43)$ MeV in the low density approximation and $U_\eta=(-52-i\,24)$ MeV in the full calculation. The SRC have no big effect on the potential, neglecting the SRC in the $s$-wave sector one obtains $U_\eta=(-49-i\,29)$ MeV. 
It follows that the iterations lead to a strong reduction of the imaginary part while hardly affecting the real part of the optical potential. This behaviour is a direct reflection of the results found for the self energy, see Fig. \ref{etamed1}.
It is interesting that the analysis of \cite{oseteta2} leads to a comparable final result, $U_\eta=(-54-i\,29)$ MeV, despite the fact that the scattering lengths in both models are different. In \cite{weiseeta} $U_\eta=(-20-i\,22)$ MeV is found, i.e. the attraction provided by $U_\eta$ is found to be only half of our value, whereas the width is comparable. Summarizing these results, most models seem to predict similar results for the imaginary part of the optical potential while uncertainties on the level of a factor of two 
persist for the real part. As far as the existence of $\eta$-mesic nuclei is concerned, the strong attraction found in our approach and in that of \cite{oseteta} is certainly encouraging.

We close the discussion of the properties of the $\eta$ meson in nuclear matter by inspecting the spectral function ${\cal A}_\eta$ shown in Fig. \ref{etamed2}. There the spectral function is displayed for three momenta ${\bf q}= 0$ GeV (solid line), ${\bf q}=0.4$ GeV (dashed line) and ${\bf q}=0.8$ GeV (dashed-dotted line). Also indicated is the position of the free $\eta$ peak. In the calculations SRC are taken into account, they have however only a small effect. One observes that at ${\bf q}=0$ GeV the coupling of the $\eta$ to $S_{11}(1535) N^{-1}$ loops is not sufficient to generate a distinct peak in the spectral function. Only a shoulder arises at invariant masses slightly above the $\eta$ peak, which is located at $q^2=0.25\,\mbox{GeV}^2$. This is in contrast to the findings both of \cite{oseteta2} and \cite{weiseeta} and is probably explained by the substantially smaller peak value of $\imag{\Pi}_{\eta}$ found in our work, which -- as mentioned above -- is due to a larger total width and a smaller branching ratio into $N\eta$. The position of the $\eta$ peak is shifted downwards as expected from the attractive nature of the interaction at small momenta.
Going to larger momenta we find that at $0.4$ GeV the $\eta$ peak is substantially broadened.
Around this momentum the energies of the $\eta$ branch and the $S_{11}(1535) N^{-1}$ are comparable (see also Fig. \ref{branches}), leading to a strong mixing and broadening of both states. At even larger momenta the resonance-hole excitation is below the $\eta$. This induces a small repulsion of the $\eta$ peak, whereas only a moderate broadening occurs.


\subsection{Baryonic Resonances}

Before presenting results for the three resonance states $P_{33}(1232)$, $S_{11}(1535)$ and 
$D_{13}(1520)$ we discuss a particular effect concerning an in-medium shift of the peak of the spectral function. As explained in Section \ref{resoself}, we obtain an in-medium mass shift of the resonance generated by the real part of the self energy. The in-medium mass is defined as the solution of the equation:
\beqa
	k^2-m_R^2-\real{\Sigma_{med}^+(k_0,{\bf k})} &=& 0 \quad.
\eeqa
On top of that, an additional shift of the peak of the in-medium spectral function -- which is of greater interest than the in-medium mass -- can arise due to the energy dependence of the width. In the case of a constant width the mass of the resonance and the location of the peak of the spectral function coincide. If, however, the width of the state rises with energy, the peak is shifted down to smaller invariant masses. This makes the interpretation of $\real{\Sigma}$ more difficult: even if that quantity indicates repulsion, the peak of the spectral function may be shifted downwards. 

\subsubsection{$P_{33}(1232)$}

It is well known that the in-medium broadening of the $P_{33}(1232)$ state is in the order of $100$ MeV at normal nuclear matter density. This value has been extracted a long time ago in an analysis of pion-nucleus scattering, \cite{hirata}. From the same analysis a slight repulsion of the $P_{33}(1232)$ of about $20$ MeV relative to the nucleon is reported. Certainly, any model trying to describe the $P_{33}(1232)$ state in nuclear matter should arrive at comparable results.
We thus use this resonance as a testing ground for our model: a reasonable description of its in-medium properties suggests that no major mechanisms are missing in our approach.

\begin{figure}[t]
\centering
\includegraphics[scale=1.3]{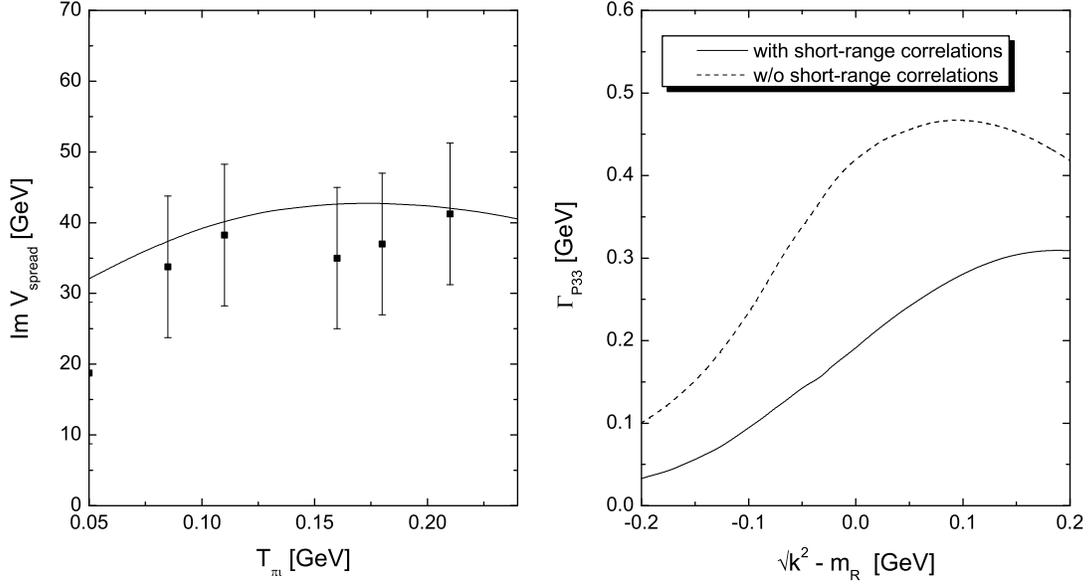} 
\caption{\label{p33med2} Left: Comparison of our model with the phenomenological spreading potential $V_{sp}$ of \cite{hirata}. The density is taken to be $\rho=0.75\,\rho_0$. Right: Influence of the SRC on the width of the $P_{33}(1232)$ at $\rho=\rho_0$. Compared are a calculation with (solid line) and w/o (dashed line) SRC. }
\end{figure}

\begin{figure}[t]
\centering
\includegraphics[scale=1.3]{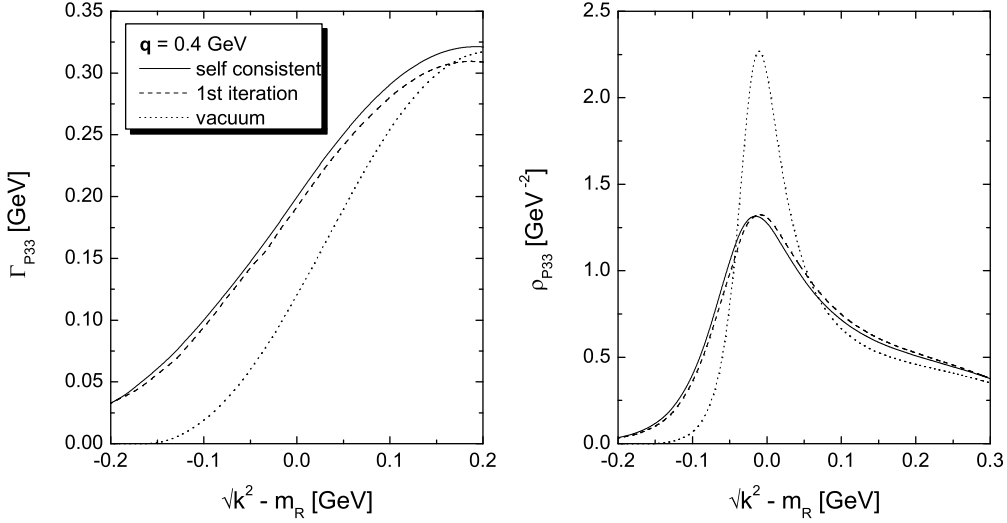} 
\caption{\label{p33med1}Width (left) and spectral function (right) of the $P_{33}(1232)$ moving with a relative momentum of $0.4$ GeV. The solid line indicates the result obtained after four iterations whereas the dashed lined is the result after the first iteration. The density is $\rho=\rho_0$.}
\end{figure}

We have adjusted the parameters of the $\pi\,N\,\Delta$ system - cutoffs, SRC - such as to achieve a reasonable description of the in-medium properties. In the left graph of Fig. \ref{p33med2} we plot the spreading potential $V_{sp}$ defined by:
\beqa
	\imag{V_{sp}(k_0,{\bf k})} &=& 
	\frac{\imag{\Sigma_{med}(k_0,{\bf k})}-\imag{\Sigma_{pauli}}(k_0,{\bf k})}{2\,\sqrt{k^2}}
\eeqa
and compare to the experimental data found in \cite{hirata}. Following an argument in \cite{salcedo} that the effective density felt in a nucleus is $0.75\,\rho_0$ rather than $\rho_0$, we perform the comparison at the lower density. The kinematics corresponds to that of a pion, which hits a nucleon with an average momentum of $\sqrt{\frac 3 5\,{\bf p}_F^2}$ and forms a $\Delta$ of energy $k_0$ and momentum ${\bf k}$:
\beqa
	k_0 &=& q_0+\sqrt{m_N^2+\frac3 5\,{\bf p}_F^2} 
	\quad,\quad {\bf k}^2 \,\,\,= \,\,\, 
	{\bf q}^2 + \frac3 5\,{\bf p}_F^2\quad,\quad T_\pi = q_0-m_\pi \quad.
\eeqa
As shown in the left plot of Fig. \ref{p33med2}, we achieve a reasonable description of $\imag{V_{sp}}$ in our model. The values needed for cutoff and short-range parameter -- given in Table \ref{param} -- lie well within the commonly accepted range, see for example \cite{helgesson1, helgesson2, salcedo}. We interpret this as a confirmation that our model contains the most relevant mechanisms. In Fig. \ref{p33med1} we show the width and the spectral function of the $P_{33}(1232)$ state for a fixed momentum of $0.4$ GeV. The width at $k^2=m_R^2$ is found to be roughly $190$ MeV, leading to a broadening of the in-medium spectral function. The coupling of the pion to the nucleon and the $P_{33}(1232)$, but also to all the other states included in our model, leads to a slight enhancement of the broadening. In our model the peak position of the spectral function of the $P_{33}(1232)$ remains more or less unchanged, while $\real{\Sigma}(k^2=m_R^2)/(2 m_R) \approx 15$ MeV, indicating a slight repulsion of the resonance. The reason that this repulsion is not observed in the spectral function is due to the energy dependent width as outlined in the introduction to this Section.

If one calculates the baryon self energy according to Eq. \ref{reswidthmed} and constructs the in-medium spectral function of the $\pi$ by coupling the pion to nucleon-hole and resonance-hole states, a dramatic overestimate of the in-medium width of the $P_{33}(1232)$ state results. Only after the inclusion of SRC according to Eq. \ref{respici} and \ref{resrhoci}, one arrives at satisfying results for the broadening \cite{salcedo,helgesson1,helgesson2}. This is shown in the right graph in Fig. \ref{p33med2}. Note that especially the vertex corrections induced in the resonance self energy diagram are important. The effect of the correlations is to systematically move strength up to the higher branches of the spectral function. Then the available phase space is reduced and a reduction of the width results. A similar argument was put forward in \cite{lutzreso}.

In \cite{salcedo} it was found that $3$-body absorption of the $P_{33}(1232)$ contributes significantly to the total width. There the $3$-body contribution was calculated based on the same diagrams that we generate in the second iteration step, i.e. $\imag{\Sigma_{med}^{2}}$ is the sum of $2$-body and $3$-body processes.
Also an additional diagram was calculated, which we do not consider here and which was claimed to be comparatively small by the authors of \cite{salcedo}. In contrast to that work, we find only small effects of about $10$ MeV from the iterations, i.e. our total broadening is essentially due to $2$-body processes contained in $\imag{\Sigma_{med}^1}$. 
A significant modification of  $\imag{\Sigma_{med}^{2}}$ relative to $\imag{\Sigma_{med}^{1}}$ can only follow if also the in-medium spectral function of the meson undergoes sizeable corrections, i.e. if ${\cal A}_{\pi}^1$ and ${\cal A}_{\pi}^2$ are very different. In particular, due to phase space arguments a large $3$-body contribution for the $P_{33}(1232)$ can only result from a substantial shift of spectral strength down to smaller invariant masses in the pion spectral function when going from ${\cal A}_{\pi}^1$ to ${\cal A}_{\pi}^2$. Such a rearrangement is not produced by the moderate resonance broadening obtained after the first iteration, see also Fig. \ref{pimed1} where the effect of iterations on the pion spectral function is explicitly displayed.
Therefore only a large additional attraction of the $P_{33}(1232)$ relative to the nucleon
might help. This however is at variance with the phenomenological spreading potential \cite{hirata}. We add that $3$-body physics also plays no important role in the iterative scheme of \cite{helgesson2}. Concerning the results of our model for the other resonances, this ambiguity of $2$ and $3$-body contributions is not important since the total width of the $P_{33}(1232)$ and therefore also the pion, which are the only quantities entering into the later calculations, are described well.

\subsubsection{$D_{13}(1520)$}
\label{resd13}

Our previous calculation \cite{postrho1} of the in-medium properties of the $D_{13}(1520)$ has been based on the effects induced by the in-medium spectral function of the $\rho$ meson. As a result a large broadening of the $D_{13}(1520)$ state was reported.
The origin of this broadening is easily explained: due to the coupling to particle-hole states, spectral strength is moved down to smaller invariant masses in the $\rho$ spectral function, thus opening up the phase space for the decay of the $D_{13}(1520)$.

This model has now been extended in three ways. On top of considering the effects of modifications in the $\rho$ spectral function - corresponding to $RN$ scattering with $\rho$ exchange - the in-medium spectral information of the pion is taken into account. Furthermore, guided by the experience with the $P_{33}(1232)$ state, effects from short-range correlations (SRC) are considered. Finally, we calculate the dispersive in-medium mass shifts.

\begin{figure}[h!]
\centering
\includegraphics[scale=1.3]{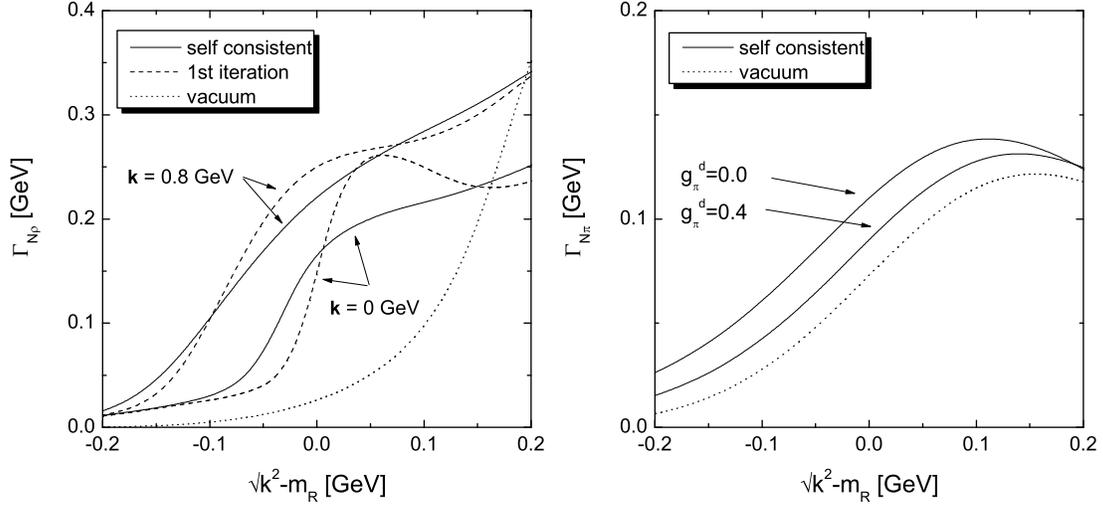} 
\caption{\label{d13med1} Decay width $\Gamma_{N\rho}$ (left) and $\Gamma_{N\pi}$(right) of the $D_{13}(1520)$. For the $N\rho$ width we show results w/o SRC at two different momenta, $0$ and $0.8$ GeV.
Also indicated is the effect of the iterations.
The $N\pi$ width $\Gamma_{N\pi}$ is shown for a momentum ${\bf k}=0.8$ GeV. Here the effects for SRC are indicated. All results are obtained at $\rho=\rho_0$. }
\end{figure}

Neglecting the SRC for the moment, the following picture emerges: the broadening induced from the $\rho$ meson is in the order of $200-250$ MeV at $k^2=m_R^2$, i.e. at the vacuum pole of the propagator for momenta around ${\bf k}=0.8$ GeV, see left plot in Fig. \ref{d13med1}. In the same figure we also show that for a $D_{13}(1520)$ at rest the broadening is about $150$ MeV only in this channel. These numbers are in approximate agreement with the results of our previous calculation \cite{postrho1}. In the language of scattering processes, most of this broadening is due to scattering $RN \to RN$ with a $D_{13}$ in the final state, which explains the smaller broadening observed for small momenta ${\bf k}$: the available phase space for the scattering process opens up with increasing $3$-momentum of the resonance. The inelastic processes $D_{13}N \to NN$ and $D_{13}N  \to NP_{33}(1232)$ play only a moderate role, accounting in total for at maximum $30-40\%$ of the total broadening. Unfortunately, this makes it difficult to obtain reliable constraints on our model from the consideration of inelastic $NN$ scattering.

\begin{figure}[t]
\centering
\includegraphics[scale=1.3]{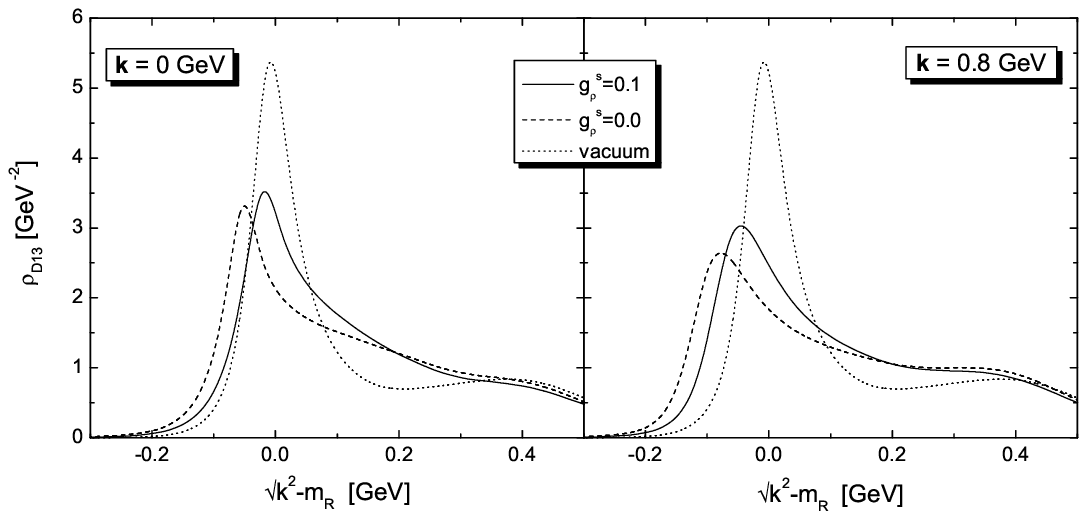} 
\caption{\label{d13specmed} Spectral function of the $D_{13}(1520)$ resonance at momenta ${\bf k} = 0$ GeV and ${\bf k}= 0.8$ GeV. The solid lines are obtained with SRC, in the dashed lines effects from SRC are not included. The results are shown for $\rho=\rho_0$.}
\end{figure}

From the pion decay we find a broadening of about $40$ MeV when SRC are neglected, see right plot in Fig. \ref{d13med1}. The momentum dependence is small, therefore we show results for only one momentum ${\bf k}=0.8$ GeV, appropriate in photonuclear reactions. In contrast to the $\rho$, here most of the broadening comes from decay into the $P_{33}(1232)$ or from absorption on the nucleon. The $D_{13}(1520)$ as a final state plays only a minor role since it couples in a $d$-wave, thus reducing the effectively available phase space. This relatively small broadening is at variance with the findings of \cite{lutzreso} where a strong broadening of several hundred MeV from this channel is reported. We recall that in \cite{lutzreso} the $D_{13}(1520)$ is dynamically generated in a coupled channel approach. Even though our results are quite sensitive on the value of the cutoff parameter $\Lambda$, we would have to relax $\Lambda$ from $1.0$ GeV - as appropriate in $\pi\,N\,\Delta$ dynamics - to a value of at least $2$ GeV in order to generate a broadening of about $200$ MeV. We conclude that within our approach a substantial softening of the $D_{13}(1520)$ state due to the decay into an in-medium pion seems unlikely. The decay mode $D_{13}(1520) \to \Delta\pi$ is not modified in the nuclear medium.

\begin{figure}[t]
\centering
\includegraphics[scale = 1.3]{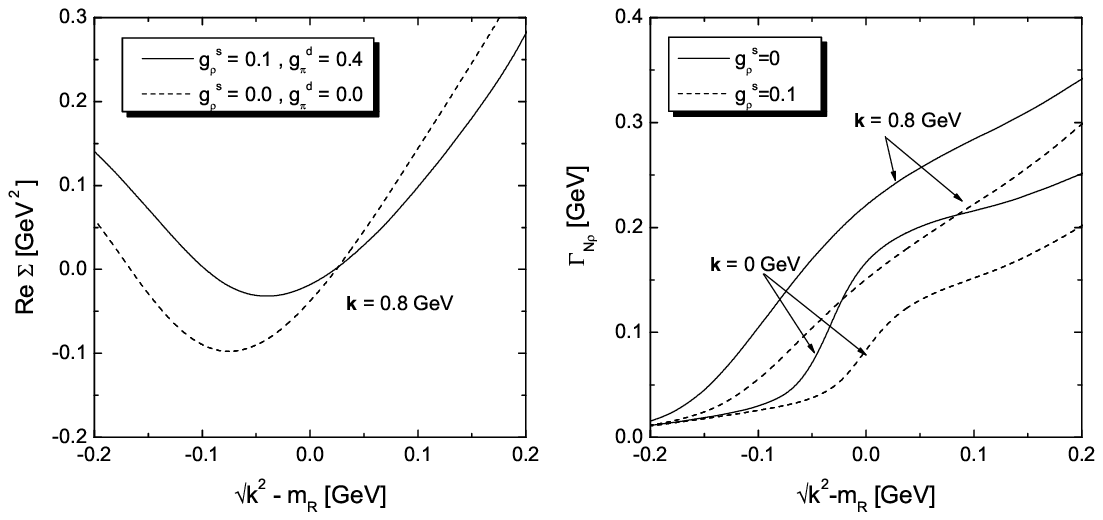} 
\caption{\label{d13realsigmed} Effect of SRC on $\Gamma_{N\rho}$ (left) and $\real{\Sigma}$ (right). The solid lines indicate the results with SRC, the dashed lines obtained w/o SRC.}
\end{figure}

\begin{figure}[t]
\centering
\includegraphics[width=17cm]{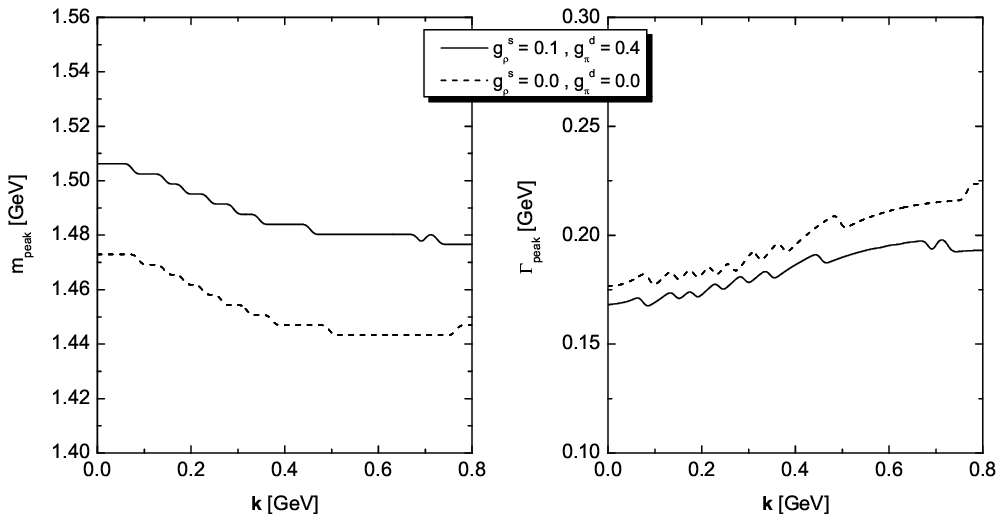} 
\caption{\label{d13med2}Left: Peak position of the spectral function of the $D_{13}(1520)$. Right: Width of the $D_{13}(1520)$ taken at the peak position. Shown are calculations with $g_\rho^s=0.1$ (solid line) and $g_\rho^s=0.0$ (dashed line). The density is $\rho=\rho_0$.}
\end{figure}

The spectral function is shown in Fig. \ref{d13specmed} in the dashed lines
for two different momenta ${\bf k}=0$ (left) GeV and ${\bf k}= 0.8$ GeV (right).
In comparison to the vacuum spectral function (dotted line) the main modifications are a shift of the peak to smaller invariant masses accompanied by a smearing of the peak.
Concerning the peak shift, we obtain an attraction of about $50-100$ MeV within our scheme. 
This is demonstrated in Fig. \ref{d13med2} where the dashed line shows the peak as a function of the $3$-momentum of the resonance.The major part of this shift is not due to the influence of $\real{\Sigma}$, however. In the left plot of Fig. \ref{d13realsigmed} (dashed line) we show $\real{\Sigma_{med}}$ for a momentum of ${\bf k}=0.8$ GeV, where the peak shift is larger than at small momenta (see Fig. \ref{d13specmed}). At $k^2=m_R^2$ one finds an attractive mass shift of $\real{\Sigma_{med}}/(2 m_R) \approx 10$ MeV. The remaining larger part of the peak shift is due to the energy dependence of the width as outlined in the introduction to this Section: Owing to the large absolute size and energy dependence of $\imag{\Sigma_{med}}$ as shown in Fig. \ref{d13med1}, the maximum of the spectral function is shifted to smaller energies. As a consequence  the resonant peak is not as broad as one might expect from the large in-medium widths at $k^2=m_R^2$. We have visualized this effect by plotting the width of the $D_{13}(1520)$ not at $k^2=m_R^2$ but rather at the true maximum of the spectral function in the right graph of Fig. \ref{d13med2} (dashed line). The fluctuations in the curves are due to the finite grid used in the calculations. Concerning the size of the peak shift, we argue that the form factor $FF1$ of Eq. \ref{ff1} leads to conservative estimates since it produces a relatively flat energy dependence. By considering the shape of the spectral function shown in Fig. \ref{d13specmed}, it becomes clear that a Breit-Wigner type parametrization in terms of mass and width is not possible, a tendency that is already visible in the vacuum. Instead, we find a structure with a rather narrow peak, but a large tail for $k^2 > m_R^2$. In comparison to the vacuum curve, the overall picture is that of a smearing of spectral strength over a much larger energy interval. In that sense the qualitative picture of a broadening of the resonance as advocated in \cite{postrho1} is not changed, although the calculations have been refined in many details.

The effect of the iterations is found to be quite small both for pion and $\rho$ meson.
This we have shown in Fig. \ref{d13med1} for $\Gamma_{N\rho}$, where the solid lines represent the results obtained after convergence has been reached. 
A tendency persists that in the second iteration the width gets somewhat reduced. This results from the effect of resonance broadening on the $\rho$ spectral function: there the $D_{13}$ peak "dies out" and less spectral strength sits at low invariant masses, leading to a relative suppression of the $\rho$ decay mode. Since the $D_{13}(1520)$ decouples more or less completely from the pion spectral function, we do not find any effects from the iterations in this sector and we do not show explicit results for this channel. Higher iterations then change only very little in the actual results.

Let us now switch on the SRC. In connection with Eq. \ref{vswave}, Chapter \ref{negpar}, we have already argued that for $s$-wave potentials these effects are supposedly small, since unlike the $p$-wave case no big additional scale (like $q_c$) is introduced. However, in a full in-medium calculation the contact interactions are iterated to all orders, see Chapter \ref{resoselfci}. This produces a correction of the form $1/(1-g_\rho^s \,\chi_s)$ for the resonance width, which is large if either $g_\rho^s$ or the coupling constant at the meson-nucleon-resonance vertex are large. Due to the large coupling of the $D_{13}(1520)$ to $N\,\rho$, a sizeable reduction might result. 

Using for the strength of the contact interaction $g_\rho^s=0.1$ as advocated in Section \ref{negpar}, a reduction of the broadening from the $N\,\rho$ channel from $250$ MeV down to around $150$ MeV results, see right plot in Fig. \ref{d13realsigmed}. This result is quite sensitive on the value chosen for $g_\rho^s$, similar to the case of the $P_{33}(1232)$. A smaller value of $g_\rho^s$ leads to more broadening of the $D_{13}(1520)$, while a larger value further enhances the suppression. Unfortunately, $g_\rho^s$ is completely unconstrained from experiment and thus some 
model dependence of our results remains. Since in our model positive and negative parity states do not couple directly to each other via SRC, the decay width into $N\Delta$ or $NN$ is left untouched and the reduction happens primarily in the channel $D_{13}N \rightarrow N D_{13}$. The effect of the correlations on the pionic decay mode, for which the results are shown in the right plot of Fig. \ref{d13med1}, is just the opposite. Since the channel $D_{13}N \rightarrow N D_{13}$ is not important, it is the mixing to nucleon- and $P_{33}(1232)$ states that leads to a reduction of the width. Again there exists some uncertainty about the correlation strength. Using $g_\pi^d=0.4$, the broadening is reduced and the width into $N\pi$ is about $20$ MeV larger than the Pauli-blocked width.

\begin{figure}[t]
\centering
\includegraphics[scale = 1.3]{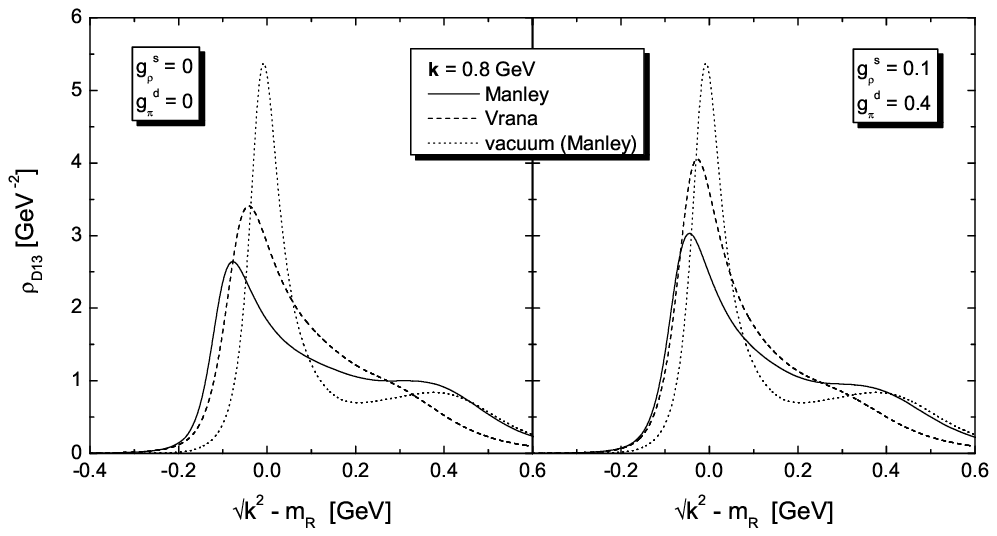} 
\caption{\label{d13med3}The spectral function $\rho$ of the $D_{13}(1520)$ as obtained using the parameters from \cite{manley1}(solid line) and \cite{vrana} (dashed lined). 
Left: no SRC are included. Right: SRC are switched on with the parameters $g_\rho^s=0.1$ and $g_\pi^d = 0.4$. The density is $\rho=\rho_0$.}
\end{figure}

The pole of the spectral function is less shifted once the correlations are switched on, compare solid and dashed curves in the left plot of Fig. \ref{d13specmed}. The total broadening is smaller and therefore the kinematic effect leading to a shift of the peak of the  spectral function is reduced. Furthermore, as can be seen in the left plot of Fig. \ref{d13realsigmed}, $\real{\Sigma}$ itself is a little bit more repulsive when the short-range interactions are switched on. The effect of the correlations on the width of the peak is not as strong as one might have expected by comparing the width at equal values of $k^2$ as shown in Fig. \ref{d13med2}. This is easy to understand: due to the smaller attraction, with SRC the width is tested at larger invariant masses, leading to a relative increase. Comparing both scenarios, even though the in-medium effects are somewhat reduced if the SRC are taken into account, the overall picture of a strongly broadened resonance structure survives.

We close this discussion by considering the spectral function as following from the parameter set of \cite{vrana}, where the branching ratio into $N\rho$ is only about half the size as compared to what is found in \cite{manley1}. In the discussion concerning the $\rho$ in-medium properties we were able to conclude that the gross features of the in-medium spectrum are left untouched by considering this parameter set. For the $D_{13}(1520)$ this statement does not necessarily hold any more. In Fig. \ref{d13med3}, we compare the results as following from the parameter sets of \cite{manley1} (solid line) and \cite{vrana} (dashed line) for two values of the short-range parameters
$g_\pi^d$ and $g_\rho^s$. The momentum of the resonance is ${\bf k} = 0.8$ GeV. As shown in the right graph of Fig. \ref{d13med3}, if one chooses the parameters of \cite{vrana} together with maximal suppression from SRC, i.e.  $g_\rho^s=0.1$ and $g_\pi^d=0.4$, a spectral function results which is already rather close to the vacuum one. In that case the total broadening does not exceed $50$ MeV, which is a small value as compared to the vacuum width of $120$ MeV. Also the shift of the peak position is very modest.
Taking the parameters of \cite{vrana} with $g_\rho^s=0=g_\pi^d$ as shown in the left plot of Fig. \ref{d13med3}, a considerable broadening of more than $100$ MeV at $k^2=m_R^2$ remains.

\subsubsection{$S_{11}(1535)$}
\label{ress11}

In this Section we discuss the results for the $S_{11}(1535)$ resonance. It will turn out that the main part of the medium modifications is due to Pauli-blocking in the $N\eta$ channel and to the coupling of the $D_{13}(1520)$ resonance to the $\rho$ meson.

\begin{figure}[t]
\centering
\includegraphics[scale = 1.3]{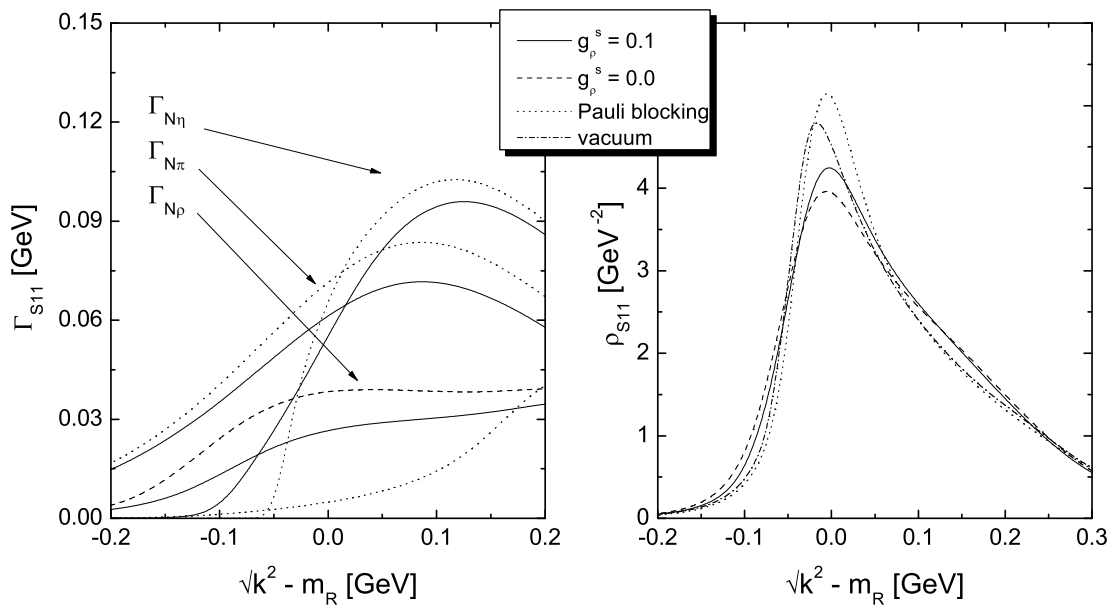} 
\caption{\label{s11med1}Left: Partial decay widths of the $S_{11}(1535)$ at a momentum of ${\bf k}=0.8$ GeV. We compare results obtained from Pauli-blocking and our full in-medium calculations with and w/o SRC. For the $N\pi$ and $N\eta$ channels the effect of SRC is very small, therefore the in-medium width is represented by the full lines only.
Right: spectral function $\rho$ of the $S_{11}(1535)$ at the same momentum. The density is $\rho=\rho_0$.}
\end{figure}

In the left plot of Fig. \ref{s11med1} we show the in-medium width of a $S_{11}(1535)$ with relative $3$-momentum ${\bf k}= 0.8$ GeV, which is approximately the momentum of an $S_{11}(1535)$ if produced from a photon scattering on a nucleon at rest. 
Shown is the decay width after the fourth iteration (dashed line) in comparison with a Pauli-blocked vacuum width (dotted line),
which is calculated on the basis of the Feynman graph Fig. \ref{resself}, taking into account Pauli-blocking of the nucleon. We find a broadening of about $20$ MeV at the point $k^2=m_R^2$ relative to the Pauli-blocked width.
Although $\pi\,N$ and $\eta\,N$ produce the main part of the total width, the in-medium modifications of $\pi$ and $\eta$ mesons do not lead to a broadening of the $S_{11}(1535)$.
In fact, for these two channels we find even a reduction of the width below the Pauli-blocked width. This is due to the renormalization of the meson decay width, as discussed in Section \ref{resoself}.
The main effect comes from the in-medium modification of the $\rho$ meson. The origin of this broadening is the same as found for the $D_{13}(1520)$ state: in the nuclear medium the $\rho$ meson couples to the $D_{13}(1520)N^{-1}$ state and various other particle-hole states which leads to shift of spectral strength to smaller invariant masses. 
Therefore also the phase space available for decay of the $S_{11}(1535)$ opens up, thus enhancing the $5$ MeV partial decay width as found in \cite{manley1} to values around $20-30$ MeV. This is a typical coupled channel effect: the coupling of the $D_{13}(1520)$ to $N\rho$ generates a broadening of the $S_{11}(1535)$ state.

In the work of \cite{oseteta} a qualitatively similar picture emerges. There the $S_{11}$ is considered to be at rest. In that formalism the particle-hole loops are not iterated and the $2\pi$ channel is treated as pure phase space with a partial decay width of $10$ MeV. Again, the broadening found from $\pi\,N$ and $\eta\,N$ is small. The broadening from the $2\pi$ channel is somewhat larger that found in our work, but this is partly due to the fact that the total $2\pi\,N$ width is taken to be larger than our $N\,\rho$ width.

The spectral function is only slightly modified as can be inferred from the right plot in Fig. \ref{s11med1} by comparing the spectral function in the vacuum (dash-dotted line), with Pauli-blocking only (dotted line) and the full results without SRC (dashed line). The peak of the spectral function is shifted upwards relative to the nucleon by about $10-20$ MeV. Such a small (dispersive) mass shift is found also in other theoretical works on the $S_{11}(1535)$ \cite{oseteta2,weiseeta}. 

\begin{figure}[t]
\centering
\includegraphics[scale = 1.3]{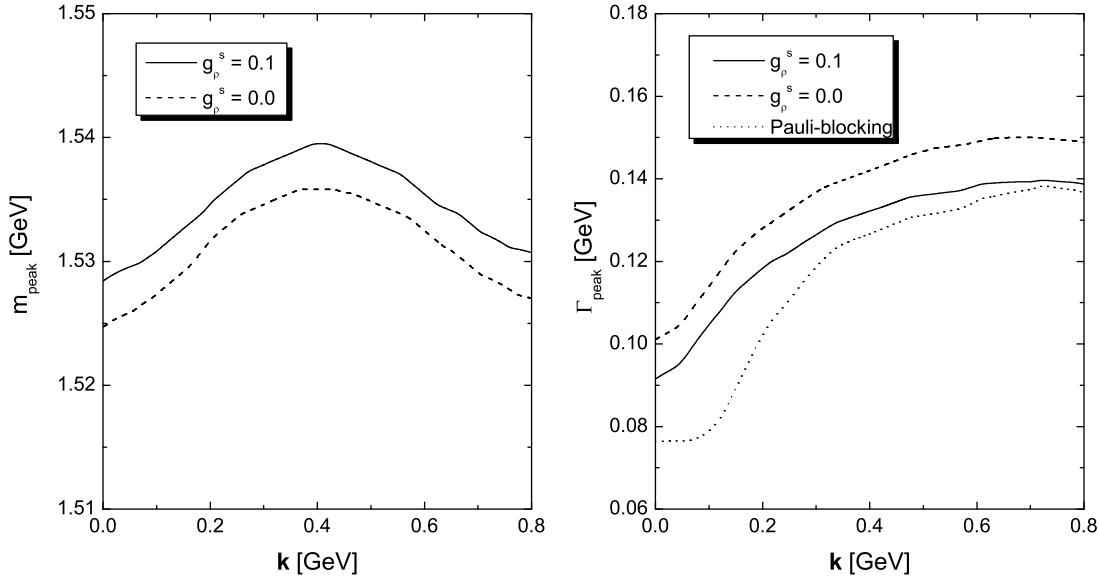} 
\caption{\label{s11med2}Left: Peak position of the spectral function $\rho$ of the $S_{11}(1535)$. Right: Width of this state taken at the peak position. Shown are calculations with $g_\rho^s=0.1$ and $g_\rho^s=0.0$. For the width we also show the results from Pauli-blocking. The calculations are carried out at normal nuclear matter density.}
\end{figure}

In Fig. \ref{s11med2} we show the position of the peak (left) and the width of the peak (right) as a function of the $3$-momentum ${\bf k}$. The peak position varies only very little. The width taken at the peak position displays some momentum dependence. This is mostly due to Pauli-blocking of the $N\eta$ channel, which completely prohibits the decay into this channel at vanishing $3$-momentum. At finite momenta the effects of Pauli-blocking are reduced, leading to an increase of the the width, which is enhanced by the fact that -- as in the case of the $D_{13}(1520)$ -- the $N\rho$ broadening receives additional support from the opening of phase space. As in the case of the $D_{13}(1520)$ the numerical fluctuations in the curves are due to finite grid effects.

Switching on the SRC reduces the broadening, see Figs. \ref{s11med1} and \ref{s11med2}. Just as in the case of the $D_{13}(1520)$ we find a reduction of the broadening in the $\rho\,N$ channel, whereas the other two channels remain essentially untouched. 
The insensitivity of the $\pi\,N$ and $\eta\,N$ channels to effects from SRC is explained from the comparatively small coupling constants at the respective resonance-nucleon-meson vertices, which prevent large corrections terms of the form $1/(1-g\chi)$. This substantiates our statement made in the introduction of Chapter \ref{nrint} that for $s$-wave states the SRC become only sizeable in the presence of large coupling constants.
The effects from the iteration are also similar to those observed for the $D_{13}(1520)$. The strong broadening of that state leads to a reduction of strength at small invariant masses and therefore the in-medium width of the $S_{11}(1535)$ slightly decreases.

We conclude from our results and those obtained in \cite{oseteta,oseteta2,weiseeta} that some consensus exists in the literature concerning the in-medium properties of the $S_{11}(1535)$. A small broadening relative to the Pauli-blocked width is expected, accompanied by a slight repulsive mass shift. In \cite{lehreta} is has been demonstrated that such medium modifications lead to a natural explanation of experimental data on $\eta$ photoproduction \cite{metageta,yorita}. There the relatively large observed mass shift hinted in the data \cite{yorita} is generated by the assumption that resonance and nucleon feel the same momentum dependent mean-field potential. As pointed out in \cite{lehreta}, a small collisional broadening of the $S_{11}(1535)$ has also been found in \cite{effeabs} based on estimates of resonance-nucleon cross sections. 

Summarizing, the broadening of the $S_{11}(1535)$ is a typical example of a coupled-channel effect. The main physical effects are generated by the $D_{13}(1520)$ via the process $S_{11}(1535)\,N \rightarrow N\,D_{13}(1520)$. Whereas we do not claim that the broadening found is accurate within $10$ MeV, we can exclude a significant in-medium modification of the $S_{11}(1535)$ within the mechanisms discussed in this work. A strong broadening would either require a much larger coupling to $N\rho$ -- which is unlikely since the total $2\pi$ width $\Gamma_{2\pi N}$ is estimated to be around $10$ MeV in the vacuum \cite{manley2,pdg} --  or effects from nuclear mean fields, which might increase the mass difference between nucleon and $S_{11}(1535)$ and thus enhance the phase space available for the decay.

\subsection{Density Dependence}

In this Section we analyze the density dependence of our results. Within the low density expansion the in-medium self energies of mesons and baryon resonances are directly proportional to the nuclear density $\rho$. Deviations from this linear scaling are already introduced from Pauli blocking and Fermi motion. More importantly, within our self-consistent scheme meson and resonance interactions with more than one nucleon are generated. They correspond to terms of higher order in the nuclear density. Finally, short-range correlations (SRC) exhibit terms of the from $1/(1-g \chi)$ which also produce deviations from a linear density dependence. It is therefore interesting to study our results as a function of the nuclear density and determine a critical density above which the low density expansion becomes unreliable.

\begin{figure}[h!]
\centering
\includegraphics[scale = 1.3]{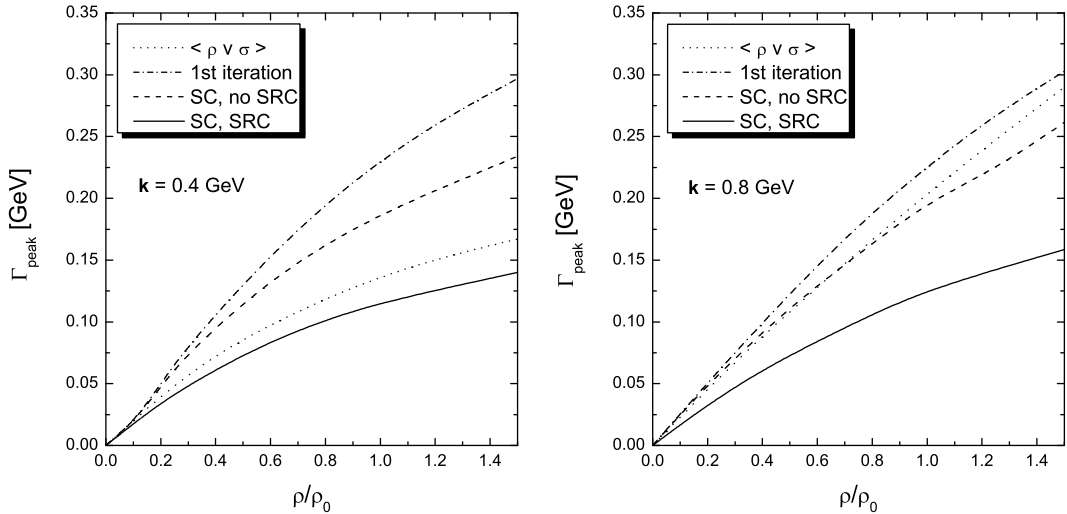} 
\caption{\label{densdep2} Collisional broadening of a $D_{13}(1520)$ with momenta ${\bf k}=0.4$ GeV (left) and ${\bf k}=0.8$ GeV (right) as a function of the nuclear density $\rho$. Compared are four calculations: low-density limit with Pauli blocking and Fermi motion (dotted), results of the first iteration (dash-dotted) and the self-consistent (SC) calculation (dashed). By the solid line we indicate the results from a self-consistent calculation where also effects from SRC are taken into account.}
\end{figure}

We begin with a study of the density dependence of our results for baryon resonances and will restrict ourselves to the case of the $N\rho$ decay of the $D_{13}(1520)$. In Fig. \ref{densdep2} we show the collisional broadening defined as 
\beqa
	\Gamma_{coll} &=& \frac{\imag{\Sigma}_{med}-\imag{\Sigma}_{pauli}}{\sqrt{k^2}},
\eeqa
in the $N\rho$ channel of the $D_{13}(1520)$ as a function of the nuclear density for two momenta, ${\bf k}=0.4$ GeV (left) and ${\bf k}=0.8$ GeV (right). The broadening is evaluated at $k^2=m_R^2$. The dotted lines indicate the results from a calculation where the particle-hole loops have not been iterated. It has been obtained by the replacement
\beqa	
	{\cal A}_\rho^{T/L}(q_0,{\bf q}) \to -\frac{1}{\pi}|D_\rho^{vac}(q)|^2\,\imag{\Pi_\rho^{T/L}(q_0,{\bf q})}
\eeqa
in Eq. \ref{reswidthmed}. This expression already goes beyond the low density theorem $\Gamma_{coll} = \rho\,v\,\sigma$ \cite{effeabs,kondrat,bugg} by including Pauli-blocking and Fermi motion. Here $\sigma$ is the total resonance-nucleon cross section and $v$ the velocity of the resonance in nuclear matter. The effects from resumming the particle-hole loops are shown by the dash-dotted lines and the impact of the self-consistent (SC) scheme is shown by the dashed line. The solid line shows the results from a self-consistent calculation which contains also the effects from SRC.

As one can see in the left plot of Fig. \ref{densdep2}, already the low density curve (dotted line) shows sizeable deviations from a linear density dependence for a resonance with momentum ${\bf k}=0.4$ GeV. This is due to Pauli blocking, which becomes more active as the density increases. The resummation of particle-hole loops in the $\rho$ propagator leads to a sizeable enhancement of the broadening already at small densities around $0.25 \rho_0$. This is a direct consequence of the fact that due to level repulsion this resummation leads to an attractive shift of the position of the $D_{13}(1520)$ excitation in the $\rho$ spectral function. Due to this shift the phase space available for the reaction $D_{13}(1520) N \to N D_{13}(1520)$ is enhanced. The effect of self consistency (dashed line) and short-range correlations (solid line) is to reduce the width as has been discussed in Section \ref{resd13}. One should not compare the results obtained with SRC and the low density curve, because the SRC change the resonance-nucleon cross section.

At ${\bf k}=0.8$ GeV (see right plot in Fig. \ref{densdep2}) the low density calculation (dotted line) displays a nearly linear density dependence, which is due to the smaller impact of Pauli-blocking at large momenta. The effect of resumming the particle-hole excitations is smaller at ${\bf k}=0.8$ GeV than at ${\bf k}=0.4$ GeV as can be seen by comparing the dotted and dash-dotted lines. This is explained as follows: The resummation is important when the particle-hole excitation and the $\rho$ peak have comparable energies. This is the case for a $D_{13}(1520) N^{-1}$ excitation at low resonance momenta. 
However, when calculating the in-medium width of a fast moving $D_{13}(1520)$ the spectral function is tested at large momenta, where the particle-hole excitation is far away from the $\rho$ peak (cf. Fig. \ref{branches}) and a smaller effect of the resummation is to be expected. The effects of self-consistency (dashed line) and SRC (solid line) are similar for both resonance momenta, both leading to a reduction of the resonance width. 
The different slope found already at very small densities for the calculation with SRC is due to a modification of the resonance-nucleon cross section from the short-range terms.
Summarizing these results, we find for the $D_{13}(1520)$ strong deviations from a low density expansion. In particular for smaller momenta the low density results are found to be unreliable already at densities around $0.25\,\rho_0$, whereas at larger momenta the low density expansion starts to work better.  This has an interesting effect on the momentum dependence of the width: in the low density limit the collisional broadening rises nearly linearly with the momentum. Since the resummation of particle-hole loops leads to an enhancement of the broadening at smaller momenta while having only a small influence on the results at large momenta, a flatter momentum dependence is expected (compare also right plot of Fig. \ref{d13med2}).

\begin{figure}[t]
\centering
\includegraphics[width=16cm]{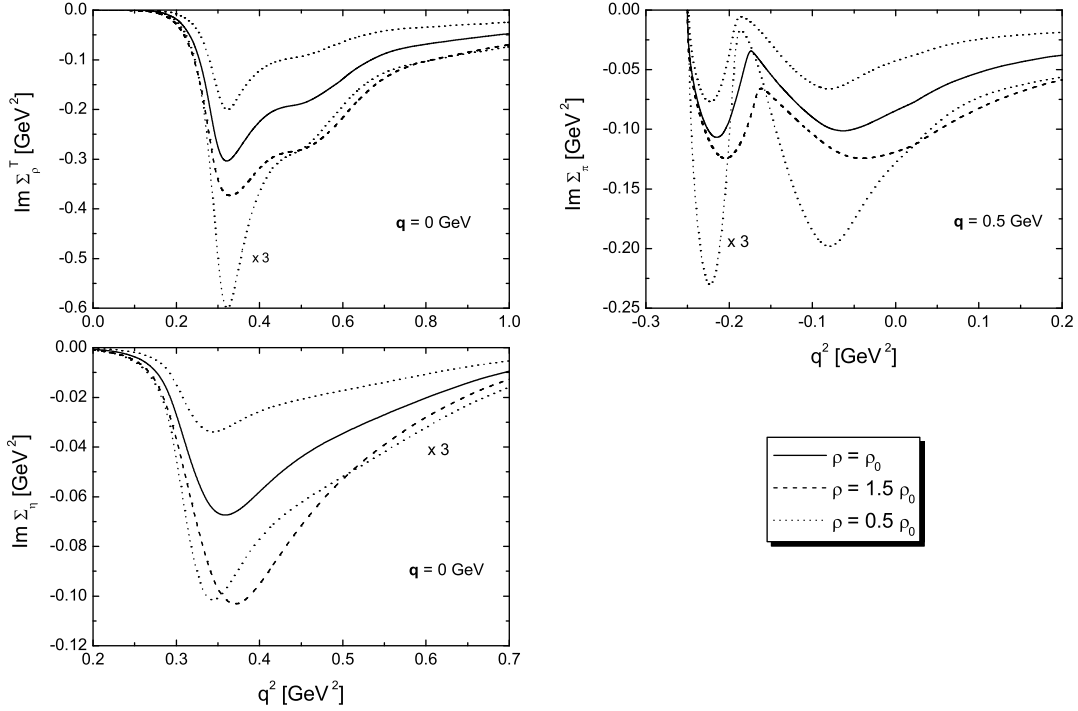} 
\caption{\label{densdep1} Influence of the density on the imaginary part of the in-medium self energy of $\rho$ meson (left, top), pion (right) and $\eta$ meson (left, bottom). The $\rho$ meson and $\eta$ meson are taken to be at rest while for the pion is moving with a momentum of ${\bf q}=0.5$ GeV. The results are shown for three different densities. In order to facilitate the comparison we have rescaled the result obtained for $\rho=0.5\,\rho_0$ by a factor of $3$. In all calculations the SRC have been taken into account.}
\end{figure}

Next we discuss the self energy of pion, $\eta$ and $\rho$ meson. 
Due to the complicated peak structure arising from the resonance-hole excitations it is difficult to summarize the information contained in the self energy at given density and momentum in one number. Therefore we show the self energy as a function of the invariant mass in Fig. \ref{densdep1}. There the imaginary part of the self energy of the $\rho$ meson (left, top), the pion (right) and the $\eta$ meson (left, botttom) are displayed for three densities $\rho=0.5\,\rho_0$ (dotted lines), $\rho=\rho_0$ (solid lines) and $\rho=1.5\,\rho_0$ (dashed lines). In order to facilitate the comparison we also show with dotted lines the result obtained at $\rho=0.5\,\rho_0$ multiplied by a factor of $3$, which would equal the result at $\rho=1.5\,\rho_0$ if the self energies would scale linearly with the density. We observe that for the pion and the $\rho$ meson a linear scaling with the density is badly violated. A more detailed investigation shows that higher order corrections are already important for densities $\le 0.5\,\rho_0$. This is not shown in Fig. \ref{densdep1}. For the $\eta$ meson we find that the height of the peak scales nearly directly with the density whereas the position of the peak is slightly shifted. As discussed in Section \ref{etares}, this shift of the peak leads to the observed strong non-linearities of the optical potential (cf. Fig. \ref{etamed2}). We already mentioned that the following sources act against a linear density dependence: Fermi-motion, Pauli-blocking, self-consistent iterations and short-range correlations. It is interesting to note that for each meson the most important deviation is generated by a different mechanism: For $\eta$ and $\rho$ the iterations act against the low density theorem by inducing a strong broadening for the $D_{13}(1520)$ and a slight repulsive mass shift for the $S_{11}(1535)$. In contrast, for the pion it is mainly the sum of Fermi motion and short-range interactions which is responsible for the non linear density dependence. 

We close this Section by noting that for channels characterized by a small coupling strength we find a nearly linear density dependence when Pauli-blocking effects are not taken into account. This is due to the fact that for such systems the forces acting against a linear density dependence are weak.

\section{Conclusions}
\label{conclusions}

We have constructed a model that allows for a combined description of in-medium modifications of hadrons. The constituents of the model are on the meson sector the pion, $\eta$ meson and $\rho$ meson and on the baryonic sector all resonances that couple at least to one of these states and the nucleon. Thus, the model assumption implicit in our analysis is that resonant terms lead to a satisfying description of the meson-nucleon scattering amplitude. Non-resonant background terms as well as contact and tadpole contributions to the meson self energy are not taken into account.

Our previous work in this direction \cite{postrho1,postrho2} has been extended and improved in several ways: The basis space of included states has been significantly enlarged. Due to the use of dispersion relations the spectral functions of all states are guaranteed to remain normalized both in the nuclear medium and in the vacuum. Special care has been taken with respect to the treatment of short-range correlations (SRC), in particular in the $s$-wave sector. This was motivated by the fact that a realistic description of the in-medium properties of the $P_{33}(1232)$ requires such repulsive mechanisms.

For the $\rho$ meson we find a significant shift of spectral strength down to smaller invariant masses generated by its coupling to the $D_{13}(1520)N^{-1}$ state. In particular at smaller momenta, the coupling to this state leads to a pronounced double-peak structure in the spectral function. In order to corroborate this finding we have tested it against variations of the coupling strength and against possible effects from SRC. As a result we find that the results remain quite stable when varying the width within ranges suggested by different analyses of pion-nucleon scattering. Also SRC do not qualitatively change the results. At finite momenta, the in-medium properties of the $\rho$ meson are influenced not only by the $D_{13}(1520)$ but also by some other higher lying resonances and our model predicts a different momentum dependence of transverse and longitudinal modes. In the transverse channel we find that the spectral function is characterized by a substantially broadened peak, whereas in the longitudinal channel the medium modifications get weaker and at momenta ${\bf q} \ge 0.8$ GeV the vacuum spectral function is recovered in this channel. The self-consistent iteration scheme mainly affects the results at low momenta and smears out the $D_{13}(1520)N^{-1}$ peak. Thus the central findings from our previous calculation \cite{postrho1} can be confirmed. On a quantitative level we find that the effects from iterations are reduced with respect to the findings in \cite{postrho1}, which -- as argued in the text -- is mainly due to a more realistic choice of the form factor.
We have also calculated the momentum integrated dilepton rate at densities and temperatures typically encountered at $SPS$ energies. The results suggest that our model is able to generate the observed shift of spectral strength down to smaller invariant masses.

For the pion we reproduce the essential features of the $\Delta$-hole model, i.e. at finite momenta around $0.3-0.6$ MeV the pion spectral function is dominated by a complicated peak structure which is derived from the coupling of the pion to $N N^{-1}$ and $P_{33}(1232) N^{-1}$ states. Going beyond the usual $\Delta$-hole model we have also investigated the effects of coupling the pion to other resonance-hole states. The coupling of these states is not sufficient to generate distinct peaks, but it nonetheless produces a smooth background that influences the detailed structure of the pion spectral function. Turning
to the $\eta$ meson, we have calculated both the optical potential -- which is of relevance for a quantitative analysis of $\eta$-mesic nuclei -- and its spectral function.
Based on reasonable predictions for the $\eta N$ scattering length, we arrive at a potential that provides rather strong binding relative to the life-time of the $\eta$. On a quantitative level the inclusion of medium modifications of the $S_{11}(1535)$ is found to be important. The spectral function of the $\eta$ meson exhibits an interesting momentum dependence, receiving attraction at small and repulsion at large momenta (${\bf q}\approx 0.8$ GeV), while in the intermediate momentum region we observe a significant broadening of the $\eta$ meson.

Turning to the baryon resonances, our model is able to reproduce the in-medium properties of the $P_{33}(1232)$ resonance and we obtain a reasonable fit of the phenomenological spreading potential if SRC are taken into account. We observe only small contributions from three-body processes.  Our results for the $D_{13}(1520)$ show some sensitivity on coupling parameters and the effects of SRC. Assuming both a large coupling of this resonance to the $N\rho$ channel (corresponding to $\Gamma_{N\rho}=26$ MeV), and small effects from SRC leads to a significant total broadening of about $250-300$ MeV. If, on the other hand, $\Gamma_{N\rho}=12$ MeV is taken in combination with a rather large value for the SRC, the in-medium broadening of the $D_{13}(1520)$ is much reduced. We do not find large contributions to the broadening from the pion sector. With these uncertainties in mind,
the experimental and theoretical challenge is to pin down the unsettled parameters -- in particular the $N\rho$ partial width and the strength of the short-range interactions -- in more detail. Finally, for the $S_{11}(1535)$ we find only modest medium effects. Even though there is some uncertainty concerning the resonance parameters and the strength of SRC, we can exclude the appearance of large medium modifications on the basis of our model. It is interesting to mention that the main body of the broadening found for the $S_{11}(1535)$ is due to typical coupled channel effect: without the rearrangement of spectral strength in the $\rho$ spectral function due to the $D_{13}(1520)$, the observed broadening would have been even smaller.

In a last step we have investigated our results as a function of the nuclear density and compared it to a low density expansion. We find that for the $D_{13}(1520)$ already at small densities around $0.25\, \rho_0$ the low density expansion breaks down and terms of higher order become important. Also the in-medium properties of the mesons deviate from a low density expansion, which is either due to effects from self consistency ($\rho$ meson, $\eta$ meson) or the effects of SRC (pion). Such effects are already important at small densities $0.25\,\rho_0 \,\mbox{--}\, 0.5\,\rho_0$.

Summarizing we have constructed a model that is able to describe or predict in-medium effects from very different areas, ranging from dilepton spectra measured in heavy ion collisions to photoproduction data on nuclei and $\eta$ mesic atoms.

\section{Acknowledgements}
The authors are grateful to L. Alvarez-Ruso, C. Greiner, A. Larionov, J. Lehr and H. Lenske for many interesting and stimulating comments and suggestions during the course of this work. Also various fruitful discussions with B. Friman, M. Lutz and G. Wolf are acknowledged.

\begin{appendix}

\section{Relations for Feynman and retarded propagators}
\label{analytic}

In this Appendix we derive some basic relations for retarded and Feynman propagators.
Starting point is the representation of these quantities in position space. In order to obtain relations in momentum space, a Fourier transformation is performed. The relations thus derived carry over directly to the self energies.

\subsection{Bosons}
For boson fields, the retarded and the Feynman propagator are defined as:
\beqa
\label{propdef}
	i\,D^+(x,y) &=& \theta(x_0-y_0)\,\expt{\left[\phi(x),\phi^\dagger(y) \right]} \\
	i\,D^F(x,y) &=& \theta(x_0-y_0)\,\expt{\phi(x)\,\phi^\dagger(y)} +
	                       \theta(y_0-x_0)\,\expt{\phi^\dagger(y)\,\phi(x)} \nonumber \quad,
\eeqa
where the expectation value can be taken either with respect to the vacuum ground state or the ground state of nuclear matter. By performing a Fourier transformation, one can prove from the definition of $D^F$ and $D^+$ that the real parts of both propagators in momentum space are identical:
\beqa
	\real{D^F(q)} &=& \real{D^+(q)}  \quad.
\eeqa
This holds in the vacuum as well as in the medium. The spectral function ${\cal A}(q)$ is defined as the imaginary part of $D^+(q)$, which allows for the following representation:
\beqa
\label{adef}
	{\cal A}(q) &\equiv& -\frac{\imag{D^+(q)}}{\pi} \,\,=\,\,	\frac{1}{2\pi}\,\int \!\!d^4\!x\,e^{i q x}\,\left[\phi(x),\phi^\dagger(0) \right]  \quad.
\eeqa 
Using the equal-time commutator relation of scalar fields, this representation leads to a sum rule for ${\cal A}(q)$:
\beqa
 \D{\int\limits_{-\infty}^{\infty}dq_0\,q_0\,{\cal A}(q)} &=& 1 \quad.
\eeqa
In vacuum and in isospin symmetric nuclear matter the spectral function is antisymmetric in the energy, i.e. ${\cal A}(-q_0,{\bf q})=-{\cal A}(q_0,{\bf q})$. One can see this by changing $q \to -q$ in the defining Eq. \ref{adef} and using that an isospin rotation transforms a boson into its own antiparticle while leaving the nuclear ground state invariant. 
This argument implies, that for charged bosons in asymmetric nuclear matter the relation between negative and positive energies is lost. Finally, by invoking the KMS relation \cite{stefankms}, one can establish a relation between the imaginary part of $D^F(q)$ and $D^+(q)$:
\beqa
	\imag{D^F(q)} &=& -(1+2\,n_B)\,\pi\,{\cal A}(q) \,\,\stackrel{T=0}{\rightarrow}\,\,   	-\,sgn(q_0)\,\pi\,{\cal A}(q) \quad.
\eeqa
Here $n_B$ is the thermal distribution factor defined in Eq. \ref{bose}. The arguments presented in this Section show that in symmetric nuclear matter both the Feynman and the retarded propagator are completely determined from the positive energy sector and a dispersion relation extending over all energies $q_0$ can easily be written down, cf. Eqs.  \ref{realrhovac} and \ref{realpi} in Chapters \ref{vacself} and \ref{itscheme}.

\subsection{Fermions}

For fermions, retarded and Feynman propagator are defined as follows:
\beqa
\label{propdef2}
	i\,{\cal G}^+(x,y) &=& \theta(x_0-y_0)\,\expt{\left\{\psi(x),{\bar \psi(y)} \right\}} \\
	i\,{\cal G}^F(x,y) &=& \theta(x_0-y_0)\,\expt{\psi(x)\,{\bar \psi(y)}} -
	                       \theta(y_0-x_0)\,\expt{{\bar \psi(y)}\,\psi(x)} \nonumber  \quad.
\eeqa
In analogy to bosons, one finds that the real parts of both propagators are identical both in the vacuum and in the medium. The matrix ${\cal A}(p)$ is introduced in analogy to Eq. \ref{adef} as:
\beqa
{\cal A}(p) &\equiv&	-\frac{\imag{{\cal G}^+(p)}}{\pi} \,\,=\,\, \frac{1}{2\pi}\,\int \!\!d^4\!x\,e^{i p x}\,\left\{\psi(x),\psi^\dagger(0) \right\}  \quad,
\eeqa
where real and imaginary part of the matrix ${\cal G}(p)$ are defined as \cite{stefanfromel}:
\beqa
 \real{{\cal G}(p) }&=& \frac12\,\left[{\cal G}(p)+\gamma_0\,{\cal G}^{\dagger}(p)\,\gamma_0   \right] \quad,\quad
 \imag{{\cal G}(p) }\,\,=\,\, \frac{1}{2i}\,\left[{\cal G}(p)-\gamma_0\,{\cal G^{\dagger}}(p)\,\gamma_0   \right]
\eeqa
Tracing ${\cal A}$ with $\gamma_0$ defines the spectral function $\rho$:
\beqa
\trace{\gamma_0\,{\cal A}} &=& 4\,p_0\,\rho(p)\,\rm{sgn}(p_0) \quad,
\eeqa
for which a sum rule is obtained by imposing the equal-time commutator relations for fermion fields:
\beqa
	\int\limits_{-\infty}^{+\infty} \frac{dp_0}{4}\,\trace{\gamma_0\,{\cal A}(p)} &=& 1 \quad.
\eeqa
In vacuum one can derive the symmetry relation $\rho(-p_0)=\rho(p_0)$ by utilizing the invariance of the ground state under charge conjugation. Isospin rotations do not help in this case since they do not relate particles and antiparticles. Due to the finite baryo-chemical potential $\mu$, charge conjugation is not a good symmetry in nuclear matter and the relation between positive and negative energies is lost. The finite $\mu$ also complicates the relation between the imaginary parts of ${\cal G}^+(p)$ and ${\cal G}^F(p)$. Imposing again the KMS relation \cite{stefankms}, we find:
\beqa
	\imag{{\cal G}^F(p)} &=& -\,(1-2\,n_F)\,\pi\,{\cal A}(p)\,\,\to\,\,\textrm{sgn}(p_0-\mu)\,\pi\,{\cal A}(p) \quad,
\eeqa
with the Fermi distribution factor 
\beqa
	n_F(p_0) &=& \frac{1}{e^{(p_0-\mu)/T}+1} \quad.
\eeqa
The above arguments show that for fermions in nuclear matter the relation between the positive and negative energy sector is non-trivial, which explains why we restrict ourselves to the positive energies when applying dispersion relations to fermion self energies. Anyway we do not expect that antibaryons are important in cold nuclear matter.


\section{Parameters}
\label{parameters}

\renewcommand{\arraystretch}{2}
\begin{table}[h!]
\bdm
\begin{array}{|l|llll|}\hline
\mbox{coupling constant} & 
f_{NN\pi}=1.0 & f_{NN\eta}=2.34 & f_{NN\rho}=7.8 & f_{\Delta N\rho}=10.5 \\ \hline
\mbox{cutoff [GeV]} & 
\Lambda_\pi = 1.0 & \Lambda_\eta = 1.5 & \Lambda_\rho=1.5 & \Lambda_g = 1.5 \\ \hline
\mbox{short range} &
g_\pi^{p,NN}=0.6 \quad\quad& g_\pi^{p,R_N R_M}=0.45 \quad\quad& g_\eta^{p,NN}=0.6 \quad\quad& \\
& g_\pi^s = (0,0.1) & g_\rho^s = (0,0.1) & g_\pi^d = (0,0.4)= g_\pi^{dp} 
& g_\eta^s=(0,0.1) \\ \hline
\end{array}
\edm
\caption{\label{param} Coupling constants, short-range and cutoff parameters of our model.}
\end{table}
\renewcommand{\arraystretch}{1}

In Table \ref{param} we give a list of coupling constants, cutoff and short-range parameters used in the calculations. The brackets denote the range within which we allow the respective parameter to vary. Note that at each vertex corresponding to short-range interactions we multiply a monopole form factor \cite{helgesson1, helgesson2}:
\beqa
	F_g(q^2) &=& \left(\frac{\Lambda_g^2}{\Lambda_g^2-q^2}\right)^2 \quad .
\eeqa
The value for also $\Lambda_g$ is also taken from this reference. Let us comment on theses choices for coupling constants and cutoff parameters. For the $\pi NN$ coupling constant a value around $f_{\pi NN}=1$ is a standard choice, see for example \cite{osetreview,helgesson1}. For $f_{NN\eta}$ and the cutoff $\Lambda_\eta$ we take the values originally suggested in \cite{bonnpot}. The values for $f_{\rho NN}$ and $f_{\rho N\Delta}$ lie within the ranges suggested in for example \cite{helgesson2,rhocoupl,osetreview} and are a obtained by a mix of quark model considerations and fits to $NN$ scattering. These fits also determine the approximately the values of the cutoff parameters, in particular the rather large value for $\Lambda_\rho$ is suggested from those data, see for example \cite{helgesson2}. The cutoff used in the form factor $F(k^2)$ of Eq. \ref{ff2} for pseudoscalar $(\pi,\,\eta)$ meson is taken to be $\Lambda=1$ GeV and for normal nuclear matter density we take a value of $\rho_0=0.15\,\textrm{fm}^{-3}$.

\section{Lagrangians and Traces}
\label{applag}

In the following paragraphs we give the Lagrangians used for the description of meson-nucleon-resonance dynamics. Of course, these Lagrangians also have an isospin part. We denote the isospin part explicitly here and omit it in the Lagrangians.
The isospin coupling of an isovector meson and nucleon forming a resonance with $I=\frac 1 2$ or $I=\frac 3 2$ is given by:
\beqa
\label{lagiso}
	\psi^\dagger \,{\boldsymbol \tau} \,\psi \,{\boldsymbol \varphi} & & \\
	\psi^\dagger \,{\bf T} \psi \,{\boldsymbol \varphi} & & \nonumber \quad.
\eeqa
Here ${\bf T}$ and ${\boldsymbol \tau}$ are the usual spin-$\frac 3 2$ transition and Pauli operators. The the isospin components of the isovector meson are denoted by ${\boldsymbol \varphi}$.

\subsection{Lagrangian Relativistic}
\label{lagrel}

In this Section we will write down the relativistic Lagrangians used for the description of the coupling of baryon resonances to nucleons and pseudoscalar mesons $\varphi$ or vector mesons/photons $V^\mu$. The guiding principle in writing down these interaction terms is that they fulfill the usual symmetry requirements Lorentz invariance, gauge invariance and parity conservation. In addition, they are required to be hermitian.
The standard coupling of a resonance with the quantum numbers $J^\pi$ for spin $J$ and parity $\pi$ to the $\varphi\,N$ channel reads:
\beq
\label{rlagpin}
\begin{array}{rcccc}
{\cal L}_{RN\varphi} &=& \D{
				\frac{f}{m_\varphi}\,{\bar 	\psi_R}\,\gamma^\mu\,
				\left\{ \begin{array}{c} i\,\gamma^5 \\ 1\!\!1  \end{array} \right\}
				\psi_N\,\partial_\mu\,\varphi}
         \quad &\mbox{for}& J^\pi = \frac 1 2 ^\pm \\ \\
{\cal L}_{RN\varphi} &=& \D{
				\frac{f}{m_\varphi}\,{\bar \psi_R^\mu}\,
				\left\{ \begin{array}{c} 1\!\!1 \\ i\,\gamma^5 \end{array} \right\}
				\psi_N\,\partial_\mu\,\varphi}
         \quad&\mbox{for}& J^\pi = \frac 3 2 ^\pm \\
\end{array}
\eeq

The standard coupling of a resonance with the quantum numbers $J^\pi$ for spin $J$ and parity $\pi$ to the $\rho\,N$ channel reads:
\beq
\label{rlagrhon}
\begin{array}{rcccc}
{\cal L}_{RN\rho} &=& \D{
				\frac{f}{m_\rho}\,{\bar 	\psi_R}\,\sigma^{\mu\nu}\,
				\left\{ \begin{array}{c} 1\!\!1 \\ i\,\gamma^5 \end{array} \right\}
				\psi_N\,\partial_\mu\,\rho_\nu}
         \quad &\mbox{for}& J^\pi = \frac 1 2 ^\pm \\  \\
{\cal L}_{RN\rho} &=& \D{
				 \frac{f}{m_\rho}\,{\bar \psi_R^\mu}\,\gamma^\nu\,
				 \left\{ \begin{array}{c} i\,\gamma^5 \\ 1\!\!1  \end{array} \right\}
				 \psi_N\,\rho_{\mu\nu}}
         \quad&\mbox{for}& J^\pi = \frac 3 2 ^\pm \\
\end{array}
\eeq
Here $\rho_\mu$ describes the $\rho$ meson and $\rho_{\mu\nu}=\partial_\mu\,\rho_\nu-\partial_\nu\,\rho_\mu$.


\subsection{Lagrangian Non-Relativistic}
\label{lagnrel}
Expect for spin-$\frac52$ resonances, the non-relativistic Lagrangians presented here are derived from the relativistic ones as given in Eqs. \ref{rlagpin} and \ref{rlagrhon}. 

The Lagrangian describing the coupling to a pseudoscalar meson and a nucleon reads:
\beq
\label{nrlagpin}
\begin{array}{rclcc}
{\cal L}_{RN\varphi} &=& i\,\D{
				\frac{f}{m_\varphi} \psi_R^\dagger
				\, \sigma_k\,\psi_N\,\partial_k\,\varphi }
         \quad &\mbox{for}& J^\pi = \frac 1 2 ^+ \\ \\
         &=& \D{
				\frac{f}{m_\varphi} \psi_R^{\dagger}\,\psi_N}\,\partial_0\,\varphi
         \quad &\mbox{for}& J^\pi = \frac 1 2 ^- \\ \\
{\cal L}_{RN\varphi} &=& \D{
				\frac{f}{m_\varphi}\,\psi_R^\dagger\,S_k^\dagger\,\psi_N}\,\partial_k\,\varphi
         \quad&\mbox{for}& J^\pi = \frac 3 2 ^+ \\ \\
         &=&  i\,\D{
				\frac{f}{2\,m_N\,m_\varphi}\,\psi_R^\dagger\,S_k^\dagger
				 \,\sigma_l\,(\partial_l\,\psi_N})\,\partial_k\,\varphi
         \quad&\mbox{for}& J^\pi = \frac 3 2 ^- \\ \\
{\cal L}_{RN\varphi} &=& \D{
				\frac{f}{m_\varphi}\,\psi_R^\dagger\,R_{i\,j}\,
				\sigma_k \psi_N}\,\partial_i\,\partial_j\,\partial_k\,\varphi
         \quad&\mbox{for}& J^\pi = \frac 5 2 ^+ 
\end{array}
\eeq
The spin transition operators $S_i$ and $R_{ij}$ can be found in \cite{fripir}.

The coupling to vector mesons and nucleons is described by:
\beq
\label{nrlagrhon}
\begin{array}{rclcc}
{\cal L}_{RN\rho} &=& \D{
				\frac{f}{m_V} \,\psi_R^\dagger
				\,\sigma_j \,\psi_N}\,\epsilon_{jkl}\,\partial_k\,\rho_l
         \quad &\mbox{for}& J^\pi = \frac 1 2 ^+ \\ \\
         &=& i\,\D{
				\frac{f}{m_V}\,\psi_R^\dagger \sigma_k\,\psi_N}\,
				\left(\partial_k\,\rho_0 - \partial_0\,\rho_k  \right)
         \quad &\mbox{for}& J^\pi = \frac 1 2 ^- \\ \\
{\cal L}_{RN\rho} &=& i\,\D{
				\frac{f}{m_V} \,\psi_R^\dagger
				\,S_j^\dagger \,\psi_N}\,\epsilon_{jkl}\,\partial_k\,\rho_l
         \quad &\mbox{for}& J^\pi = \frac 3 2 ^+ \\ \\
         &=& \D{
				\frac{f}{m_V}\,\psi_R^\dagger\,S_k^\dagger\,\psi_N}\,
				\left(\partial_k\,\rho_0 - \partial_0\,\rho_k  \right)
         \quad &\mbox{for}& J^\pi = \frac 3 2 ^- \\ \\
{\cal L}_{RN\rho} &=& \D{
				\frac{f}{m_V} \psi_R^\dagger\,R_{ij}
				\,\psi_N } \partial_j\,\rho_i^T
         \quad &\mbox{for}& J^\pi = \frac 5 2 ^+ 
\end{array}
\eeq
In the last line the notion $\rho_i^T$ is meant to imply that only transversely polarized vector particles couple to spin-$\frac 5 2$ resonances.

For the coupling of spin-$\frac 3 2$ resonances to the $\Delta\pi$ channel we use the following Lagrangian:
\beq
\label{nrlagpidel}
\begin{array}{rclcc}
	{\cal L}_{R\Delta\pi} &=& \D{
				\frac{f}{m_\Delta} \psi_R^{\dagger}\,S_k^\dagger\,S_k\,
			  \,\psi_\Delta}\,\pi \quad &\mbox{for}& J^\pi = \frac32^- \quad. \\ \\
			  {\cal L}_{R\Delta\pi} &=& \D{
				\frac{f}{m_\Delta} \psi_R^{\dagger}\,S_i^\dagger\,S_j\,
			  \,\psi_\Delta}\epsilon_{ijk}\partial_k\,\pi \quad &\mbox{for}& J^\pi = \frac32^+ \quad.
\end{array}
\eeq
For the coupling of spin-$\frac 1 2$ resonances the appropriate Lagrangians of Eq. \ref{nrlagpin} can be applied.


\renewcommand{\arraystretch}{2}
\begin{table}
\begin{equation*}
\begin{array}{|l|c|l||l|c|l|}\hline 
{\bf \Omega^{\varphi}_{1/2} } \quad  &\pi=+1& 
\quad \D{8\,m_N\,m_R\,{\bf q}^2 } &
{\bf \Omega^{\varphi}_{3/2} } &\pi=+1&
\quad \D{\frac{16}{3}\,m_N\,m_R\,{\bf q}^2 }\\ 
&\pi=-1& 
\quad \D{8\,m_N\,m_R\,q_0^2 } & 
&\pi=-1& 
\quad \D{\frac{16}{3}\,m_N\,m_R\,\frac{{\bf q}^4}{4\,m_N^2}}\\
\hline
{\bf \Omega^{\varphi}_{5/2} } &\pi=+1&
\quad \D{\frac{16}{5}\,m_N\,m_R\,{\bf q}^2 } & & & \\ \hline \hline
{\bf \Omega^T_{1/2} } &\pi=+1& 
\quad 8\,m_N\,m_R\,{\bf q}^2 &
{\bf \Omega^T_{3/2} } &\pi = +1& 
\quad  \D{\frac{16}{3}\,m_N\,m_R\,{\bf q}^2} \\ 
&\pi=-1& 
\quad  8\,m_N\,m_R\,{\bf q_0}^2 &
&\pi = -1&
\quad  \D{\frac{16}{3}\,m_N\,m_R\,{\bf q_0}^2}\\  \hline
{\bf \Omega^{L}_{1/2} } &\pi = +1&  \quad \D{0} &
{\bf \Omega^{L}_{3/2} } &\pi = +1&  \quad \D{0} \\
&\pi = -1& \quad  8\,m_N\,m_R\,q^2 & 
&\pi = -1& \quad  \D{\frac{16}{3}\,m_N\,m_R\,q^2}\\  \hline 
{\bf \Omega^T_{5/2} } &\pi=+1&
\quad \D{\frac{12}{5}\,m_N\,m_R\,{\bf q}^2 } & & &\\ \hline
\end{array}
\end{equation*}
\caption{\label{ntraces} Non-relativistic traces for the resonance decay and the meson self energy.}
\end{table}
\renewcommand{\arraystretch}{1}

\renewcommand{\arraystretch}{2}
\begin{table}
\begin{equation*}
\begin{array}{|l|c|l||l|c|l|}\hline 
{\bf \Omega^{\varphi,{red}}_{1/2} } \quad  &\pi=+1& 
\quad \D{8\,m_N\,m_R} &
{\bf \Omega^{,{red}}_{3/2} } &\pi=+1&
\quad \D{\frac{16}{3}\,m_N\,m_R}\\ 
&\pi=-1& 
\quad \D{8\,m_N\,m_R} & 
&\pi=-1& 
\quad \D{\frac{16}{3}\,\,m_N\,m_R\,\frac{{\bf q}^2}{4\,m_N^2}}\\
\hline
{\bf \Omega^{T,{red}}_{1/2} } &\pi=+1& 
\quad 8\,m_N\,m_R &
{\bf \Omega^{T,{red}}_{3/2} } &\pi = +1& 
\quad  \D{\frac{16}{3}\,m_N\,m_R} \\ 
&\pi=-1& 
\quad  8\,m_N\,m_R &
&\pi = -1&
\quad  \D{\frac{16}{3}\,m_N\,m_R} \\  \hline
{\bf \Omega^{L,{red}}_{1/2} } &\pi = +1&  \quad \D{0} &
{\bf \Omega^{L,{red}}_{3/2} } &\pi = +1&  \quad \D{0} \\
&\pi = -1& \quad  8\,m_N\,m_R& 
&\pi = -1& \quad  \D{\frac{16}{3}\,m_N\,m_R }\\  \hline 
{\bf \Omega^{T,{red}}_{5/2} } &\pi=+1&
\quad \D{\frac{12}{5} \,m_N\,m_R} & & &\\ \hline
\end{array}
\end{equation*}
\caption{\label{ntracesred} Reduced traces $\Omega^{red}$, which arise from the contact interactions.}
\end{table}
\renewcommand{\arraystretch}{1}


\subsection{Traces}
\label{tracenonrel}
In this work, the Lagrangians are used to find analytic expressions for the decay width of a resonance and to calculate the meson-nucleon forward scattering amplitude, which is closely related to the in-medium self energy of the meson. In both cases one needs to calculate a trace which is of the generic form
\beqa
\label{omega_def}
\Omega^\varphi_{1/2}&=&\trace{\Gamma \Gamma^\dagger} \quad,\\	
\Omega^\varphi_{3/2}&=&\trace{\Gamma_i\,P_{3/2}^{ij}\,\Gamma_j^\dagger} \nonumber \quad.
\eeqa
for pseudoscalar mesons and
\beqa
\label{omega_def2}
\Omega^{T/L}_{1/2}&=&P^{T/L}_{\mu\nu}\,\trace{\Gamma^\mu\,\Gamma^{\nu\dagger}} \quad,\\
\Omega^{T/L}_{3/2}&=&P^{T/L}_{\mu\nu}\,
\trace{\Gamma^\mu_i\,P_{3/2}^{ij}\,\Gamma^{\nu\dagger}_j} 	\nonumber \quad.
\eeqa
for vector mesons. As usual $T$ and $L$ denote the polarization of the vector meson. The vertex factors $\Gamma$ are obtained from the above non-relativistic Lagriangians.
We display the results for these traces in Table \ref{ntraces}. In the actual calculations, energy $q_0$ and momentum ${\bf q}$ of the meson are taken in the rest frame of the resonance.

In calculating these traces we have used the projectors onto spin-$\frac 3 2$ and spin-$\frac 5 2$ fields, which are given by \cite{fripir,ericsonweise}:
\beqa
\label{project32nr}
		P_{3/2}^{ij}
		&=& \delta^{ij}-\frac 1 3 \,\sigma^i\,\sigma^j  \\
	P_{5/2}^{ij,kl}&=& \frac 1 2 (\delta_{ik}\delta_{jl} + \delta_{il}\delta_{jk}) - 								\frac 1 5 \, \delta_{ij}\delta_{kl} - \nonumber \\ && -\frac{1}{10}\, 
	(\delta_{ik}\,\sigma_j\,\sigma_l+ \delta_{il}\,\sigma_j\,\sigma_k+ 
	 \delta_{jk}\,\sigma_i\,\sigma_l+ \delta_{jl}\,\sigma_i\,\sigma_k ) \nonumber \quad.
\eeqa


\section{Details of the Derivation of Short-Range Interactions}
\label{srcdetails}

This Appendix is concerned about the derivation of the contact interactions given in Eqs. \ref{lagconplus}, \ref{lagconminus} and \ref{lagconmix}. We also present details of how we obtain estimates for the strength of the short-range correlations by matching contact interactions and correlation potential.

\subsection{Positive Parity States}
\label{pospar}

We begin with a discussion of positive parity states. Reading the hadronic current $J_\mu$ off the Lagrangians Eq. \ref{rlagpin} in Appendix \ref{applag} and performing
a non-relativistic reduction, the following contact interactions ${\cal L}_C^\pi$ result for the nucleon-resonance interactions.

$J^\pi=\frac1 2^+$:
\beqa
\label{cpi12p}
	{\cal L}_C^\pi &=& c_\pi\,\left(\frac{f}{m_\pi}\right)^2\,
	\left({\bar \psi}_R\,\gamma_5\,\gamma^\mu\,\psi_N \right)
	\left({\bar \psi}_N\,\gamma_5\,\gamma_\mu\,\psi_R \right) \\ &\Rightarrow&
	c_\pi\,\left(\frac{f}{m_\pi}\right)^2\,
	\left({\psi}_R^\dagger\,\sigma^i\,\psi_N \right)
	\left({\psi}_N^\dagger\,\sigma_i\,\psi_R \right) \quad, \nonumber
\eeqa

$J^\pi=\frac3 2^+$:
\beqa
\label{cpi32p}
	{\cal L}_C^\pi &=& c_\pi\,\left(\frac{f}{m_\pi}\right)^2\,
	\left({\bar \psi^\mu}_R\psi_N \right)
	\left({\bar \psi}_N \psi_{R,\mu} \right) \\ &\Rightarrow&
	c_\pi\,\left(\frac{f}{m_\pi}\right)^2\,
	\left({\psi}_R^\dagger\,S^{i\,\dagger}\,\psi_N \right)
	\left({\psi}_N^\dagger\,S_i \,\psi_R \right)
	\quad , \nonumber
\eeqa 
Here the spin-$\frac{3}{2}$ transition matrix $S$ has been introduced, which contains the Clebsch-Gordan coefficients for the spin coupling $\frac 1 2 \oplus 1 = \frac 3 2$ \cite{ericsonweise}. Note that in the lower line of each of the equations $\psi_N$ and $\psi_R$ denote non-relativistic two-component spinor fields. 

In order to build ${\cal L}_C^\rho$, we first decompose the baryonic tensor $B^{\mu\,\nu}$ into its spatial $(j,k)$ and time $(j,0),\,(0,k)$ components. The non-relativistic interaction is then obtained by keeping only the leading terms in $p_N/m_N$ of $B^{\mu\nu}$. 

$J^\pi=\frac1 2^+$: 
\beqa
\label{crho12p}
{\bar \psi}_R\,\sigma^{\mu\,\nu} \,\psi_N &=& \left\{ 
\begin{array}{rcl}
(j,k) &:& \D{\epsilon^{jkl}\,\psi_R^\dagger \,\sigma_l\,\psi_N} \\ && \\
(j,0) &:& \D{-i\,\psi_R^\dagger \,\sigma^j\,\frac{\sigma_m\,\partial_m}{2m_N}\,\psi_N} 
\\ && \\ (0,k) &:& -(j,0)
\end{array}
\right. \eeqa
Dropping now the terms $\frac{{\boldsymbol \sigma}\cdot {\bf p}_N}{2 m_N}$ gives:
\beqa
	{\cal L}_C^\rho &=& 2\, c_\rho\,\left(\frac{f}{m_\rho}\right)^2\,
	\left({\psi}_R^\dagger\,\sigma^{i}\,\psi_N \right)
	\left({\psi}_N^\dagger\,\sigma_i\,\psi_R \right)
	\quad . \nonumber
\eeqa

$J^\pi=\frac3 2^+$:
\beqa
\label{crho32p}
i\,\left({\bar \psi^\mu}_R\,\gamma^\nu - {\bar \psi^\nu}_R\,\gamma^\mu \right)\,\gamma^5\,\psi_N &=& 
\left\{ 
\begin{array}{rcl}
(j,k) &:& \D{\psi_R^\dagger \,\epsilon^{jkl}\,S^\dagger_l\,\psi_N} \\ && \\
(j,0) &:& \D{i\,\psi_R^\dagger \,S^{j\,\dagger}\,\frac{\sigma_m\,\partial_m}{2m_N}\,\psi_N} 
\\ && \\ (0,k) &:& -(j,0)
\end{array}
\right. \eeqa
Again dropping terms $\frac{{\boldsymbol \sigma}\cdot {\bf p}_N}{2 m_N}$ leads to:
\beqa
{\cal L}_C^\rho &=& 2\, c_\rho\,\left(\frac{f}{m_\rho}\right)^2\,
	\left({\psi}_R^\dagger\,S^{i\,\dagger}\,\psi_N \right)
	\left({\psi}_N^\dagger\,S_i\,\psi_R \right)
	\quad , \nonumber
\eeqa

Both $\pi$ and $\rho$ induced interactions have the same spin-structure
$\propto {\boldsymbol \sigma}_1\cdot{\boldsymbol \sigma}_2$ for spin-$\frac 1 2$ states and 
$\propto {\bf S}_1\cdot{\bf S}_2$ for spin-$\frac 3 2$ states, as required from Eq. \ref{vcorr2}. 
The form of ${\cal L}_C^\pi$ and ${\cal L}_C^\rho$ implies, that in any calculation where ${\cal L}_C^\pi$ contributes, also ${\cal L}_C^\rho$ has to be considered.
Therefore it is advisable to consider the sum of both terms with a new parameter $g^p$, where the index $p$ refers to the $p$-wave coupling of the underlying interaction.
To give an example, the short-range interactions for $J^\pi=\frac 1 2^+$ states then read:
\beqa
\label{gprmatch0}
	g^p\,\left(\frac{f_{\pi}}{m_\pi}\right)^2 {\boldsymbol\sigma}_1\cdot{\boldsymbol\sigma}_2&=&
	\left[c_\pi\,\left(\frac{f_{\pi}}{m_\pi}\right)^2  +
	2\,c_\rho\,\left(\frac{f_{\rho}}{m_\rho}\right)^2 \right] 
	{\boldsymbol\sigma}_1\cdot{\boldsymbol\sigma}_2 \quad.
\eeqa
Now we need to determine the strength parameter $g^p$. To this end we subject the $p$-wave potential corresponding to $P=+1$ states to the correlation integral Eq. \ref{corrint}, as motivated in the introduction of Chapter \ref{nrint}.

As an example let us discuss the $\pi$ exchange in the $NN$ potential, Eq. \ref{vcorr1}.
Plugging in the central part of the potential one obtains the following result \cite{osetweisecorr}:
\beqa
V_C(q_0,{\bf q}) &=& \int \frac{d\Omega_{q^\prime}}{4\pi} \,\left(\frac{f_{NN\pi}}{m_\pi}\right)^2 F^2({\bf q}+{\bf q^{\prime}})\, \frac{{\boldsymbol\sigma}_1\cdot{\boldsymbol\sigma}_2\,\delta_{ij}}{q_0^2-({\bf q^{\prime}}+{\bf q})^2-m_\pi^2} \,\times \nonumber 
\\ && \times
\left[q_i\,q_j+q^{\prime}_i\,q^{\prime}_j+q^{\prime}_i\,q_j+ q^{\prime}_j\,q_i\right]_{|q^{\prime}|=q_c}   						
\\ &\approx&
\left(\frac{f_{NN\pi}}{m_\pi}\right)^2 
{\tilde F^2({\bf q})} \, {\tilde D_\pi}(q_0,{\bf q})
\,{\boldsymbol\sigma}_1\cdot{\boldsymbol\sigma}_2\,\frac{1}{3}\,q_c^2 \nonumber \quad.
\eeqa
The form factor $F$ introduced in \cite{osetweisecorr} is the monopole form factor of Eq. \ref{fft}. The new quantities ${\tilde D_{\rho/\pi}}$ and ${\tilde F_{\rho/\pi}}$ are defined like the usual propagators and form factors with the replacement ${\bf q}^2 \rightarrow  {\bf q}^2 + q_c^2$ \cite{osetweisecorr}. 

A similar term is found for $\rho$ exchange, and the sum of both yields for the central part of the correlation potential in the limit ${\bf q}=0$, see also Eq. \ref{vcorr2}:
\beqa
\label{vcorr3}
	V_C(q_0,0) &=& \frac {q_c^2}{3} \,\left[ 
\left(\frac{f_{NN\pi}}{m_\pi}\right)^2 \,{\tilde F_\pi}^2\,
{\tilde D_\pi}(q_0,0) + 2\,
\left(\frac{f_{NN\rho}}{m_\rho}\right)^2 \,{\tilde F_\rho}^2\,
{\tilde D_\rho}(q_0,0)   
								\right] \,\boldsymbol{\sigma_1}\cdot\boldsymbol{\sigma_2}  \quad.
\eeqa
Note that the correlation induces a large correction to the free potential Eq. \ref{vcorr1} due to the presence of the large scale set by $q_c$. For spin-$\frac 3 2$ states the potential is obtained with the replacement ${\boldsymbol \sigma} \rightarrow {\bf S}$.

The correlation potential Eq. \ref{vcorr3} and the contact interaction Eq. \ref{gprmatch0} can now be matched, yielding the following results for $g^p$:
\beqa
\label{gprmatch1}
	g^p\,\left(\frac{f_{\pi}}{m_\pi}\right)^2 &=&
	\frac{q_c^2}{3}\left[\left(\frac{f_{\pi}}{m_\pi}\right)^2 \, 
	{\tilde F_\pi}^2\,{\tilde D_\pi} \,+\,
	2\,\left(\frac{f_{\rho}}{m_\rho}\right)^2 \,
	{\tilde F_\rho}^2\,{\tilde D_\rho}\right]  \quad.
\eeqa 
It has been known for a long time that this sort of matching leads to reasonable guesses for $g^p$ with the kinematical matching point chosen to be $q=(0,0)$ \cite{osetweisecorr,osetreview}. Phenomenology requires values for $g^p$ to be in the order of $0.6$ for the $NN$ potential and somewhat less for $N\Delta$ and $\Delta\Delta$ transition potentials. Similar results are found by using Eq. \ref{gprmatch1}. Depending on one's favourite values for cutoff parameters and coupling constants, one obtains for the $NN$ potential values in the range of $0.4-0.7$, owing to a large extent to the $\rho$ exchange. Somewhat smaller values are found for the  $\Delta N$ transition potential.

We have now achieved a description of SRC from two different starting points, namely contact interactions of Eqs. \ref{crho12p}, \ref{crho32p} and the correlation approach Eq. \ref{vcorr3}. Matching both approaches gives reasonable results for the strength parameter $g^p$ of the SRC. This success in setting up a model for the short-range correlations for $P=+1$ states motivates us to proceed along the same lines for the $P=-1$ sector and thus obtain reasonable parameter ranges describing the strength of the SRC.
At this point we iterate that in the actual calculations we do not use the values for $g^p$ as obtained from Eq. \ref{gprmatch1} but vary $g^p$ within accepted boundaries in order to obtain a reasonable description of the in-medium properties of the $P_{33}(1232)$.


\subsection{Negative Parity States}
\label{negpar}

Let us now turn to the discussion of negative parity states. We proceed along the same lines as for $P=+1$ states. For the contact interactions derived in the $\pi$ sector the following Lagrangians are obtained.

$J^\pi=\frac 1 2^-$:
\beqa
\label{cpi12m}
	{\cal L}_C^\pi &=& c_\pi\,\left(\frac{f}{m_\pi}\right)^2\,
	\left({\bar \psi}_R\,\gamma^\mu\,\psi_N \right)
	\left({\bar \psi}_N\,\gamma_\mu\,\psi_R \right) \\ &\Rightarrow&
	c_\pi\,\left(\frac{f}{m_\pi}\right)^2\,
	\left({\psi}_R^\dagger\,\psi_N \right)
	\left({\psi}_N^\dagger\,\psi_R \right) \quad . \nonumber
\eeqa 

$J^\pi=\frac 3 2^-$:
\beqa
\label{cpi32m}
	{\cal L}_C^\pi &=& c_\pi\,\left(\frac{f}{m_\pi}\right)^2\,
	\left({\bar \psi^\mu}_R\,\gamma^5\,\psi_N \right)
	\left({\bar \psi}_N\,\gamma^5\,\psi_{R,\mu} \right) \\ &\Rightarrow&
	\D{c_\pi\,\left(\frac{f}{m_\pi}\right)^2\,
	\bigg({\psi}_R^\dagger\,S^{i\,\dagger}\,\frac{\sigma_k\,\partial_k}
	{2 m_N}\,\psi_N \bigg) \bigg({\psi}_N^\dagger\,
	\frac{\stackrel{_\gets}{\partial_k} \,\sigma_k}{2 m_N}\,S_i\,\psi_R \bigg) }\quad . \nonumber
\eeqa 

For $J^\pi=\frac 3 2^-$ resonances, the contact interaction is of the order $\frac{{\boldsymbol \sigma}\cdot{\bf p_N}}{2 m_N}$ and up to now we have dropped such terms. Here these terms should be kept for consistency, since they also arise in the non-relativistic reduction of the $\pi N R$ interaction (see Eqs. \ref{rlagpin} and \ref{nrlagpin} in Appendix \ref{applag}), where they produce the necessary $d$-wave coupling.

In order to obtain the non-relativistic contact interaction ${\cal L}_C^\rho$, we first decompose the tensor $B_{\mu\nu}$ into spatial $(j,k)$ and time $(j,0),\,(0,k)$ components. 
Keeping only the leading non-relativistic terms, this leads to the following expressions for the nucleon-resonance interaction:

$J^\pi=\frac 1 2^-$:
\beqa
\label{crho12m}
i\,{\bar \psi}_R\,\sigma^{\mu\,\nu}\,\gamma^5 \,\psi_N &=& \left\{ 
\begin{array}{rcl}
(j,k) &:& \D{i\,\psi_R^\dagger \,\epsilon^{jkl}\,\sigma_l\,\frac{\sigma_m\,\partial_m}{2m_N}\,\psi_N} \\ && \\
(j,0) &:& \D{\psi_R^\dagger \,\sigma^j\,\psi_N} 
\\ && \\ (0,k) &:& -(j,0)
\end{array}
\right. \\ && \nonumber \\
{\cal L}_C^\rho &=& 2\, c_\rho\,\left(\frac{f}{m_\rho}\right)^2\,
	\left({\psi}_R^\dagger\,\sigma^{i}\,\psi_N \right)
	\left({\psi}_N^\dagger\,\sigma_i\,\psi_R \right)
	\quad . \nonumber
\eeqa

$J^\pi=\frac 3 2^-$:
\beqa
\label{crho32m}
\left({\bar \psi^\mu}_R\,\gamma^\nu - {\bar \psi^\nu}_R\,\gamma^\mu \right)\,\psi_N &=& 
\left\{ 
\begin{array}{rcl}
(j,k) &:& \D{-i\,\psi_R^\dagger \,\epsilon^{jkl}\,S^\dagger_l\,\frac{\sigma_m\,\partial_m}{2m_N}\,\psi_N} \\ && \\
(j,0) &:& \D{\psi_R^\dagger \,S^{j\,\dagger}\,\psi_N} 
\\ && \\ (0,k) &:& -(j,0)
\end{array}
\right.\\ &&  \nonumber \\
	{\cal L}_C^\rho &=& 2\,c_\rho\,\left(\frac{f}{m_\rho}\right)^2\,
	\left({\psi}_R^\dagger\,S^{i\,\dagger}\,\psi_N \right)
	\left({\psi}_N^\dagger\,S_i\,\psi_R \right)
	\quad . \nonumber
\eeqa
As in the $P=+1$ case, these interactions are the simplest contact interactions leading to non-vanishing contributions in typical diagrams such as Fig. \ref{resholegpr}. 

As mentioned already in the introduction of Chapter \ref{nrint}, there is a clear difference between the contact interactions for $P=+1$ and those for $P=-1$ states. In the negative parity sector ${\cal L}_C^\pi$ and ${\cal L}_C^\rho$ are not equivalent. Considering, for example, a correction to the meson self energy  according to Fig. \ref{resholegpr}, the pion self energy will not receive contributions from ${\cal L}_C^\rho$ and neither will the $\rho$ self energy receive contributions form ${\cal L}_C^\pi$. For spin-$\frac12$ resonances this can be motivated as follows: since the coupling to pions is $s$-wave and the pion is a pseudoscalar particle, there is no vector available that could couple to a $\sigma$ matrix and consequently the leading non-relativistic term of $J_\mu$ contains no spin-flip terms. For the $\rho$ meson, albeit coupling in an $s$-wave as well, there is still the polarization vector and therefore the leading non-relativistic terms of $B_{\mu\nu}$ produce spin-flip contributions. 
Doing the spin-summation as appropriate for calculations in nuclear matter, the decoupling of $\pi$ and $\rho$ sector follows. A similar argument holds for $J^\pi=\frac32^-$ states.
It is therefore necessary to keep both ${\cal L}_C^\pi$ and ${\cal L}_C^\rho$ independently and fix the respective parameters $c_\pi$ and $c_\rho$. We will from now on denote $c_\pi$ and $c_\rho$ by $g_\pi^d$, $g_\pi^s$ and $g_\rho^s$, where $s$ and $d$ indicate the angular momentum of the underlying meson-nucleon interaction.

We now subject the $\pi$ and $\rho$ exchange potentials to the correlation integral Eq. \ref{corrint}, which allows us to determine the strength of the contact interaction.
Let us start off with the $\rho$ exchange for $J^\pi=\frac 3 2^-$ states.
Constructing the potential from the corresponding non-relativistic Lagrangian as given in the Appendix \ref{applag}, Eq. \ref{nrlagpin}, leads to the following result:
\beqa
\label{vs}
	V_\rho^s(q_0,{\bf q}) &=& -\left(\frac{f}{m_\rho} \right)^2\,F_\rho^2\,
	V^{\mu\nu}\,g_{\mu\nu}\,D_\rho(q_0,{\bf q}) \nonumber \\
	&=& \left(\frac{f}{m_\rho} \right)^2\,F_\rho^2\,D_\rho(q_0,{\bf q})
	\,\left(q_0^2\,{\bf S}_1\cdot{\bf S}_2
	-{\bf S}_1\cdot{\bf q}\,{\bf S}_2\cdot{\bf q}  \right) \\
	&\Rightarrow& \left(\frac{f}{m_\rho} \right)^2\,F_\rho^2\,D_\rho(q_0,{\bf q})\,
	\left(q_0^2 -{\bf q}^2  \right) \,{\bf S}_1\cdot{\bf S}_2 \nonumber \\ \mbox{with} && \nonumber \\
	V^{\mu\nu} &=& S^i_1\,S^j_2\,\D{\left(\begin{array}{cc}
									q_i\,q_j & q_0\,q_j \\ & \\
									q_i\,q_0 & q_0^2
									\end{array}\right)} \nonumber \quad.
\eeqa
Here we have projected out the spin-central part ${\bf S}_1\cdot{\bf S}_2$, which is generated by the contact interactions. The energy dependence of the $s$-wave potential is a direct consequence of current conservation, which we impose on the coupling of the baryon resonances to the vector meson-nucleon channel. 

The correlation potential $V_C$ obtained via Eq. \ref{corrint} reads:
\beqa
\label{vcorrs}
	V_C(q_0,{\bf q}) &=& \left(\frac{f}{m_\rho} \right)^2\,{\tilde F}_\rho^2\,
	{\tilde D}_\rho(q_0,{\bf q})\,
	\left(q_0^2 -\frac 1 3\,q_C^2  \right) \,{\bf S}_1\cdot{\bf S}_2 \quad.
\eeqa
By matching this with Eq. \ref{crho32m}, we estimate the strength of the contact interaction to be
\beqa
\label{gprsmatch}
	g_\rho^s &\equiv& 2\,c_\rho \,\,=\,\, \left(q_0^2 -\frac 1 3\,q_c^2  \right)\,{\tilde F}_\rho^2\,{\tilde D}_\rho(q_0,{\bf q}) \quad.
\eeqa
This expression should be evaluated at the energy which a $\rho$ meson needs to excite a resonance:
\beqa
\label{q0s}
	q_0({\bf q}) &=& \sqrt{m_R^2+{\bf q}^2}-m_N \quad.
\eeqa
Comparable energies are also encountered in the reaction $R\,N \rightarrow N\,R$.
For the $D_{13}(1520)$ this amounts to $q_0 \approx 0.5$ GeV at vanishing 3-momentum. Plugging this value into Eq. \ref{gprsmatch} we find $g_\rho^s \approx 0.05 - 0.1$.
In the calculations we vary the parameter $g_\rho^s$ within the boundaries $(0,0.1)$, 
and take the same value for $g_\rho^s$ for all other $\frac 3 2^-$ resonances.

The non-relativistic Lagrangian describing the coupling of $J^\pi=\frac3 2^-$ states to $N\pi$ is given in Appendix \ref{applag} and leads to a $d$-wave potential:
\beqa
\label{potd}
	V(q_0,{\bf q}) &=& \left(\frac{f}{m_\pi}\right)^2\,F_\pi^2\,\frac{1}{4\,m_N^2}\,
	\frac{({\boldsymbol S_1}\cdot{\bf q})\,({\boldsymbol \sigma_1}\cdot{\bf p_N})\,({\boldsymbol S_2}\cdot{\bf q})({\boldsymbol \sigma_2}\cdot{\bf p_N})}{q_0^2-{\bf q}^2-m_\pi^2} \quad.
\eeqa
Just as for $J^\pi=\frac32^+$ states, the interaction of pions and nucleons is spin-longitudinal in the $D_{13}$ channel. The only difference is an additional momentum dependence coming from the nucleon. 
From Eq. \ref{corrint} we obtain for the central part of the  correlation potential:
\beqa
\label{vcorrd}
 V_C(q_0,{\bf q})
 &=& \frac 1 3 \,q_c^2\,{\tilde F}_\pi^2\,\frac{({\boldsymbol \sigma_1}\cdot{\bf p_N})\,({\boldsymbol \sigma_2}\cdot{\bf p_N})}{4\,m_N^2}\,
 \left(\frac{f}{m_\pi}\right)^2  \,({\bf S}_1\cdot{\bf S}_2)\,
{\tilde D_\pi}(q_0,{\bf q})  \quad.
\eeqa
While $V_C$ scales with $\frac 1 3\,q_c^2\,\frac{{\bf p}_N^2}{4\,m_N^2}$, comparison with Eq. \ref{potd} shows that the meson exchange potential scales like ${\bf q}^2\,\frac{{\bf p}_N^2}{4\,m_N^2}$. Therefore the relative correction from $V_C$ is large (in fact as large as for $p$-wave states). The spin-structure of $V_C$ is the same as found in ${\cal L}_C^\pi$, Eq. \ref{cpi32m}, which in principle allows for the matching:
\beqa
\label{gprdmatch}
	g_\pi^d &=& \frac 1 3 \,q_c^2\,{\tilde F}_\pi^2\,{\tilde D}_\pi(q_0,{\bf q}) \quad.
\eeqa
Taking as a matching energy the solution of Eq. \ref{q0s}, a large value of about $0.6$ for $g_\pi^d$ results when the $D_{13}(1520)$ is considered.
However, this value has to be interpreted with some care since at these comparatively large energies one approaches an pole of ${\tilde D_\pi}$ when $q_0^2-{\bf q}^2-q_c^2=m_\pi^2$.
Such a pole already is found in scattering processes in the vacuum where the exchange particle can go on-shell, as detailed in \cite{barz,peierls,salcedo}. We therefore prefer to utilize the above mentioned similarity of $d$-wave potentials Eq. \ref{vcorrd} and the $p$-wave potentials Eq. \ref{vcorr3}, which suggests that the strength of the respective SRC should be comparable. Taking into account the uncertainties for $g_\pi^d$, we vary this parameter in the interval $(0,0.4)$. Note that due to the larger mass of the $\rho$ meson this problem does not arise when matching $g_\rho^s$.

Turning finally to $J^\pi=\frac1 2 ^-$ states, not much changes in the $\rho$ sector.
With the usual replacement ${\bf S} \rightarrow {\boldsymbol \sigma}$, we find exactly the same expressions as for $J^\pi=\frac3 2 ^-$ states, Eqs. \ref{vcorrs} and \ref{gprsmatch}. Here the state we are most interested in is the $S_{11}(1535)$ and we take the same value for $g_\rho^s$ as for $J^\pi=\frac3 2 ^-$ resonances. This is suggested from Eqs. \ref{gprsmatch}, \ref{q0s} and the fact that the $S_{11}(1535)$ and the $D_{13}(1520)$ have similar masses. We take this value also for all other $\frac 1 2^-$ resonances, thus reducing the amount of parameters. 
For the pion one runs again into the pole problem. We therefore take $g_\pi^s = 0.1$ as in the $\rho$ sector. As it will turn out, our results are not sensitive to this parameter, since there are no $s$-wave states with a large coupling to $N\,\pi$ and consequently the effects from the contact interactions are small regardless of the precise value for $g_\pi^s$. In our model only $J^\pi=\frac12^-$ states couple to the $\eta$ meson. For the corresponding short-range parameter $g_\eta^s$ we take in accordance to the pion a value of $g_\eta^s=0.1$.


\subsection{Mixing}
\label{mixing}

Up to now we have considered processes of the type $R N \rightarrow N R$. An interesting question is whether the contact interactions allow for mixing of different resonance states.
On the level of Feynman diagrams, this corresponds to a situation where $R_1$ and $R_2$ in Fig. \ref{resholegpr} resemble different resonances. The terms driving these processes are determined by the Lagrangian:
\beqa
	{\cal L}_C^\pi &=& c_\pi\,J^\mu_{R1}\,J_{\mu,R2} \\
	{\cal L}_C^\rho &=& c_\rho\,B^{\mu\nu}_{R1}\,{B_{\mu\nu,R2}} \nonumber \quad.
\eeqa
Clearly, processes like $R_1 N \rightarrow N R_2$ are possible if $R_2$ has the same quantum numbers as $R_1$. The situation is more complicated if resonance $R_2$ has different quantum numbers. Then for pions mixing is allowed in the non-relativistic reduction, if the leading terms of both $J^\mu_{R1}$ and $J^\mu_{R2}$ are derived either from the vector $(i)$ or the time $(0)$ component. Similarly, in ${\cal L}_C^\rho$ the leading terms of $B^{\mu\nu}_{R1}$ and $B^{\mu\nu}_{R2}$ must both be either the spatial $(j,k)$ or the time $(j,0)$, $(0,k)$ components.

For the pion sector, it follows that mixing is allowed between $J^\pi=\frac 3 2^+$, $J^\pi=\frac 3 2^-$ and $J^\pi=\frac 1 2^+$ states. In all three cases the leading non-relativistic contributions come from the vector components of $J_\mu$. In contrast, there is no mixing to $J^\pi=\frac 1 2^-$ states, which derive their leading behaviour from the 0-th component of $J_\mu$. For the remainder of this work the most important finding is the possibility that nucleon, $P_{33}(1232)$ and $D_{13}(1520)$ states can mix. 

In the $\rho$ sector the situation is slightly different. For positive parity states the leading components come from the spatial $(j,k)$ components and for negative states from the
time  $(j,0)$ or $(0,k)$ components of $B^{\mu\nu}$, such that the contraction vanishes if states of different parity are considered. This implies, that unlike the pion case, there is no mixing of the $D_{13}(1520)$ state to the $P_{33}(1232)$ or the nucleon. However, mixing between states of the same parity is possible and taken into account in this work.

The mixing of negative parity and positive parity states in the pion sector is proportional to the momentum ${\bf p}_N$ as is evident from Eq. \ref{lagconmix}. This is to be expected from parity conservation: consider a scattering process of a $P=+1$ state and a nucleon into a $P=-1$ state and a nucleon. The switch in the internal parities of the involved states must then be compensated by an odd angular momentum, leading to the observed momentum dependence. 
\end{appendix}

\bibliography{reference}
\bibliographystyle{h-elsevier3}

\end{document}